\documentclass[%
 aip,
 amsmath,amssymb,
 reprint,%
]{revtex4-1}

\usepackage{graphicx}
\usepackage{dcolumn}
\usepackage{bm}

\usepackage[utf8]{inputenc}
\usepackage[T1]{fontenc}
\usepackage{mathptmx}
\usepackage{etoolbox}

\makeatletter
\def\@email#1#2{%
 \endgroup
 \patchcmd{\titleblock@produce}
  {\frontmatter@RRAPformat}
  {\frontmatter@RRAPformat{\produce@RRAP{*#1\href{mailto:#2}{#2}}}\frontmatter@RRAPformat}
  {}{}
}%
\makeatother

\usepackage{bm}
\usepackage[version=4]{mhchem}

\usepackage{graphicx} 
\usepackage{subfig}
\usepackage{multirow}

\usepackage{array,booktabs}
\usepackage{float}
\usepackage{caption}
\usepackage[none]{hyphenat}
\usepackage{textcomp}
\usepackage{gensymb}
\usepackage{wrapfig}
\usepackage{xcolor}

\usepackage{relsize}
\usepackage{filecontents}

\usepackage{xr}
\makeatletter
\newcommand*{\addFileDependency}[1]{
  \typeout{(#1)}
  \@addtofilelist{#1}
  \IfFileExists{#1}{}{\typeout{No file #1.}}
}
\makeatother
\newcommand*{\myexternaldocument}[1]{%
    \externaldocument{#1}%
    \addFileDependency{#1.tex}%
    \addFileDependency{#1.aux}%
}
\myexternaldocument{anc/anc1}

\begin{document}

\title{Building Robust Machine Learning Models for Small Chemical Science Data: The Case of Shear Viscosity}
\author{Nikhil V. S. Avula}
\altaffiliation[Authors for Correspondence: ]{ Email: nikhil@jncasr.ac.in, bala@jncasr.ac.in}
\affiliation{Chemistry and Physics of Materials Unit, Jawaharlal Nehru Centre for Advanced Scientific Research, Bangalore 560064, India}

\author{Shivanand K. Veesam}
\affiliation{Chemistry and Physics of Materials Unit, Jawaharlal Nehru Centre for Advanced Scientific Research, Bangalore 560064, India}

\author{Sudarshan Behera}
\affiliation{Chemistry and Physics of Materials Unit, Jawaharlal Nehru Centre for Advanced Scientific Research, Bangalore 560064, India}

\author{Sundaram Balasubramanian}
\altaffiliation[Authors for Correspondence: ]{ Email: nikhil@jncasr.ac.in, bala@jncasr.ac.in}
\affiliation{Chemistry and Physics of Materials Unit, Jawaharlal Nehru Centre for Advanced Scientific Research, Bangalore 560064, India}

\date{\today}

\begin{abstract}
Shear viscosity, though being a fundamental property of all liquids, is
computationally expensive to estimate from equilibrium molecular dynamics
simulations. Recently, Machine Learning (ML) methods have been used to augment
molecular simulations in many contexts, thus showing promise to
estimate viscosity too in a relatively inexpensive manner. However, ML methods
face significant challenges - like overfitting, when the size of the data set
is small, as is the case with viscosity. In this work, we train seven ML models
to predict the shear viscosity of a Lennard-Jones (LJ) fluid, with particular
emphasis on addressing issues arising from a small data set.
Specifically, the issues related to model selection, performance
estimation and uncertainty quantification were investigated. First, we show
that the widely used performance estimation procedure of using a single unseen
data set shows a wide variability - in estimating the errors on – small data
sets. In this context, the common practice of using Cross validation (CV) to
select the hyperparameters (model selection) can be adapted to estimate the
generalization error (performance estimation) as well. We compare two simple CV
procedures for their ability to do both model selection and performance
estimation, and find that k-fold CV based procedure shows a lower variance of
error estimates. Also, these CV procedures naturally lead to an ensemble of
trained ML models. We discuss the role of performance metrics in training and
evaluation and propose a method to rank the ML models based on multiple
metrics. Finally, two methods for uncertainty quantification - Gaussian Process
Regression (GPR) and ensemble method - were used to estimate the uncertainty on
individual predictions. The uncertainty estimates from GPR were also
used to construct an applicability domain using which the ML models provided
more reliable predictions on another small data set generated in this work.
Overall, the procedures prescribed in this work, together, lead to robust
ML models for small data sets.
\end{abstract}

\maketitle 

\section{Introduction}
Shear viscosity is a fundamental transport property of all fluids.\cite{tosi_march2002} 
Understanding its molecular underpinnings would advance
our scientific understanding of supercooled liquids,\cite{egami_prl} magma
transport,\cite{dingwell_epslett_2008} mixing of fluids, etc.
For example, a good estimate of shear
viscosity is crucial to model the earth’s outer core which is believed to be a
liquid form of Iron based alloys.\cite{first_aimd_visc_1998,lidunka_2007}
However, in the absence of direct measurements, its estimates from different
methods differ by about fourteen orders of magnitude \cite{secco_1995}. Hence,
a better understanding of the behavior of viscosity from simulations can help
address some of these issues. Further, from the point of view of applications,
predicting the viscosity of industrially relevant fluids (such as hydrocarbons
and carbonates) in-silico would accelerate the progress in energy storage,
petroleum, lubricants, chemical processing, pharmaceutical, and many other
sectors.\cite{hydrocarbon_viscosity_fuel_2018,industry_transport_iecr_2021} 

\subsection{Viscosity from Molecular Dynamics simulations}

Atomistic Molecular Dynamics (MD) simulations with \textit{ab initio} or empirical force
fields can be used to estimate the viscosity of any liquid, however complex,
\textit{in silico}.
\cite{livejournal,berkhess2002,earthcore_prl,vlugt_finiteD_visc_2018,lidunka_2021,baroni_npjcomput_2022,turq_JPCM_2012,Maginn_moltensalts_JCP_2020,karniadakis_PNAS_2011}
While there exists many methods to estimate viscosity from MD simulations, they
largely fall into two categories – Equilibrium MD (EMD)
\cite{maginn2015,livejournal} and Non-Equilibrium MD (NEMD) based methods.\cite{NEMD_plathe_PRE_1999,NEMD_Heyes_Friction_2018,NEMD_Heyes_JCP_2018} A
comparison between them is beyond the scope of this work and the readers are
directed to some excellent works in this area.\cite{berkhess2002,livejournal}

Despite the progress in this area,\cite{berkhess2002,stillinger_JPCB_2005,mandadapu_JCP_2012,borodin_jcc_2015,maginn2015,oliveira_PRE_2017,livejournal,heyes_JCP_2019,heyes_JCP_2020,heyes_JCP_2021,heyes_JCP_2022,baroni_npjcomput_2022}
the state-of-the-art methods to estimate viscosity accurately from MD
simulations require huge computing time especially for viscous fluids,
\cite{maginn2015,bala_JCTC_2021,highvisc_pisarev_FPE_2021,highvisc_lopes_JCTC_2021}
as it is a collective quantity. This drawback precludes the use of MD
simulations in viscosity-based high throughput screening processes in the
industry.\cite{hydrocarbon_viscosity_fuel_2018} Also, force field refinement
strategies which use the experimental viscosity as a benchmark require
significant effort for the same reason.\cite{teamff_jced_2019} Another
important difficulty in estimating viscosity from MD is its sensitive
dependence on primary and ancillary simulation setup parameters. Hess showed
that ancillary MD run parameters such as the number of independent replicas,
numerical precision of the MD engine, neighbor list cutoff, and Ewald sum
related parameters have a significant effect on the estimated viscosity value.\cite{berkhess2002}
Hence, reporting meaningful confidence intervals of
viscosity estimates is crucial and is a topic of ongoing research.\cite{berkhess2002,livejournal,kim_JCP_2018,mandadapu_JCP_2012,borodin_jcc_2015}
To address these problems related to estimating shear viscosity from MD
simulations, we look at alternate approaches using Machine Learning (ML)
methods. 

\subsection{Machine Learning methods}

Recently, Machine Learning and Deep Learning (DL) models have started to
augment various aspects of MD simulations.\cite{MLMD_tiwary_review_2020,MLMD_franknoe_annrev_2020,deva_JCS_2021,MLMD_slowmodes_2021,MLMD_torchMD_2021}
More specifically, in the context of using ML methods to predict slowly
converging properties of liquids - of which shear viscosity is one - some
initial advances have been made. Alam \textit{et al}.  used Neural Networks (NN)
and Random Forests (RF) to predict the self diffusion coefficients of particles
interacting via Lennard-Jones (LJ) interaction and showed them to be performing
better than empirical models.\cite{alam_jcp_2020} They also used ML methods to
predict the finite size corrections to self diffusion coefficients in binary LJ
fluids.\cite{alam_jpcl_2020} Also, several ML models were also used to predict
the experimental viscosity and ionic conductivity of ionic liquids
using only molecular features.\cite{pfaedtner_rsc_2018,welton_chemsci_2021,Valderrama_firstILvisc_2011,rani_ChemEngComm_2013,domanska_JCIM_2014,mowla_jmolliq_2017,habib_JMolLiq_2017,rampi_JCP_2022,greaves_JCP_2022}
However, to our knowledge, ML methods have not been used to predict shear
viscosity derived purely from MD simulations.

The protocols for developing and comparing ML models - especially for
chemical science applications - are not yet completely established.\cite{viswakarma_metrics} 
There exist a plethora of supervised learning
algorithms,\cite{bishop_book,Goodfellow_book} model selection rules,\cite{CV_survey_arlot_StatSurvey_2010,cawley_2010,CV_yang_JEcon_2015,modelselection_book1,goodacre_2018}
performance metrics,\cite{viswakarma_metrics,collopy_1992,gneiting_2011} and uncertainty
quantification methods
\cite{errorbars_muller_JCIM_2007,ulilssi_MLSciTech_2020,UQ_DL_green_JCIM_2020,UQ_review_ceriotti_JCP_2021,UQ_tavazza_ACSOmega_2021}
which are used to create ML models; yet, there are no clear protocols on which
combination should be chosen.  For example, among the many performance metrics
that are used to train and compare ML models, the best choice is still a matter
of debate.\cite{viswakarma_metrics,kolassa_2019,m4_competition}
Similar conclusions hold true for model selection,\cite{goodacre_2018} and uncertainty quantification
\cite{nogood_UQ_NN_coley_JCIM_2020} as well.

In this context, 'No-Free-Lunch' theorems by Wolpert and Mcready imply that there
is no single algorithm that has the best performance across all possible
optimization problems.\cite{nofreelunch1} These theorems when applied to ML
indicate that any single ML algorithm cannot be expected to perform well across
all possible ML tasks. Though there is still a debate on the applicability of
these theorems to practical ML problems, it is considered common knowledge that
the ML algorithm should be tailored to the specific task at hand.\cite{Goodfellow_book} 
Exploiting the idiosyncrasies of the data set can lead
to significant improvements in the performance of ML models. For example,
M$\ddot{\rm u}$ller \textit{et al}. were able to improve the performance (MAE) of ML models used to predict
atomization energies from 10 kcal/mol to 3 kcal/mol (70 \% improvement) by
exploring a number of ML techniques and molecular representations.\cite{muller_2013}
In this work, we use this approach to develop ML models
tailored to viscosity data set generated from equilibrium MD simulations. 

\subsection{ML models for small data}

As it is computationally costly to get reliable (including standard errors)
estimates of viscosity from atomistic MD, generating large data sets (like GDB-17 with
nearly 150 billion data points \cite{gdb_17}) is practically infeasible with
the current computing resources. The largest MD-derived viscosity data set (with
1061 data points) we could find was  the work of Vlugt \textit{et al}. in which the MD
computed viscosity was used to predict the box size corrections to diffusion
coefficients of particles in binary Lennard-Jones (LJ) systems.\cite{vlugt_jctc_2018}
Such small data sets pose unique challenges to the ML
methods \cite{smalldata_rodolphe_JACS_2022} - (1) they are hard to generalize, (2) they are susceptible to
overfitting, and (3) they tend to  underestimate the generalization error.\cite{cawley_2010,smallsample_cvfail} 
For example, Neural Networks (NN) which
generally outperform other ML models, struggle in the low data regime.\cite{kernel_vs_dl_chemsci2021,muller_2013,alam_jpcb_2021}
Also, Casson \textit{et al}., on surveying over 50 articles on machine learning
for autism, have shown that ML models tended to produce overoptimistic results
when the sample sizes are small.\cite{nestedcv_plosone} All these issues
exacerbate the reproducibility of the results which is already a fast growing
challenge in all fields using ML methods.\cite{walters_JCIM_2013,hicks_NatMeth_2021,reprod_crisis_narayanan_arxiv_2022}

In this work, we train seven ML models to predict the shear viscosity
of binary LJ fluids, with particular emphasis on addressing issues arising from
a small data set. Specifically, the issues related to model selection,
performance estimation, performance metrics, uncertainty quantification and
applicability domain were investigated. First, we show that the common
practice of estimating the performance of
the ML models on a single unseen data set shows wide variability for small data
sets. The consequences of using individual unseen data set on the
hyperparameter optimization landscape are demonstrated. Then, we compare
two simple CV procedures for their ability to do both the model selection and performance estimation together. We discuss the
role of performance metrics in selecting and evaluating ML models. We discuss
some general principles for comparing different metrics and use them to choose
a suitable set of metrics relevant to the viscosity data set. We propose a
holistic ranking method based on multiple metrics to choose the best performing
ML algorithms. To complement the traditional ML models, we train a
probabilistic model to capture the inherent uncertainty in the data set and
compare its performance with that of other models. The performance of the ML
models developed here is shown to be better than empirical models for
viscosity. Finally, the applicability domain of ML models is also constructed (and tested)
to assist the decision making of the end user. We believe that the techniques
adopted herein to train the ML models, combined with the uncertainty quantification
and applicability domain can lead to accurate, reliable and \textit{reproducible} ML models
to predict shear viscosity of binary LJ mixtures and can help researchers to develop ML
models for small data sets in general. Also, we hope that the detailed descriptions 
and the codes attached in the supporting information would help with the reproducibility of the results presented in this work.

The paper is organized as follows: the next section describes the background
theory and empirical evidence that aids in the discussion on model selection
and performance metrics. It is followed by sections on computational details and results.
The final section presents our conclusions and suggestions for developing ML
models for small data sets.


\section{Background}

\subsection{The structure of the problem} \label{sec:struct}

We assume that there exists a joint probability density $p(\mathrm{x},y)$ that
generated the data set.\cite{bishop_book,Goodfellow_book} Here, $\mathrm{x}$ is
a vector of input features and $y$ is the target variable, also called as the
label. In the context of shear viscosity prediction, the feature vector can be
constructed from quantities like  $x_1$ , $\sigma_2$ , $\epsilon_2$ ,
k$_{12}$, $\zeta$, $\rho^*$ and the target variable is the shear viscosity
$\eta$ (see section \ref{sec:comp-method-data}). We focus on the regression
task which aims to determine a function $f^*(\mathrm{x})$ that is an
\textit{optimal} representation of the data set. The sense in which the function
$f^*(\mathrm{x})$ is \textit{optimal} is often taken to be the one that
minimizes the expected loss (also called as \textit{risk}) $\mathbb{E}[L]$ (Eq
\ref{eq:exploss}). Most common ML models like Kernel Ridge Regression (KRR),
Support Vector Regression (SVR), Neural Network (NN), etc. fall under this
category.

\begin{equation} \label{eq:exploss}
    f^*(\mathrm{x}) = \arg \min_{f(\mathrm{x})} \mathbb{E}[L] = \arg \min_{f(\mathrm{x})} \iint L(y,f(\mathrm{x})) \, p(\mathrm{x},y) \,\mathrm{dx}\,\mathrm{d}y
\end{equation}

Where $L(y,f(\mathrm{x}))$ is the user-defined loss function. The choice of the
loss function has a direct relation to the kind of function $f^*(\mathrm{x})$ obtained.\cite{kolassa_2019}
The most common loss function, the squared loss, where
$L(y,f(\mathrm{x})) = (y-f(\mathrm{x}))^2$ yields the conditional mean
$\mathbb{E}_y[y|\mathrm{x}]$ (Eq \ref{seq:conditionalmean}) as the
$f(\mathrm{x})$.\cite{bishop_book} Discussion on other loss functions and the
consequent effect on the properties of $f^*(\mathrm{x})$ is presented in
section \ref{ssec:metrics}.

However, to compute the expected loss/risk, the underlying joint probability
density $p(\mathrm{x},t)$ has to be known which is hard to do in practice.
Hence, the expected loss is approximated by the empirical loss,
$\mathbb{E}_{emp}[L]$ 

\begin{equation} \label{eq:emploss}
    f^*(\mathrm{x}) = \arg \min_{f(\mathrm{x})} \mathbb{E}_{emp}[L] =  \arg \min_{f(\mathrm{x})} \left( \frac{1}{N} \sum_{i=1}^{N} L(y_i^{true},f(\mathrm{x_i})) \right )
\end{equation}

where, $N$ is the number of data points, $y_i^{true}$ are the target values
corresponding to the feature vector $\mathrm{x}_i$. Generally, the empirical
loss tends to be much lesser on the data set used to infer $f^*(\mathrm{x})$
(called the training set) than on new/unseen data set(s). This is because the
minimization of empirical loss (\textit{per se}) incentivizes the learning
machine to learn the peculiarities (like noise) of the particular training
data sample rather than the trends in the underlying model that generated that
data set.\cite{muller_intro_2001,bishop_book,Goodfellow_book} Hence, the goal
of the learning protocol should be to minimize the error on new/unseen
data set(s) called the generalization error. This phenomena of ML methods having
significantly lesser training error than the generalization error is called
overfitting and is especially relevant for models on small data sets.\cite{muller_intro_2001,bishop_book,Goodfellow_book}

The most common way to alleviate the problem of overfitting is to reduce the
complexity/capacity of the learning machine, thereby reducing its ability to
learn the noise associated with the training data sample. However, the
complexity should not be reduced to such an extent that the general trends in
the data are lost, resulting in underfitting. Hence, the ML model should choose
an \textit{optimal} complexity corresponding to the general trends in the data.
A popular method to control the complexity of the models is called
regularization in which a penalty term (called the regularizer, Eq
\ref{eq:regularizer}) which penalises complex models is added to the empirical
loss.\cite{muller_intro_2001,Goodfellow_book} The common forms of the
regularizer are based on the norm of the weights ($w$) of the model like -
$L^2$ norm (called as ridge regression or Tikhonov regularization), $L^1$ norm
(for example in LASSO model), or a combination of both  .\cite{Goodfellow_book}
We also note that there are many other regularization techniques that are
specific to Deep Learning (DL) methods such as - early stopping, dropout, soft
weight sharing, etc.\cite{dropout_2014}

\begin{equation} \label{eq:regularizer}
  f^*(\mathrm{x}) = \arg \min_{f(\mathrm{x})} J = \arg \min_{f(\mathrm{x})} \left( \frac{1}{N} \sum_{i=1}^{N} L(y_i^{true},f(\mathrm{x_i}))  + \sum_{j} \lambda_j \, \Omega_j(f) \right)
\end{equation}
Where $\Omega_j(f)$ are the regularizers and $\lambda_j$ are the parameters
that control the amount of regularization. Now that the ML models have a
mechanism to control the complexity through regularization, the natural next
step would be to choose the values of regularization parameters. This task
falls under the purview of model selection.\cite{Goodfellow_book} In the
following section, we discuss various model selection criteria and a closely
related topic of performance evaluation.

\subsection{Model Selection and Performance Evaluation} \label{sec:background-model-selection}
It is a common practice to distinguish the parameters of ML and DL models into
model parameters and hyperparameters.\cite{bishop_book,Goodfellow_book} The
model parameters are learnt during the training phase on the training data.
Examples of model parameters include - slope and intercept in linear
regression, coefficients of kernel expansion in kernel methods (like KRR),
weights of neurons in Neural Networks (NNs), etc. The hyperparameters are
generally the high level settings of ML algorithms which are either set by the
user or inferred during the model selection procedure. Examples of hyperparameters
include - regularization parameters, the degree of polynomial in polynomial
regression, choice of the kernel in kernel methods, choice of activation
function in NNs, number of neurons in NNs, etc.

The task of selecting the model with the \textit{optimal} complexity is
reduced to the estimation of values of hyperparameters; the criteria used
for such selection are called model selection criteria. As stated earlier, the
goal of ML models is to minimize the generalization error which is the average
error over \textit{all} unseen data. However, generalization error cannot be
obtained in most practical situations and hence estimators on finite data sets
are constructed to approximate it. The process of estimating the generalization
error by using estimators on finite data sets is called performance evaluation
and is a prerequisite for model selection. It is crucial to note that the error
estimates are obtained over finite data sets and hence depend on the size of the
data set, especially for small data sets.  A simple example of such an estimator
is the split sample estimator where the whole data set is split into two parts
(generally unequal) and the error is computed on the split that was not used
for training.\cite{cawley_2010} Split sample estimator is known to be unbiased
i.e., the average split sample error over multiple independent realisations of
unseen data asymptotically converges to the generalization error. Hence,
minimizing the split sample error can in principle reduce the generalization
error. However, it was recently shown that the unbiasedness \textit{per se} is
not as important as the variance of the estimator when it is used for model
selection.\cite{cawley_2010} When an estimator has high variance (occurs with
small data sets\cite{cawley_2010}) the value of the estimated error on any one
particular unseen data sample can be very different from the generalization
error; hence the hyperparameters that minimize the estimated error can be far
off from the \textit{optimal} ones. Cawley \textit{et al}. showed (on a synthetic
data set) that hyperparameters selected based on split sample estimators can
severely overfit or underfit the data.\cite{cawley_2010} In practice, the
users rarely have the capability of generating multiple independent
realizations of the data and hence the variance of the estimator plays a major
role. Therefore, for small data sets it is not considered a good practice to
estimate the error on a single realisation of the data set.\cite{cawley_2010}
In order to mitigate this problem, various cross-validation (CV) schemes are
generally used.

The core idea of k-fold cross-validation (CV) is to split the entire
data set into k equal disjoint sets, train the ML models on k-1 sets and
estimate the error on the remaining one set. This process is repeated k times,
each time with a different hold-out set.\cite{bishop_book,Goodfellow_book} The
average error over k folds is used as the estimate for the generalization
error. It is a common practice to use 5 or 10 folds during CV.\cite{Goodfellow_book}

The error estimates from k-fold CV are often used for model selection by
searching over the space of hyperparameters and choosing the one that yields
minimum CV error. But once the k-fold CV error is used to optimize the
hyperparameters, it is no longer unbiased.\cite{nestedcv_ex_2006,nestedcv_ex_2014,Goodfellow_book} 
Typically another
unseen data set (called the test set) is used to estimate the generalization
error of the models with optimized hyperparameters.\cite{Goodfellow_book}
Using a single realization of the test set, however, suffers from the high
variance issue discussed above. Nested cross-validation or double
cross-validation improves upon k-fold CV by doing performance evaluation and
model selection in two nested loops.\cite{cawley_2010,muller_2013,nestedcv_plosone,nestedcv_ex_2006,nestedcv_ex_2014}
The outer loop is used to estimate the generalization error and the inner loop
is used to select the hyperparameters. Also, we note that there are many
methods of splitting the data set into train/validation/test sets such as -
Monte-Carlo CV, bootstrapping, Kennard-Stone splitting, and combinations
thereof.\cite{goodacre_2018,Guyon_2006} Xu and Goodacre compared the
performance (in terms of their ability to predict the generalization error) of
various data splitting methods including k-fold CV, Monte-Carlo CV,
bootstrapping, etc. and found that a single best method could not be found
\textit{a priori} and suggest that the choice of the method should be tuned to
the kind of data (No Free Lunch again).\cite{goodacre_2018}

Finally, we note that model selection and performance evaluation are big and
unsolved challenges on small data sets.\cite{cawley_2010,Guyon_2006,Goodfellow_book,nestedcv_ex_2006,nestedcv_ex_2014,smallsample_cvfail,Robinson2020}
Guyon \textit{et al}. organized a
performance prediction challenge in which the participants (more than 100) were
asked to predict the generalization error on finite data sets of real world
importance like medical diagnosis, speech recognition, text categorization, etc.\cite{Guyon_2006} 
They observed that most submissions were overconfident about
their ML models i.e., their prediction of generalization error is less than the
true generalization error. They also noted that the performance of the ML
models truly improved in the first 45 days of the 180 day challenge after which
overfitting set in. It is now a common belief that when a data set is worked
upon repeatedly, even careful performance prediction protocols can result in
optimistic performance predictions over time.\cite{Goodfellow_book} 

\subsection{Performance Metrics for Regression}
In this section, we summarize some of the principles that can be used to choose
a relevant metric to the particular ML task at hand and also consider the
particular case of viscosity data set. Performance metrics are generally used in
two critical areas of ML model development workflow - model training and model
comparison. Though the choice of the metric can significantly alter the
\textit{kind} of ML model developed and consequently its real-world
performance, there is no clear consensus on this topic.\cite{viswakarma_metrics,kolassa_2019,collopy_1992} 
As is the case with model
selection criterion, there is no single best metric \textit{for model training} that can be used across all
ML tasks.\cite{collopy_1992} Further discussion on this topic can be found in
section \ref{ssec:metrics}. 

Another area in which loss functions are used in ML workflow is model
comparison, in which models are ranked based on their generalization
performance. Ideally, the generalization performance of ML models should also
be measured using the same metric used in their training phase.\cite{kolassa_2019} 
For example, an ML model trained by minimizing MSE should
be compared to other models using MSE generalization error. However, in many
cases, the choice of the loss functions cannot be controlled by the model
developers and hence it is difficult to choose just one metric to compare such
models. For example, Makridakis \textit{et al}.  use a weighted average of sMAPE
and MASE to compare the models in the M4 forecasting competition citing a lack
of agreement on the advantages and drawbacks of various metrics.\cite{m4_competition}
 Hence, it is generally recommended to report the
estimates of generalization error using multiple metrics.\cite{viswakarma_metrics,muller_2013,alam_jpcb_2021,collopy_1992} 
Also, given the proliferation of various metrics, it is important to choose the set of
metrics that are relevant to the ML task at hand and preferably containing
complementary information to each other. Armstrong and Callopy compared six
commonly used metrics and ranked them qualitatively (good,fair,poor) according
to five characteristics - reliability, construct validity, sensitivity, outlier
protection, and their relationship to decision making.\cite{collopy_1992} They
conclude that there is no single metric \textit{for model comparison} that can be considered the best in all
situations and that they should be selected based on the kind of data set. 

We use some of the arguments presented in their work to identify metrics suitable to the viscosity data set. See section
\ref{sec:comp-method-data} for a discussion about the characteristics of the
viscosity data set used in this study. First, we look at the compatibility of
metrics to a data set that spans many orders of magnitude. All metrics that have
units i.e., are not scaled, tend to be dominated by the error from the highest
order of magnitude and hence do not give information about the contributions of
the errors from low orders of magnitude.\cite{collopy_1992} Metrics based on
scaled error like MAPE are more suited to such a situation. Next, we look at
the level of outlier protection of various metrics. All metrics that take an
average of individual errors suffer from outlier problem because the mean
itself is sensitive to large outliers. Median based metrics like MedAE are
better suited to such a situation. However median based metrics are not
sensitive to small changes in the errors and also do not have clearly defined
gradients with respect to model parameters.  Finally, we look at metrics that
can capture systematic biases (over or underestimation) in the ML models.
Metrics based on error function with strictly positive range like SE, AE, APE, 
etc., cannot distinguish between systematic over or under prediction by the ML
models. Metrics based on Mean Error (ME) or Mean Percentage Error (MPE) can be
used to gauge the bias in the models. Therefore, we rank the ML models
developed in this work based on the following metrics - MSE, MAE, MAPE, MedSE,
MedAE, MedAPE, ME, MPE, MedE, MedPE, and R$^2$ (coefficient of determination).

\section{Computational Methods}

\subsection{Data} \label{sec:comp-method-data}
Viscosity is one of the few properties that can span many orders of magnitude
($>$10), depending on the complexity of the system and the thermodynamic
conditions. In this work, we restrict ourselves to studying systems with simple
interaction parameters (Lennard-Jones only). This has the twin advantage that
the liquid part of the phase diagram is well understood and also being a simple fluid, the viscosity
computation is relatively easy. However, even for such simple systems, a
consistent data set with a large number (several thousand) of systems is not
yet available in the literature. In the absence of a coherent data set, smaller
data from multiple sources is generally collated to build a larger data set.
However,  due to the sensitivity of viscosity (especially the confidence
interval) to the ancillary MD parameters, this procedure can result in
unreliable models.

\begin{figure}[h!]
\centering
    \includegraphics[width=3in]{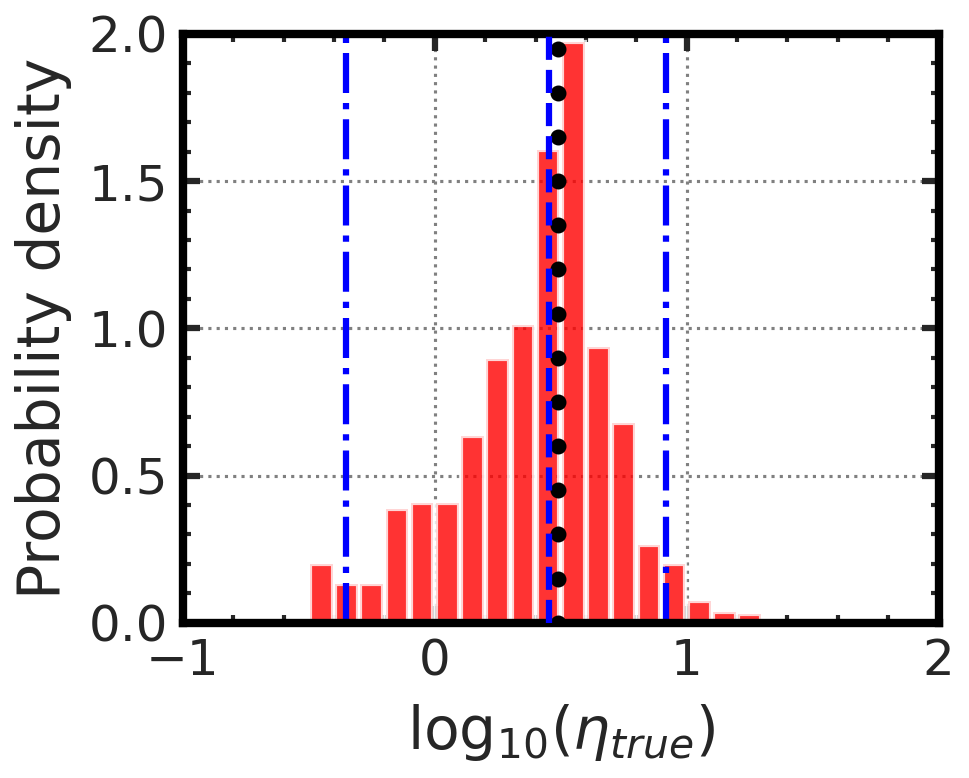}
    \caption{\textbf{Data Distribution:} The distribution of the viscosity ($\eta_{true}$) 
    from the Vlugt data set~\cite{vlugt_jctc_2018} across decades of viscosity. The blue 
    vertical dashed line represents the median and the two blue 
    vertical dash-dot lines represent the 95 percentile range around the 
    median. The black dotted line represents the mean of the data. }
\label{fig:viscdist}
\end{figure}

\subsubsection{Vlugt data set} \label{sec:comp-method-vlugt-data}

Recently, Vlugt and co-workers simulated 250 binary Lennard-Jones fluids to
study the system size dependence of the self-diffusion coefficients.\cite{vlugt_jctc_2018} In order
to test the analytical expression for the system size corrections to the
self-diffusion coefficients, they also computed the viscosity using the
Einstein-Helfand equation.\cite{allen2017,grest_jcp_1997,livejournal} We found that their work reported the largest
consistent data set that used multiple long independent runs to compute
viscosity and crucially, its confidence interval. 
In view of these attributes, our ML models were built using
this data set.

The data set contains a total of 1061 points, all of them at the same
temperature and pressure of 0.65 and 0.05 respectively. All the quantities are
reported in dimensionless units with interaction parameters of the first
component as the base units i.e., $\sigma_1$ = 1, $\epsilon_1$ = 1 , mass$_1$ = 1. The
state space is spanned by varying three interaction parameters ($\sigma_2$,
$\epsilon_2$, k$_{12}$) and one compositional parameter (X$_1$). We call these parameters
as 'pre-MD' features to be consistent with ML nomenclature where the independent
variables are called as features. Further, each state point was studied at four
different system sizes (quantified in terms of the total number of particles in
the simulation box). In sum, about 250 state points were simulated, each at
four different system sizes, giving a total of about 1000 data points. Unlike the 
self diffusion coefficient, shear viscosity does not have a strong dependence on 
system size.\cite{kim_JCP_2018,kim_JCP_2019,yeh_hummer_JPCB_2004,turq_JPCM_2012,steinhauser_JCP_2012,vlugt_finiteD_visc_2018,Maginn_moltensalts_JCP_2020,petravic_JCP_2004} 
Hence, we use only the data points at the system size of 2000 particles 
(273 out of 1000 data points) to develop the ML models in this work.

In the raw data, viscosity values span four decades, from 10$^{-1}$ to 10$^3$, but
only two data points had a value greater than 20. These two
data points ($\eta_{true}$ $>$ 20) were identified as outliers and were not considered
during the ML modeling. Figure \ref{fig:viscdist} shows the distribution of data points by their
viscosity values indicating that most of the data points are populated around
the mean viscosity value of around 3 and the extremal decades are sparsely
populated. Due to the uneven distribution of viscosity values across decades,
the models trained subsequently can be biased towards values around the mean. The
distribution of the standard error relative to the corresponding mean is shown
in Figure \ref{sfig:viscerrdist}. The relative standard error seems to be uncorrelated to the
viscosity value itself indicating that the data across decades is of similar
quality. The standard error on the mean was used to calculate the
irreducible minimum value of various loss functions \cite{bishop_book} (also
called as the Bayes error \cite{Goodfellow_book}). The irreducible errors are
incurred by all non-probabilistic ML models because of their approximation of
the conditional density $\mathbb{E}_y[y|\mathrm{x}]$ by point estimates.\cite{bishop_book} 
The irreducible MSE is the average variance in the data.
The irreducible MAE and MAPE were estimated by sampling from a Gaussian
conditional density.\cite{bishop_book} We also note that, in general, metric
value obtained from the average standard error would be different from the
irreducible loss. For example, the MAPE value obtained from the average
standard error(\%) of the data is about 2\%, whereas the irreducible MAPE is
0.8\%. Unless explicitly mentioned, the metric values obtained from the average
standard errors are used to compare the corresponding metrics from the ML
models. Hence, we consider ML models with MAPE metric lower than 2 \% to be
successful models.

Furthermore, to get a preliminary understanding about the underlying
correlations in the data, viscosity is plotted against other features - X$_1$,
$\sigma_2$, $\epsilon_2$, k$_{12}$, box length, packing fraction ($\zeta$), and self-diffusion
coefficients (D1 and D2). These features can be divided into two sets - 
preMD and postMD features. As their names suggest, the preMD feature set 
consists of all those features that are fixed before running the MD simulation
and postMD features are obtained only after the MD simulations. In this case, 
there are four preMD features -  X$_1$, $\sigma_2$, $\epsilon_2$, k$_{12}$ and 
six postMD features - number density, $\zeta$ and the four preMD features.
As expected, self-diffusion coefficients are inversely correlated
to viscosity, consistent with the Stokes-Einstein relation. Apart from
self-diffusion constant, only the packing fraction seems to be well correlated
with viscosity, with higher packing fractions corresponding to higher viscosity.
Rest of the plots show a wide spread of viscosity values at any given
feature value indicating that no single feature can predict viscosity
accurately. See section \ref{ssec:data-features} for more details.

Two different sets of models, using postMD and preMD features respectively,
were developed for each ML algorithm. Unless otherwise mentioned, the results
presented are from the models developed using postMD features.  All the
features were scaled using Min-Max scaler before training the ML models. Also,
the logarithm of viscosity was used as the label. However, all the metrics
presented in the subsequent sections were computed on the untransformed
viscosity values.

\subsubsection{Interpolation data set} \label{sec:comp-method-vlugt-data}
In order to test the predictive performance of the ML models away from the
Vlugt data grid, a complementary data set called the interpolation data set was
created. As the name suggests, the data set is created in the interpolation
region of the preMD feature space of the Vlugt data set. The interpolation set
consists of a total of 17 points at several interpolation distances. We note
that the interpolation space is not entirely in the equilibrium liquid region
of the binary LJ phase diagram at the thermodynamic conditions
studied by Vlugt \textit{et al}.\cite{vlugt_jctc_2018} Hence, it is difficult to
generate a "representative sample" of the interpolation space, which is
required to obtain quantitative estimates of predictive performance. In this
context, the current data set of 17 points (though small) can be used to
understand the predictive performance in a qualitative sense. Details of the MD
simulation procedure used to create the interpolation data set are given in the
Supporting Information.

\subsubsection{Applicability Domain} \label{sec:comp-method-applicability-domain}
Given that the interpolation space is not entirely in the liquid region, the ML
models cannot be expected to perform well over the entire interpolation space,
especially far away from the Vlugt data grid. One way to tackle this issue is
to define an Applicability Domain (AD) within which the ML models are expected
to perform well.\cite{AD_similarity_simon_JCICS_2004,AD_QSAR_stepwise_jay_JCIM_2005,AD_GPR_muller_JCAMD_2007,AD_kernels_zell_JChemInfo_2010,AD_QSPR_review_IJMS_2020}
There are many methods to construct an AD and a detailed comparison is beyond
the scope of this work.\cite{AD_GPR_muller_JCAMD_2007,AD_kernels_zell_JChemInfo_2010,AD_QSPR_review_IJMS_2020}
The Applicability Domain (AD) used in this work is described in section
\ref{sec:results-uq}. The interpolation data set is divided into two parts
called In-AD and Out-AD based on whether the points fall within or outside the
AD respectively. 

\subsection{Machine Learning Models} \label{sec:comp-method-ml-methods}

\begin{figure*}
\centering
    \includegraphics[width=0.9\linewidth]{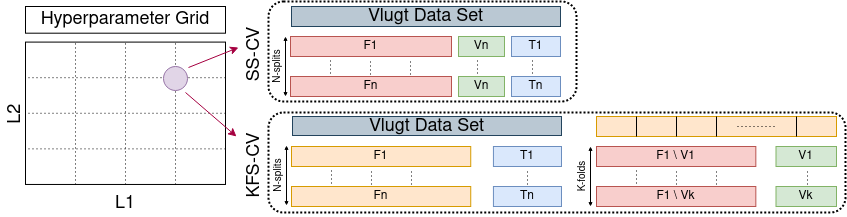}
    \caption{\textbf{Model selection and performance estimation:} A schematic
  representation of the two model selection procedures used in this study
  - Shuffle Split Cross validation (SS-CV) and K-fold Split Cross
  validation (KFS-CV). L1, L2 represent the hyperparameters to be tuned
  in the model selection procedure. F$_i$, V$_i$, T$_i$ represent the
  train, validation and test sets respectively. See section
  \ref{ssec:results-model-sel} for more details.} \label{fig:flowchart}
\end{figure*}

A total of seven ML models were tested for their ability to predict shear viscosity -
Kernel Ridge Regression (KRR), Artificial Neural Network (ANN), Gaussian
Process Regression (GPR), Support Vector Regression (SVR), Random Forest (RF), 
k-Nearest Neighbors (KNN), and Least Absolute Shrinkage and Selection Operator (LASSO).
In the current work, GPR is the probabilistic model (level 2 in section \ref{ssec:struct}) and 
all others are non-probabilistic in nature (level 2 in section \ref{ssec:struct}). 
Except ANN, all other the ML models used in this work were from the
scikit-learn implementation.\cite{scikit-learn2011} The ANN models were built
using the keras library in a python environment.\cite{KERAS2015} Many helper
functions from numpy,\cite{NUMPY2020} scipy,\cite{SCIPY2020} pandas
\cite{PANDAS2020} and scikit-learn \cite{scikit-learn2011} were also used in
the model construction, model selection and performance estimation steps.

\subsection{Model Selection and Performance Estimation} \label{sec:comp-method-model-sel}
In this work, we compare two popular model selection and performance estimation
methods called - Shuffle Split Cross validation (SS-CV) and K-fold Split Cross
validation (KFS-CV). A precise algorithmic description of these methods is
given in section \ref{ssec:comp-details-model-sel}. Briefly, SS-CV splits the
data set into three parts (named train/validation (val)/test) multiple times. Each
time the data is shuffled and hence independent random realisations of the data
can be obtained by SS-CV. KFS-CV is a two step procedure in which firstly
entire data is split into two parts (named train/test) and later the train part
is again split into k folds of roughly same size. The k folds (obtained in the
second step) are used to obtain validation score and the test sets are used to
do performance evaluation. Figure \ref{fig:flowchart} summarizes the two
procedures. Finally, we note that the procedures outlined above are referred to
by slightly different names in the literature
\cite{pfaedtner_rsc_2018,goodacre_2018,bishop_book,cawley_2010} and the ML
software packages.\cite{scikit-learn2011} Hence we recommend using the
algorithmic description of these methods given in section
\ref{ssec:comp-details-model-sel}. 

\subsection{Interpolation Grid}
The interpolation capabilities of the models can be qualitatively tested by
plotting the predicted viscosity values at the grid of interpolated feature
values. As the feature space is generally high dimensional (four in this case),
only projections onto 1D/2D sub-spaces can be visualized. To keep the
visualization uniform across the features, interpolation was done in the
scaled feature space i.e., after the min-max scaler is applied. The Vlugt data
set was generated at discrete values of each feature -  X$_1$ : (0.1, 0.3, 0.5,
0.7, 0.9) , $\sigma_2$ : (1.0, 1.2, 1.4, 1.6), $\epsilon_2$ : (1.0, 0.8, 0.6,
0.5), k$_{12}$ : (0.05, 0.0, -0.3, -0.6). However, the final data set consists
of only 250 unique combinations of preMD features as opposed to 320, had all 
the combinations been studied.

We have constructed four different interpolation grids to test the interpolation
capability across each feature individually. The interpolation grid for a
particular feature is generated at values uniformly spaced in the range of 0 to
1 while holding all other features at the values from the Vlugt data set. For
example, in order to generate the X$_1$ interpolation grid, 19 uniformly
distributed values between 0 to 1 were used for X$_1$, while the values for
$\sigma_2$,  $\epsilon_2$ and k$_{12}$ were taken from their scaled values in
the Vlugt data set. We also define the interpolation distance of any
interpolated point as its Euclidean distance from the nearest training data
point.

\section{Results and discussion}
In this section, we first present evidence related to aspects of model
selection and performance evaluation that are pertinent to ML models at the low
data regime. We then compare different ML models based on their performance
metrics and interpolation behavior. We finally compare the predicted
uncertainties of ensemble models and that of a probabilistic ML model (GPR).  

\subsection{Model Selection and Performance Estimation}

\subsubsection{Understanding the Hyperparameter Optimization Landscapes}

\begin{figure*}
\centering
    \includegraphics[width=0.9\linewidth]{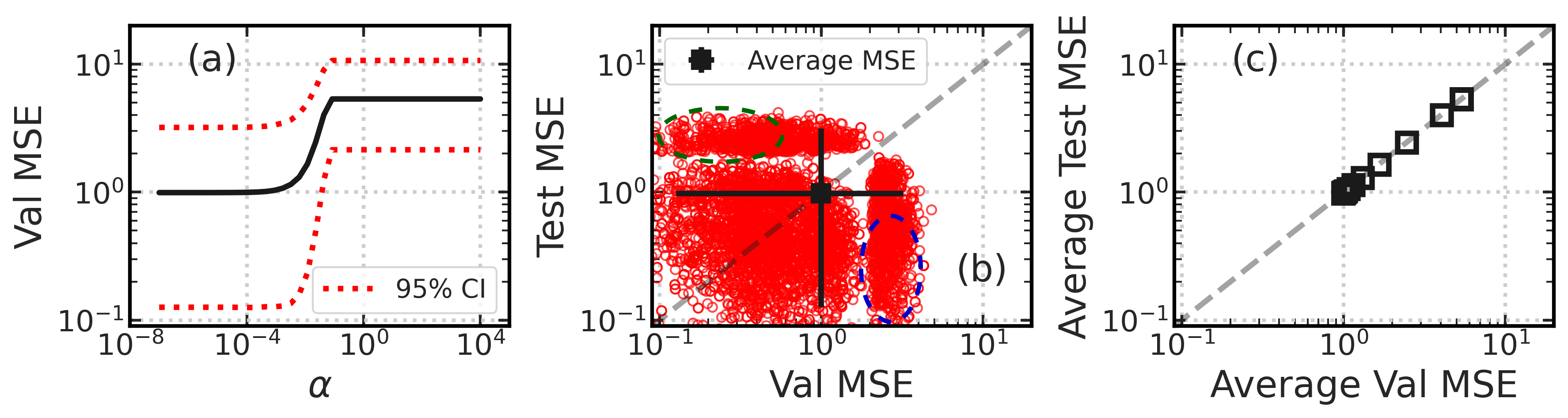}
    \caption{\textbf{Model selection:} (a) Average validation MSE plotted against the
  $\alpha$ hyperparameter of the LASSO model. The red dotted lines show
  the 95\% confidence interval of the validation MSE. (b) A scatter plot of the
  \textit{individual} test MSE and validation MSE at $\alpha=5*10^{-6}$. The black square shows
  the test MSE vs validation MSE averaged over the 5000 splits. The error
  bars correspond to the 95 percentile about the median of the respective MSE.
  The green and the blue ellipses are marked to indicate the data splits in which test MSE can be much greater or lower than validation MSE respectively.
  (c) Average test vs Average validation MSE at different values of the
  $\alpha$ hyperparameter of the LASSO model. The test and validation MSE
  were obtained using LASSO model and shuffle-split data sampling scheme
  (See Computational Methods).} \label{fig:modelsel-LASSO}
\end{figure*} 

In order to understand the optimization landscape of the hyperparameters, we
use LASSO and KRR models. They were chosen because they have less than three
hyperparameters and are hence conducive for the visualization of the
optimization landscape in 2D plots. Moreover, both models have analytic
solutions and are hence much faster when trained on small data sets. Both these
models were also studied in the context of model selection (albeit on synthetic
data sets) and hence allow a close comparison wherever possible.\cite{cawley_2010}

In this section, we elucidate the dependence of the performance of
these ML models on the particulars of the data splitting procedure, which is an
essential step for model selection and performance evaluation. The entire data
set is randomly split into three parts - train, validation (val) and test sets with
60/20/20 ratio. This procedure is repeated  $N_{split}$ times thereby creating
multiple random realisations of the train, validation and test splits. These
$N_{split}$ train sets are used to train $N_{split}$ ML models at each
hyperparameter value. The trained models are then evaluated on their
corresponding validation and test sets.  

Figure \ref{fig:modelsel-LASSO} (a) shows the \textit{average} test MSE of the
LASSO model across a wide hyperparameter ($\alpha$) range. Clearly, $\alpha$
values less than $10^{-3}$ are suited for the viscosity data set. However, no
single optimal hyperparameter can be selected as there is no discernible
change in the MSE values for $\alpha$ less than $10^{-3}$. A similar
"flat-minima" hyperparameter landscape was also observed by Pfaendtner \textit{et al}.
on an ionic liquid experimental viscosity data set.\cite{pfaedtner_rsc_2018}
Hence, strictly applying the common model selection criteria of selecting the
hyperparameter with the best performance on the validation set (in this case,
minimizing MSE) belies the flat-minima nature of hyperparameter landscape.
Also, the wide confidence interval around the average test MSE indicates that
there is significant variability in the performance (MSE) of the ML models
across different data splits.

\begin{figure*}
\centering
    \includegraphics[width=\linewidth]{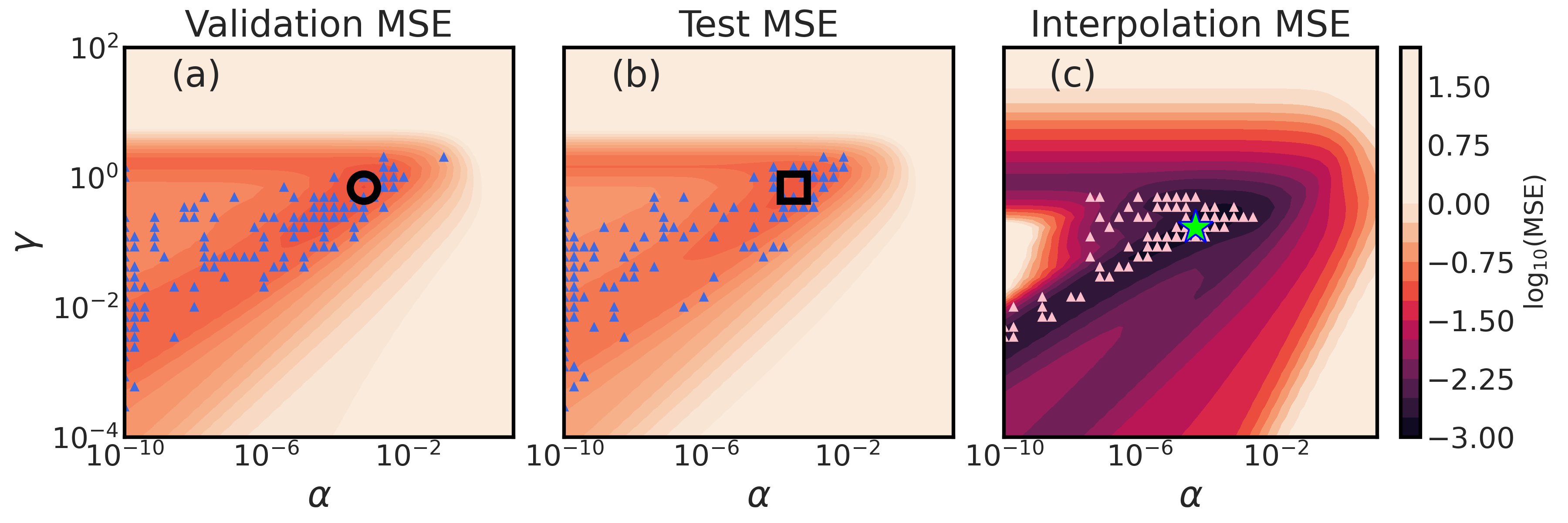}
    \caption{\textbf{Model selection:} 2D contour plots of (a) average
  validation MSE, (b) average test MSE and (c) Average Interpolation MSE
  of KRR model. The minima in the 2D contour plots are marked as follows
  -  black-unfilled-circle: average validation MSE, black-unfilled-square:
  average test MSE, green-filled-star: average Interpolation MSE. The
  blue triangles in (a) and (b) mark the optimal hyperparameters that minimize
  \textit{individual} validation and test MSE respectively. The pink
  triangles in (c) mark the optimal hyperparameters that minimize
  \textit{individual} Interpolation MSE. The validation, test, and
  Interpolation MSE were obtained using KKR model and shuffle-split data
  sampling scheme (See Computational Methods).} \label{fig:modelsel-KRR}
\end{figure*} 

Interestingly, the wide scatter of points in Figure \ref{fig:modelsel-LASSO}
(b) indicates that the \textit{individual} test MSEs are not correlated to
\textit{individual} validation MSEs. Hence model selection criteria based on
optimization of the performance of \textit{individual} validation sets need not
necessarily result in a good generalization performance. However, the average
validation MSE  (over $N_{split}$ splits) is perfectly correlated to the
average test MSE as shown in Figure \ref{fig:modelsel-LASSO} (c). Consequently,
the data splitting procedures that reserve a single "unseen" data set (often
called as test set \cite{domanska_JCIM_2014,alam_jpcb_2021}) for evaluating the
generalization performance should be discouraged as they suffer from wide
variability. We note that this problem is unique to small data sets and the
variability decreases rapidly with increase in data set size.\cite{cawley_2010} 

The same procedure was applied to the KRR model to elucidate its hyperparameter
optimization landscape. In addition to validation and test sets, the
performance of the KRR models was also evaluated on a single Interpolation set
containing 17 data points. Figure \ref{fig:modelsel-KRR} shows the 2D contour
plots of the average validation, test and Interpolation MSE over a wide range
of KRR hyperparameters $\alpha$ and $\gamma$. The landscapes of the average
test and validation MSE are similar, consistent with the LASSO results (Figure
\ref{fig:modelsel-KRR}). The Interpolation MSE landscape is slightly different from others
near the minima while still retaining the overall features. Importantly, it is
more rugged than the validation and test MSE landscapes, possibly because the
same Interpolation set was used across all the $N_{split}$ splits.  

\begin{figure}[h!]
\centering
    \includegraphics[width=\linewidth]{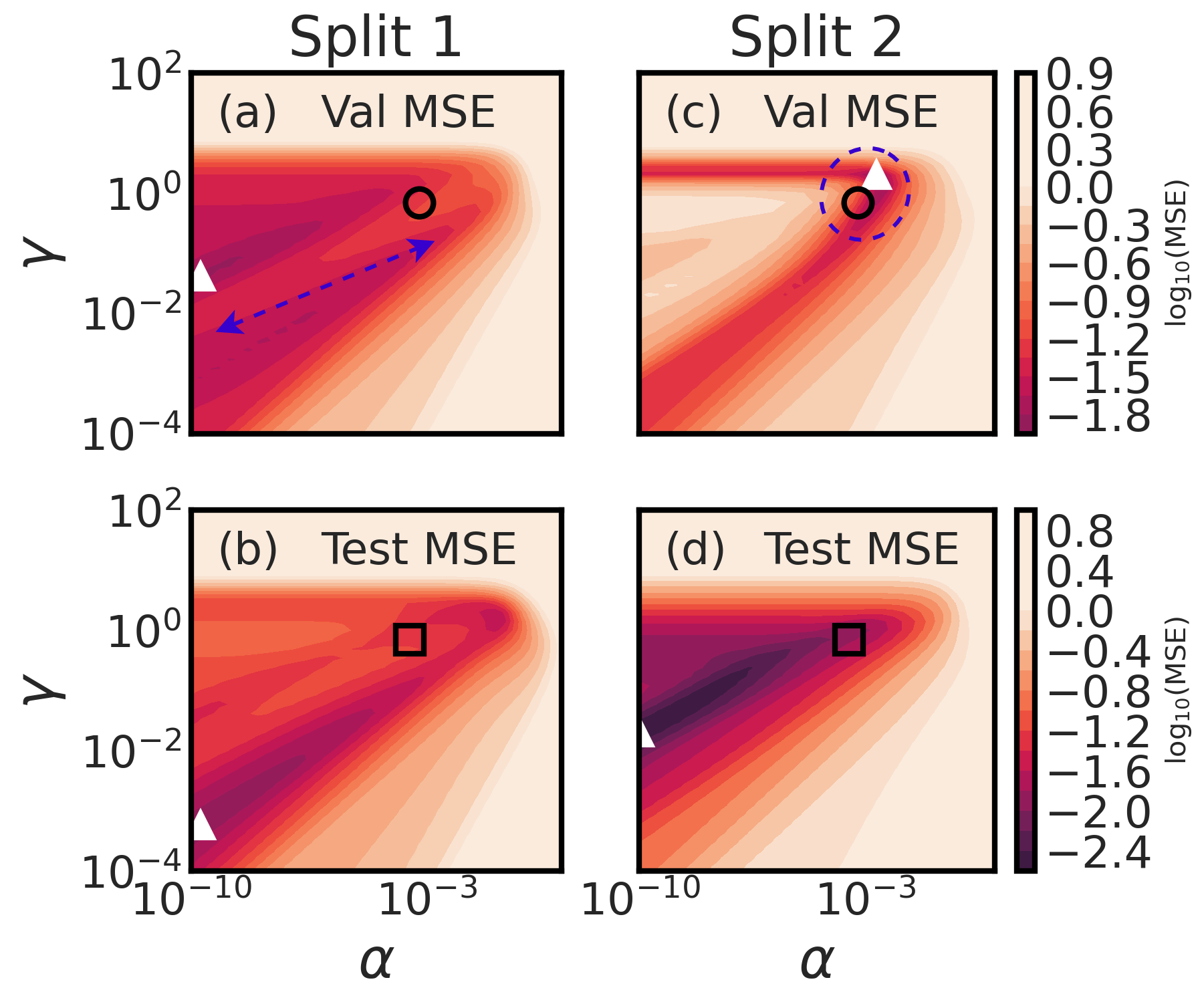}
    \caption{\textbf{Model selection:} (a) and (b) show 2D contour plots of
  the \textit{individual} validation and test MSE
  respectively of a randomly chosen realisation (called split 1) of the train-val-test
  split. (c) and (d) show the same for another random realisation (called split 2)
  of the train-val-test split. In all panels, the minima of the average
  MSE are marked as follows: black-unfilled-circle: average validation MSE,
  black-unfilled-square: average test MSE. The minima of these particular realisations are
  marked with a white triangle symbol. The blue dashed arrow in (a) is drawn as a 
  guide to the eye to highlight the large difference between the individual (white triangle)
  and the the average (black circle) minima. The blue dashed ellipse in (c) is drawn as a
  guide to the eye to indicate that the individual and the the average minima can be
  close by chance. The validation and test were obtained using KRR model and shuffle-split data 
  sampling scheme (See Computational Methods).} \label{fig:modelsel-KRR2}
\end{figure} 

 The wide scatter of points in Figure \ref{fig:modelsel-KRR} (a,b,c) show the
 optimal hyperparameter selected by minimizing the \textit{individual}
 validation, test and Interpolation MSE respectively. This demonstrates that
 using a single realisation of data split to model selection or performance
 evaluation can result in wide variability, again consistent with LASSO
 results. Cawley \textit{et al}. used KRR on a small synthetic data set to demonstrate
 this issue of wide variability when single realisations of data are used.\cite{cawley_2010} 
 Further, while the average MSE landscapes are smooth,
 those corresponding to individual realisations of the data splits are rugged
 as shown in Figure \ref{fig:modelsel-KRR2}. The validation and test landscapes
 of these individual splits also show significant differences. For example,
 Figure \ref{fig:modelsel-KRR2} (d-f) corresponds to validation, test and
 Interpolation landscapes of a randomly chosen realisation and their optimal
 hyperparameters vary by more than seven orders of magnitude.  
 
 These results demonstrate that there is a wide variability in choosing the
 optimal hyperparameter values (model selection) and also in estimating the
 generalization performance (performance evaluation) of the ML models on small
 data sets. Hence, it is crucial to do both the model selection and performance
 evaluation tasks by training an ensemble of models on different random splits
 of the data set. 
 
 \subsubsection{Comparison of CV Procedures}
 
 The model selection criterion is based on the \textit{average} validation score
 obtained from the CV procedures. We observe that the \textit{average} validation score 
 landscapes are almost identical in both SS-CV and KFS-CV irrespective of the kind 
 of metric used to construct the landscape (Figure \ref{sfig:ss-vs-kfs-mean}). 
 However, the two procedures differ in the variance
 of the validation landscape, with KFS-CV yielding a lower variance than SS-CV
 (Figure \ref{sfig:ss-vs-kfs-relerr}). Cawley \textit{et al}. show that the
 estimators with lower variance can do a better job of selecting the optimal
 hyperparameters.\cite{cawley_2010} Hence we use KFS-CV to do model selection
 and performance evaluation on rest of the ML models - SVR, RF, KNN, ANN.
 
 Also, in both the CV procedures, the variance of the validation landscapes was 
 strongly dependent on the error metric used to construct the landscapes. In general,
 we observe that MSE shows the highest variance followed by MAE, MAPE and R$^2$.
 The variance in MSE validation landscape is often so high ($> 100\%$), that 
 unambiguous selection of optimal hyperparameters is difficult. Hence, the use of 
 other metrics that that have lesser variance such as MAE, MAPE, R$^2$ can help rectify
 the issue. In this work, we use the MAE validation landscape to choose the optimal
 hyperparameter values. Tables in section \ref{ssec:results-hyperpar-sel} list the 
 optimal hyperparameters and the corresponding values of metrics for all the ML models studied in this work.

 Finally, the performance estimation is done by evaluating the trained models
 (with the optimal hyperparameters) on the test sets. As both CV
 procedures result in nearly identical performance scores, we use the KFS-CV method
 to do the performance estimation for the other ML models - SVR, RF, KNN, ANN. 
 In the following section, we compare the performance of
 various ML methods and rank them using multiple metrics.

\subsection{Model Comparison and Ranking}

\begin{figure*}
\centering
    \includegraphics[width=\linewidth]{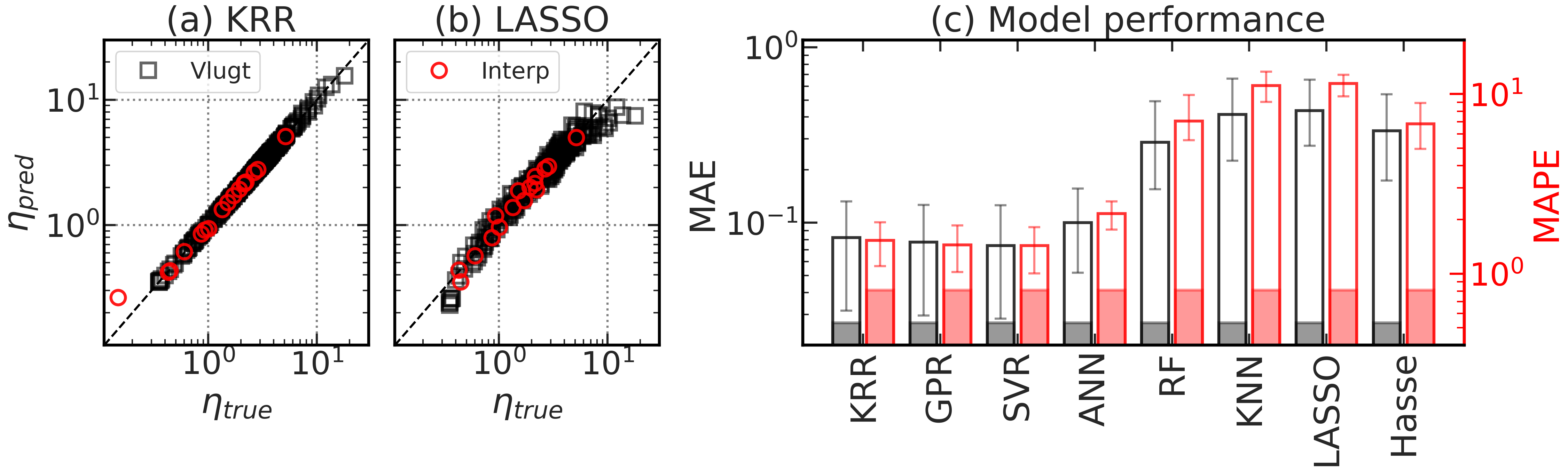}
    \caption{\textbf{Model performance:} (a),(b) Predicted vs the true viscosity 
    values from KRR and LASSO ensemble models respectively. The black squares are for the Vlugt data set
    and the red circles are for the interpolation data generated in this work. The black dashed line 
    represents the y=x line. (c) Comparison of the Mean Absolute Error (MAE) and
  Mean Absolute Percentage Error (MAPE) of all the seven ML models studied in this work.
  The performance of the empirical correlation by Hasse \textit{et al}. \cite{hasse_FPE_2019} is also shown for comparison.
  The black bars show the MAE and the red bars show the MAPE. 
  The error bars indicate the 95\% confidence interval region. 
  The filled regions of the bars indicate the irreducible errors 
  (0.027 for MAE and 0.8\% for MAPE) in the Vlugt data set
  (section \ref{sec:comp-method-vlugt-data}). See the computational methods
  section for model description and performance estimation.}
\label{fig:modelperf}
\end{figure*} 

The predicted viscosity values from all the models, except KNN and LASSO, agree
well with the true viscosity values both for test and interpolation data sets as
shown in Figure \ref{fig:modelperf}. The agreement is seen to be good across decades of
viscosity values indicating, at least to the naked eye (Figure
\ref{fig:modelperf} (a)), that models do not bias any particular decade of
viscosity values. A detailed discussion on the model bias is presented in the
Supporting Information (section \ref{ssec:model-bias}).
Figure \ref{fig:modelperf} (c) compares the MAE and MAPE of all
the models evaluated using the KFS-CV performance estimation procedure 
(section \ref{sec:comp-method-model-sel}). The MAPE of KRR, GPR,
ANN and SVR models are below the average standard error (\%) of the
data (called as threshold henceforth) and hence can be considered as successful
models. On the other hand, KNN and LASSO models have both their test and train MAPE much above the
threshold and can hence be considered as unsuccessful models. While the RF model
performs well, it is still considered unsuccessful, due to its
peculiar interpolation behavior (see discussion below). Also, the successful models outperform
the empirical model developed by Hasse \textit{et al}. for pure LJ fluids (see section \ref{ssec:emp-model}).\cite{hasse_FPE_2019}

\begin{figure}[h!]
\centering
    \includegraphics[width=\linewidth]{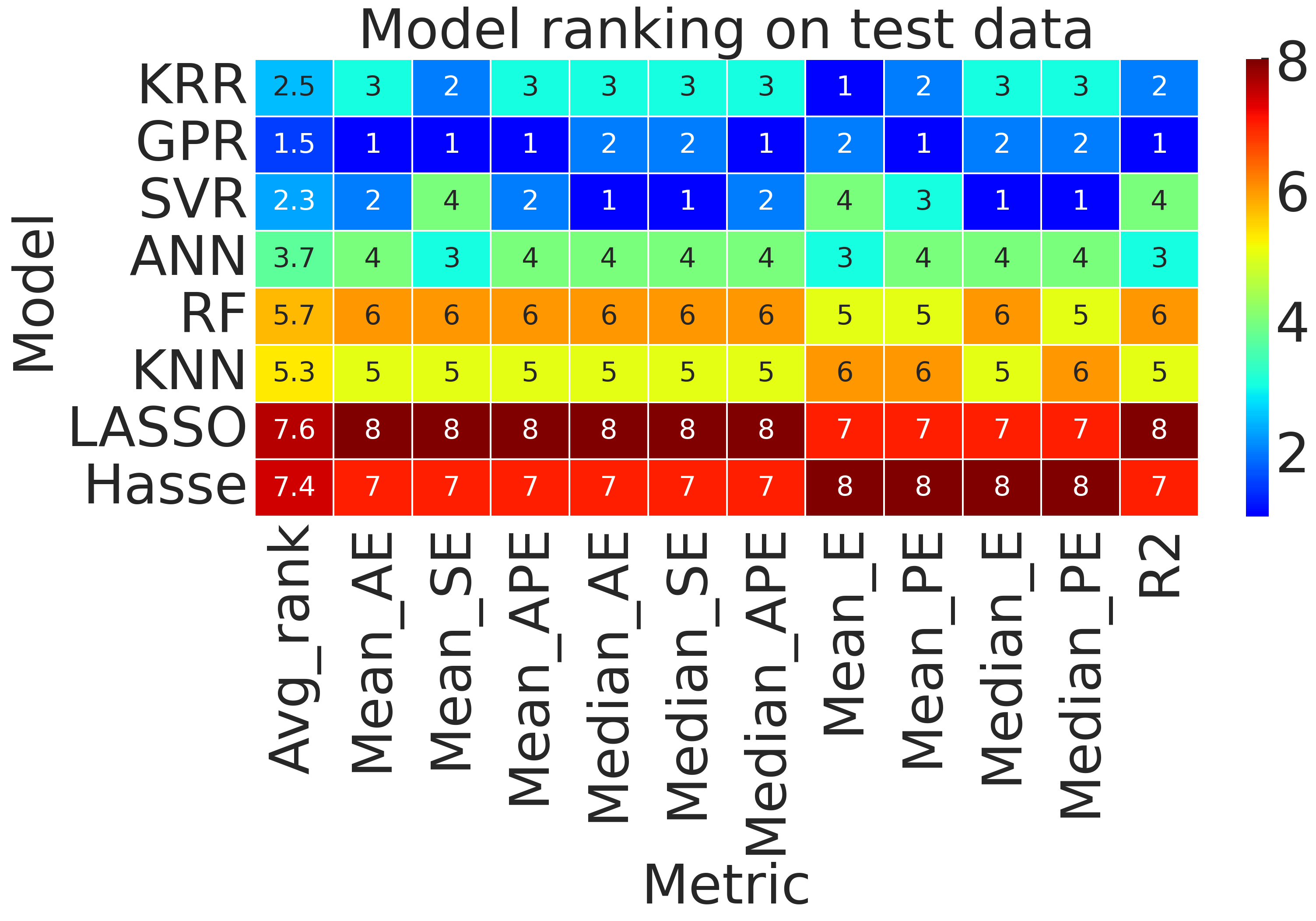}
    \caption{\textbf{Model ranking:} Grid plots of model's ranking on the entire 
  Vlugt data based on various metrics. The text and color of each square
    represent the relative ranks of models. Each row shows the ranks of the model
    labeled on the y-axis and each column corresponds to the metric based on
    which the ranking was done. The metric values are computed using the ensemble predictions
  of the ML models on the entire Vlugt data.\cite{vlugt_jctc_2018} Lower model rank is better.}
\label{fig:modelrank}
\end{figure} 

The MAE and MAPE of the successful models - i.e., KRR, GPR, ANN, and SVR - are
very close to each other as seen in Figure \ref{fig:modelperf} (c) and hence
more information is required to unambiguously rank them. In Figure
\ref{fig:modelrank}, we show the relative ranks of all the models based on seven
(MAE, MSE, MAPE, MedAE, MedSE, MedAPE and R$^2$)performance based metrics 
and four (ME, MPE, MedE, MedPE) bias based metrics. The mean, median values of
the Absolute Error (AE), Squared Error (SE), and
the Absolute Percentage Error (APE) along with the coefficient of determination (R$^2$) are the performance
metrics, while mean, median Error (E) and the Percentage Error (PE) constitute the bias
metrics. An average rank (averaging done across the metrics with uniform
weights) is also shown for each model. The average rank follows closely the
MAE rank with four successful models - KRR, GPR, SVR, and ANN - having an
average rank less than four and three unsuccessful models having a rank greater
than four. According to the average rank, GPR is the best performing model,
followed closely by SVR and KRR. These three models are followed by ANN, RF, KNN,
and LASSO respectively with the last two consistently ranked sixth and seventh.
Interestingly, there is considerable mixing of ranks based on MAE vs MSE, 
indicating that a holistic approach using a
combination of metrics needs to be used to objectively evaluate the models.
Another noticeable trend is the disconnect between the ranking based on
performance and those based on bias metrics. These findings highlight the need
to evaluate models based on metrics beyond the simple loss functions used to
train the models themselves to get a complete picture of the models accuracy,
bias and generalizability. Now that the models have been validated 
for performance and bias, we discuss their interpolation capabilities in the next section.

\subsubsection{Interpolation Behaviour}

\begin{figure*}
\centering
    \includegraphics[width=0.9\linewidth]{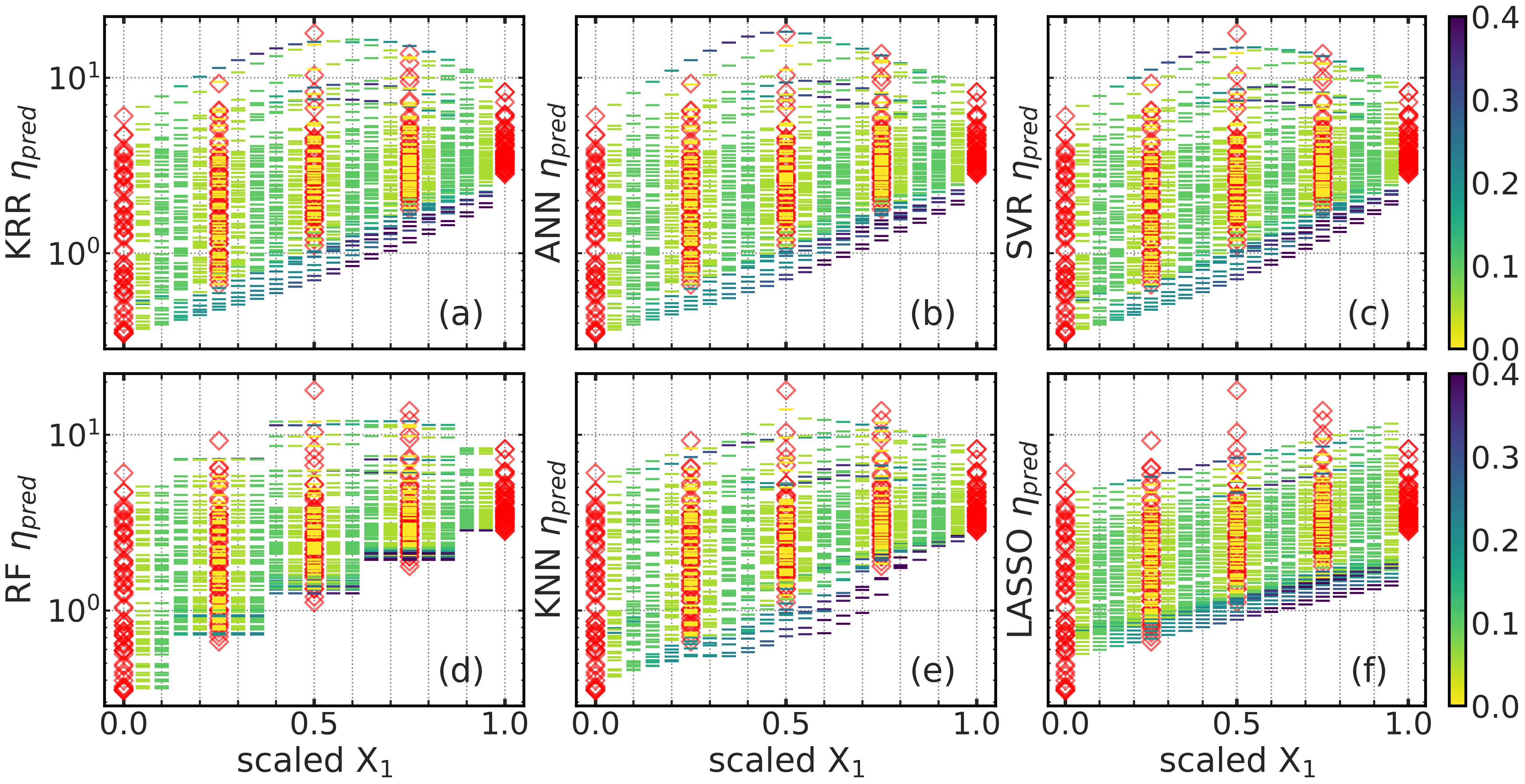}
    \caption{\textbf{Model interpolation:} (a) to (f) show the predicted
    viscosity values from the ensembles of KRR, ANN, SVR, RF, KNN, and LASSO 
  respectively in the X$_1$ interpolation range. Red diamonds are viscosity values from the Vlugt data set. The 
    color of the dashes indicate the distance from the nearest Vlugt data
    point in the scaled feature space (abscissa). See Computational Methods section for 
    details on feature scaling and the construction of the interpolation grid.}
\label{fig:modelinterp2}
\end{figure*} 

Figure \ref{fig:modelinterp2} shows the predicted viscosity values from KRR,
ANN, SVR, RF, KNN, and LASSO models at the interpolation points plotted against X$_1$ feature values. 
The color of the points (represented as dashes in the figure) indicates the distance
from the nearest training data point that the models have seen, with darker
shades being farther away. The viscosity values from the Vlugt data set
are also shown (as red diamond symbols) for comparison. KRR, SVR, and ANN models show a smooth variation as the
feature values move farther away from their corresponding values in the
Vlugt data set. On the other hand, RF and KNN models show sudden 
discontinuities in the predicted viscosity values at some specific X$_1$ values.
These discontinuities are probably due to the presence of decision boundaries in RF and
a sudden change of nearest neighbors in the case of KNN. Such sudden discontinuities are 
incompatible with viscosity which is expected to be continuous (at least as long as there
is no phase transition).
 
\subsection{Uncertainty Quantification} \label{sec:results-uq}
As demonstrated in the previous sections, the performance of ML models trained 
on small data sets can have wide variations. In this regard, models that can 
estimate the uncertainty on individual predictions can help alleviate this issue.
The uncertainty estimates can be used as a guide to end-users about the reliability 
of a given prediction and thus of the models. More generally,
uncertainty quantification has many other applications 
\cite{errorbars_muller_JCIM_2007,UQ_khorshidi_PCCP_2017,UQ_tavazza_ACSOmega_2021,UQ_DL_green_JCIM_2020,UQ_review_ceriotti_JCP_2021}
such as - setting the applicability domain of 
the ML models,\cite{AD_GPR_muller_JCAMD_2007} 
active learning for generating data on the fly,\cite{UQ_active_boris_NatCompMat_2020,UQ_activelearning_boris_arxiv_2022} etc.

\begin{figure}[h!]
\centering
    \includegraphics[width=0.9\linewidth]{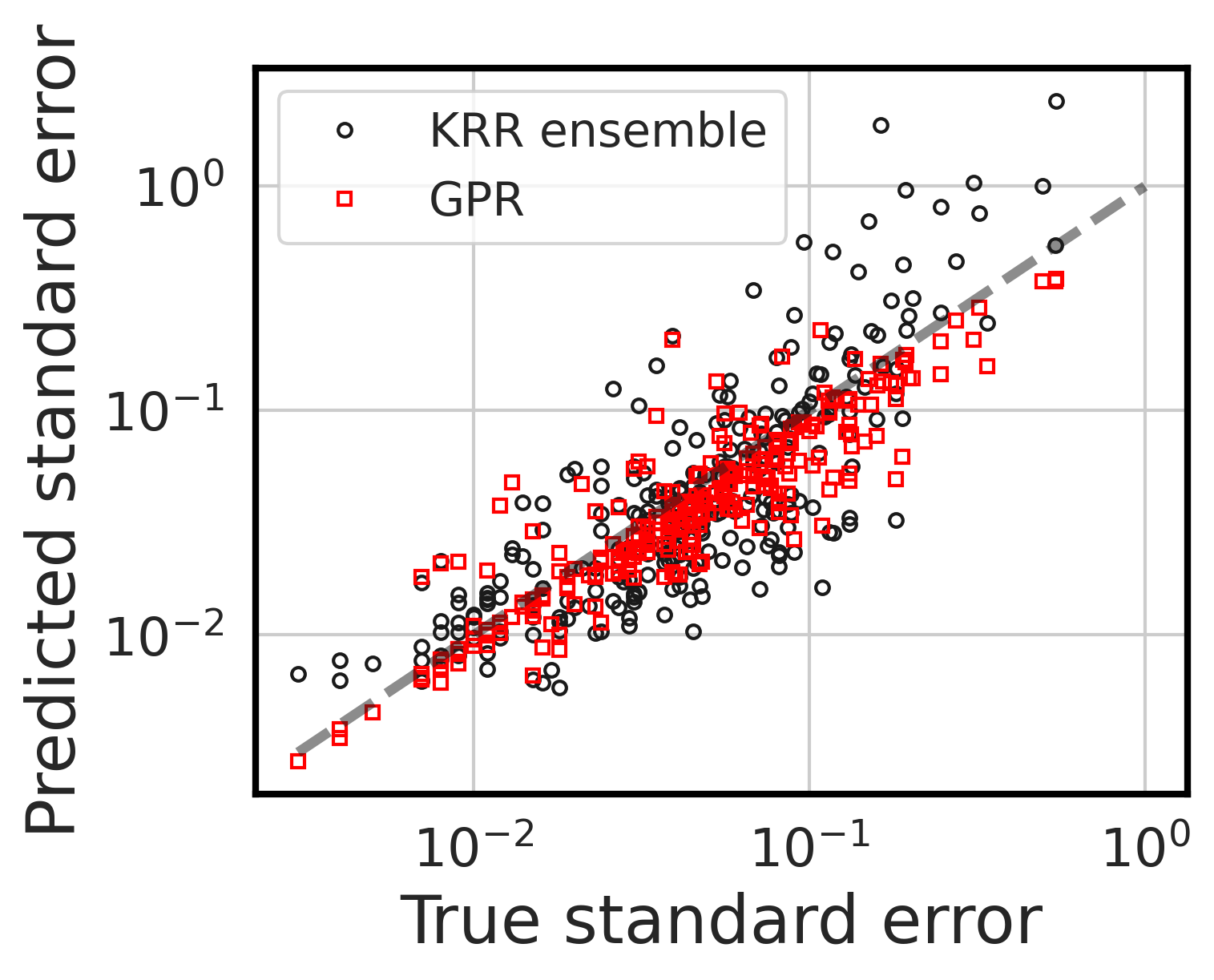}
    \caption{\textbf{Uncertainty Comparison:} The standard errors of the
  predicted viscosity values plotted against their corresponding true
  standard errors. The standard errors from the ensemble of KRR models
  are shown as black circles and those from the GPR model are shown as
  red squares. The grey dashed line corresponds to y=x and is drawn as a
  guide to the eye.} \label{fig:uq-compare}
\end{figure} 

In this work, we used two approaches to estimate the uncertainty -
probabilistic ML method and ensemble ML method. The probabilistic ML methods
inherently capture the uncertainty through their model architecture, whereas
the ensemble ML approach uses an ensemble of ML models on several random
realisations of the data. Gaussian Process Regression (GPR) was the choice of
probabilistic ML method due to its simplicity (few hyperparameters), wide
applicability and near universality. KRR was used to test the ensemble approach
again due to its simplicity (two hyperparameters), training speed (analytical
solution) and wide applicability. 

Figure \ref{fig:uq-compare} compares the standard error estimated from the ML
models and the true standard error of the data. Both GPR and ensemble KRR
methods show good agreement with the true data. The standard errors predicted
by GPR show a slightly better agreement with that of the data than KRR ensemble does,
especially at high error values. This is because the ensemble methods
capture the uncertainty in the data indirectly by repeated sampling from the
training data set. On the other hand, the standard error values of the data are
directly fed into the GPR training. A similar observation was made by M$\ddot{\rm u}$ller \textit{et al}.
on their comparison of GPR and ensemble methods to predict error bars on the solubility data.\cite{AD_GPR_muller_JCAMD_2007}

\begin{figure*}
\centering
    \includegraphics[width=0.8\linewidth]{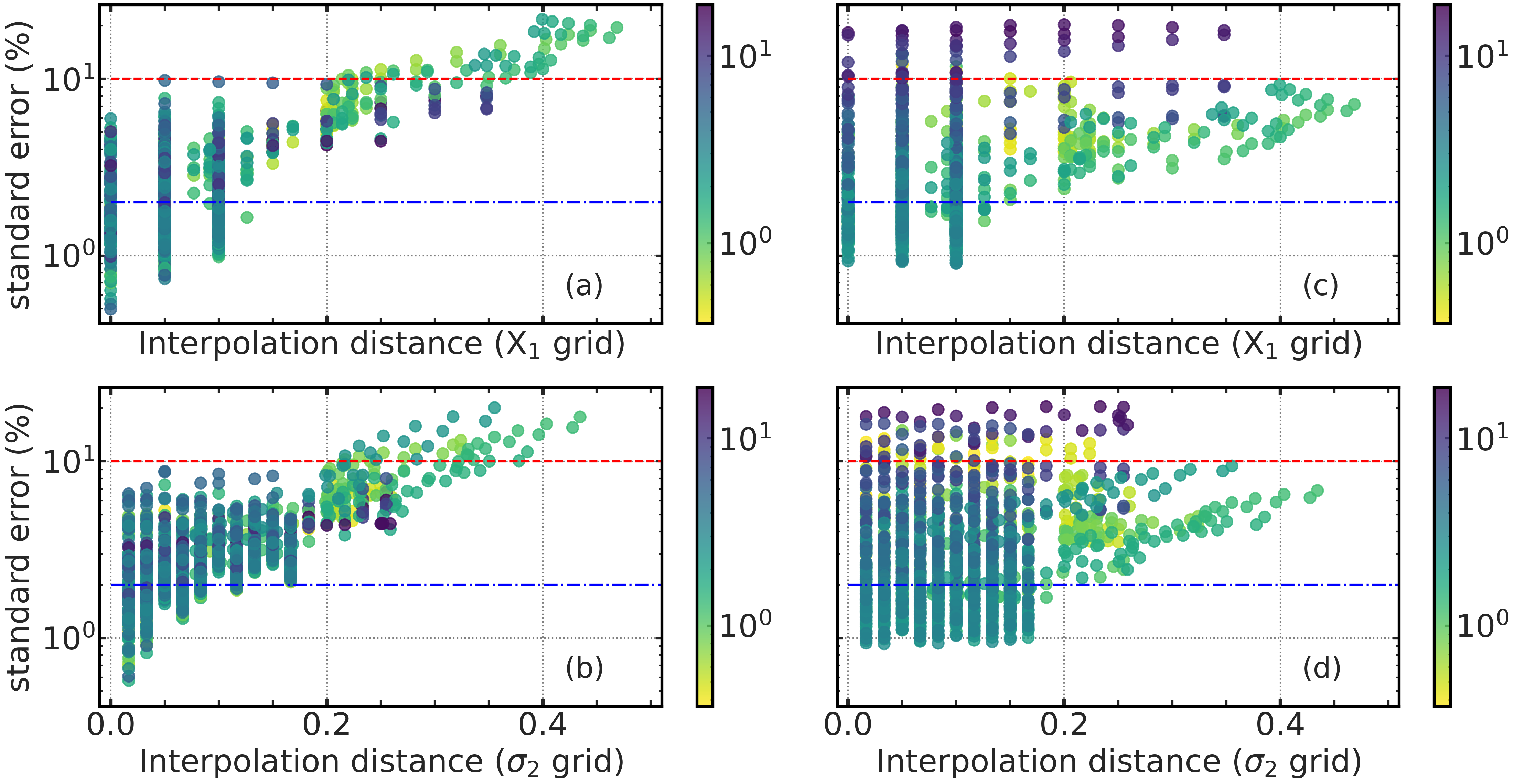}
    \caption{\textbf{Interpolation Uncertainty:} (a), (b) show the GPR
  predicted relative standard error (\%) plotted against the
  interpolation distance in the X$_1$, $\sigma_2$ interpolation grids
  respectively. (c), (d) show the KRR ensemble predicted relative
  standard error (\%) plotted against the interpolation distance in the
  X$_1$, $\sigma_2$ interpolation grids respectively. The colors indicate
  the predicted mean viscosity value corresponding to the interpolation
  point. The blue horizontal dash-dot line indicates the average standard error(\%) (2\%) 
  of the Vlugt data set and the red horizontal dashed line indicates 10\% standard error(\%)
  which is then used to define a qualitative cut-off for the Applicability Domain.
  The interpolation distance is computed in the scaled feature
  space.}
\label{fig:interp-uncertainty}
\end{figure*} 

Information about closeness of a new query point to the training data would be
useful to decide whether or not to trust the values predicted by the model.
This information can be naturally encoded into GPR in the form of epistemic uncertainty.\cite{UQ_active_boris_NatCompMat_2020}
Ideally, the predicted
standard error should increase beyond the natural uncertainty in the data as
the query point moves away from the training data set. Figure
\ref{fig:interp-uncertainty} (a),(b) show the predicted relative standard error
from the GPR model plotted against the interpolation distance with the colors
indicating the predicted mean viscosity value. The predicted relative standard
errors clearly increase with increasing interpolation distance. However, this
behavior could not be seen in the case of relative standard errors estimated
by the ensemble method as seen in Figure \ref{fig:interp-uncertainty} (c),(d).
Hence, the uncertainties estimated by GPR represent the true uncertainties
better and also systematically increase when the query points move far from the
training set. Additionally, GPR needs to be trained only once when compared to
ensemble methods which need to be trained over multiple realisations of the
training set. 

\begin{figure}[h!]
\centering
    \includegraphics[width=0.9\linewidth]{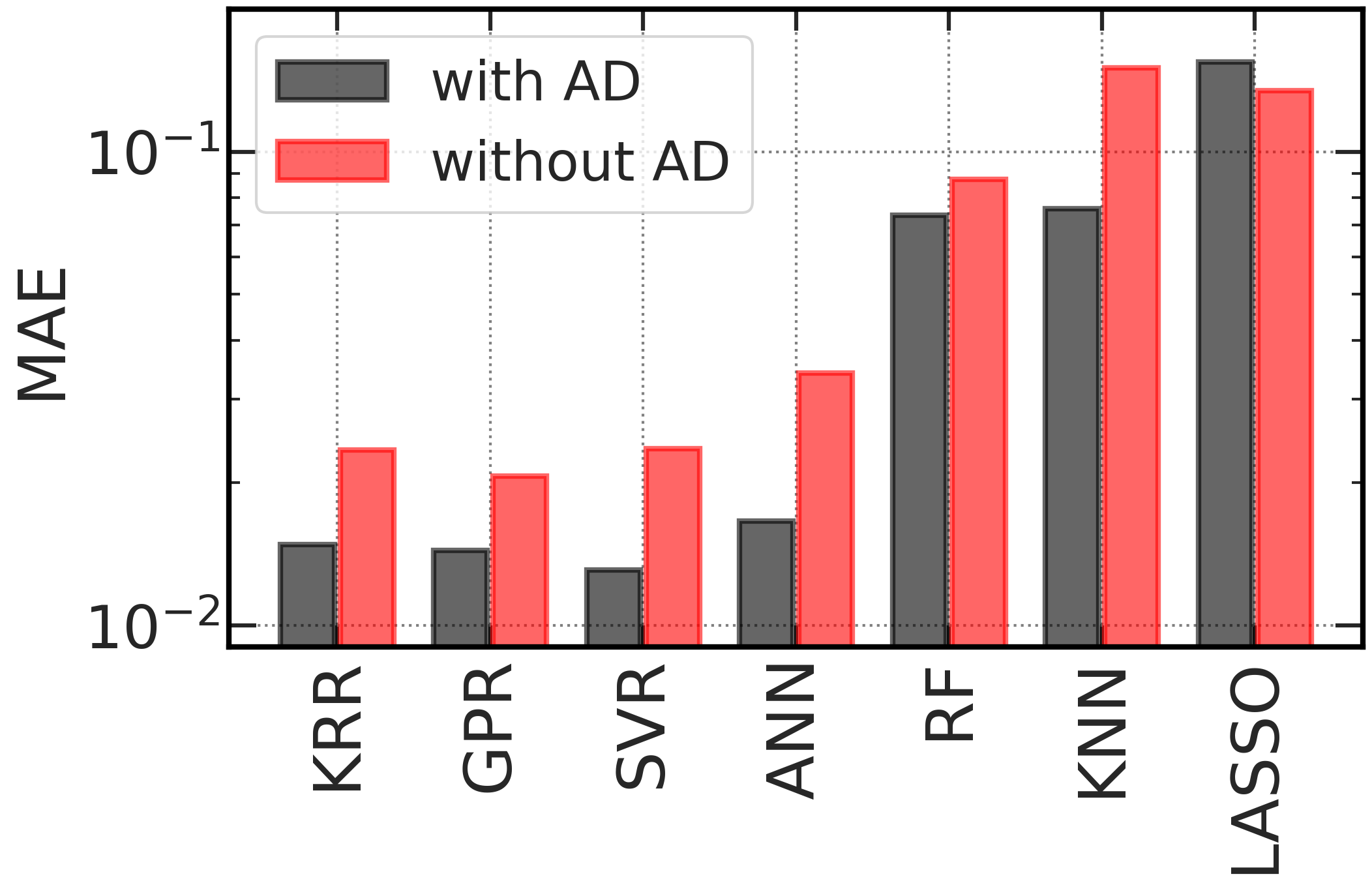}
    \caption{\textbf{Applicability Domain (AD):} MAE values of the
  ensemble ML models estimated on the interpolation data set. The red
  bars indicate the MAE values on the entire interpolation data (without
  Applicability Domain) and the black bars indicate the MAE values on the
  part of the interpolation data that falls within the Applicability
  Domain (In-AD). Due to the small size of interpolation data set, these
  MAE values should be compared in a qualitative sense.} \label{fig:AD}
\end{figure}

Finally, the standard errors predicted by the GPR can be used to construct
applicability domain of the ML models.\cite{AD_GPR_muller_JCAMD_2007} For example, for queries which have a
distance less than 0.2 (in the scaled feature space) from the nearest training
point, a relative error of less than 10\% can be expected (Figure
\ref{fig:interp-uncertainty}). Hence, the scaled distance of 0.2 can be set as
a limit for the Applicability Domain, beyond which the predictions
from the ML models need to be treated with caution. Though such a
distance based approach is simple, it can be applied to any ML model thereby
justifying its use.\cite{AD_GPR_muller_JCAMD_2007} Also, such distance based
AD methods have been successfully implemented for QSAR
\cite{AD_similarity_simon_JCICS_2004} and ML models.\cite{AD_GPR_muller_JCAMD_2007}

The interpolation data set was split into In-AD and Out-AD based on whether the
points fell within the AD or otherwise. The performance of all the ML models on
the In-AD data set is much better than that on Out-AD set, demonstrating that
the application of AD can be used to detect and remove the outliers (Figure
\ref{fig:AD}). Also, the improvements from the AD (though constructed from GPR
only) were observed across all the ML models and performance metrics.

\section{Conclusions}
In this work, we trained and evaluated several successful ML models to predict
the shear viscosity of binary LJ fluids. Being a collective property, shear
viscosity is expensive to predict from equilibrium atomistic MD simulations and
hence only small data sets can be found in the literature. The major challenges
posed by the small data sets on ML methods are discussed. Specifically, we
focus on - 1. model selection and performance estimation, 2. performance
metrics, and 3. uncertainty quantification.

ML models are prone to overfitting on small data sets at both the model
parameters and hyperparameter levels. We discuss various model selection
methods - K-fold CV, nested K-fold CV, Monte Carlo CV, etc. - that are generally
used to address this issue. While these methods are commonly used for selecting
hyperparameters, the generalization error (performance evaluation) is estimated
on a single unseen data set (test set). We demonstrate that such estimates are
prone to wide variability because of the small data set size. The
hyperparameter optimization landscapes of LASSO, KRR models are shown to have
"flat-minima" thereby making it difficult to unambiguously select the optimal
hyperparameters. We compared two simple CV procedures - SS-CV and 
KFS-CV and found that while their mean error estimates were almost identical,
KFS-CV showed lower variance. Hence, it was chosen to both the model selection
and performance estimation tasks \textit{simultaneously}.

We discuss the role of performance metrics in model training and model
evaluation, both from theoretical and empirical standpoints. We compare several
commonly used metrics like MSE, MAE, MAPE and R$^2$ and discuss their relevance
to the viscosity data set. We propose a holistic model ranking procedure based
on inputs from multiple complementary metrics. The interpolation behavior of
the ML models are compared qualitatively. While the KRR, ANN and SVR models
showed smooth interpolation behavior, RF and KNN models showed sudden
discontinuities and are hence considered unsuccessful. The successful models
are also shown to outperform the best-in-class empirical model of Hasse \textit{et al}.\cite{hasse_FPE_2019}

We present two methods to estimate uncertainty in individual predictions from
the ML models - 1. Gaussian Process Regression (GPR) and 2. ensemble of KRR ML
models. The uncertainty (in terms of standard error) estimated by the methods
showed overall agreement with the the true uncertainty of the data, with GPR
faring slightly better. The behavior of the estimated uncertainty by both the
methods in the interpolation feature range is also compared. The GPR's
uncertainty steadily increased as the query data points moved away from the
training data, while no discernible pattern could be identified in the
uncertainty from the ensemble method. The relative standard error estimated
from the GPR model can be used to set distance limits for the query points
thereby defining the applicability domain (AD) in which the results from the ML
models are reliable. We found that the points of the interpolation set
that fell within the AD were better estimated by the ML models than the ones
that fell outside the AD, thereby demonstrating the utility of AD. Finally,
the principles discussed in this work can be
applied to develop ML models of viscosity for more complex fluids. However, in
such fluids, the identification of the features that are most relevant to shear
viscosity would also be non-trivial and constitutes a part of our future work.

\section*{Supplementary Information}
See the Supplementary Information for additional information on theoretical background, model description, Vlugt data set, empirical model description, computational details of model selection, model performance, bias, etc.

\begin{acknowledgements}
The authors thank the Department of Science and Technology, India,
for support. This work is a part of National Supercomputing Mission (NSM)
project entitled "Molecular Materials and Complex Fluids of Societal Relevance:
HPC and AI to the Rescue" (grant no.
DST/NSM/R$\&$D\_HPC\_Applications/2021/05). The support and the resources
provided by “PARAM Yukti Facility” under the National Supercomputing Mission,
Government of India, at the Jawaharlal Nehru Centre For Advanced Scientiﬁc
Research are gratefully acknowledged.
\end{acknowledgements}

\section*{AUTHOR DECLARATIONS}

\subsection*{Conflict of Interest}
The authors have no conflicts to disclose.

\subsection*{Author Contributions}


\textbf{Nikhil V. S. Avula} Formal Analysis (lead);  Methodology (lead);  Software (lead); Visualization (lead); Writing/Original Draft Preparation (lead); Writing/Review \& Editing (equal);  Validation (equal); Data Curation (equal); Conceptualization (supporting); Funding Acquisition (supporting);  \textbf{Shivanand K. Veesam} Data Curation (equal); Software (supporting);  Validation (equal); \textbf{Sudarshan Behera} Formal Analysis (supporting); Methodology (supporting); Software (supporting); Writing/Review \& Editing (supporting); \textbf{Sundaram Balasubramanian} Conceptualization (lead); Funding Acquisition (lead); Project Administration (lead); Resources (lead); Supervision (lead); Writing/Review \& Editing (equal);

\section*{Data Availability Statement}
The data that supports the findings of this study are available within the article and its supplementary material.

\appendix
\section{Performance Metrics}

\begin{equation} \label{eq:error}
   e_i = y_i^{pred} - y_i^{true}
\end{equation}

\begin{equation} \label{eq:rel-error}
  r_i = (y_i^{pred} - y_i^{true})/y_i^{true}
\end{equation}

\begin{equation} \label{eq:samp-mean}
  \Bar{y} = \frac{1}{N}\sum_{i=1}^{N} y_i^{true}
\end{equation}

\begin{equation} \label{eq:me}
  ME = \frac{1}{N}\sum_{i=1}^{N} e_i
\end{equation}

\begin{equation} \label{eq:mpe}
  MPE = \frac{1}{N}\sum_{i=1}^{N} 100 \times \, r_i
\end{equation}

\begin{equation} \label{eq:mse}
  MSE = \frac{1}{N}\sum_{i=1}^{N} e_i^2
\end{equation}

\begin{equation} \label{eq:medse}
  MedSE = median( { e_i^2 })
\end{equation}

\begin{equation} \label{eq:mae}
  MAE = \frac{1}{N}\sum_{i=1}^{N} |e_i|
\end{equation}

\begin{equation} \label{eq:medae}
  MedAE = median( { |e_i| })
\end{equation}

\begin{equation} \label{eq:mape}
  MAPE = \frac{1}{N}\sum_{i=1}^{N} 100 \times \, |r_i|
\end{equation}

\begin{equation} \label{eq:medape}
  MedAPE = 100 \times \, median( { |r_i| })
\end{equation}

\begin{equation} \label{eq:rsq}
  R^2 = 1 - \sum_{i=1}^{N} \frac{|e_i|^2}{|y_i^{true} - \Bar{y}|^2}
\end{equation}

\bibliography{ms}

\end{document}


\tableofcontents
\clearpage
\newpage

\clearpage
\section{Background}

\subsection{The structure of the problem} \label{ssec:struct}

We assume that there exists a joint probability density $p(\mathrm{x},y)$ that
generated the data set \cite{bishop_book,Goodfellow_book}. Here, $\mathrm{x}$ is
a vector of input features and $y$ is the target variable also called as the
label. In the context of shear viscosity prediction, the feature vector can be
constructed from quantities like  $x_1$ , $\sigma_2$ , $\epsilon_2$ ,
k$_{12}$, $\zeta$, $\rho^*$ and the target variable is the shear viscosity
$\eta$ (see section \ref{ssec:data-features}). The task of the ML algorithm is
to infer the joint probability density (or properties thereof) from the finite
data set generated from running the MD simulations. This inference task can be
classified into three levels with progressively lesser complexity
\cite{bishop_book}: \begin{enumerate}
    \item Determine the joint density $p(\mathrm{x},y)$. It is the most
    demanding of the three and generally needs huge amount of data especially
    when $\mathrm{x}$ is of high dimensionality. But if this task is achieved,
    $p(\mathrm{x},y)$ can be used to generate new data. We do not attempt this
    task in this work due to the sparsity of the viscosity data set. 
    \item Determine the conditional density $p(y|\mathrm{x})$. It is much
	simpler than the above because we bypass the difficult task of
	estimating $p(\mathrm{x})$. Gaussian Process Regression (GPR)
	method falls under this category.
    \item Determine a function $f^*(\mathrm{x})$ that is an \textit{optimal}
	representation of the data set. It is the simplest and the most common ML
	approach of the three. The sense in which the function $f^*(\mathrm{x})$ is
	\textit{optimal} is often taken to be the one that minimizes the expected loss
	(also called as \textit{risk}) $\mathbb{E}[L]$ (Eq \ref{seq:exploss}). Most
	common ML models like Kernel Ridge Regression (KRR), Support Vector Regression
	(SVR), Neural Network (NN), etc. fall under this category.
\end{enumerate}

\begin{equation} \label{seq:exploss}
    f^*(\mathrm{x}) = \arg \min_{f(\mathrm{x})} \mathbb{E}[L] = \arg \min_{f(\mathrm{x})} \iint L(y,f(\mathrm{x})) \, p(\mathrm{x},y) \,\mathrm{dx}\,\mathrm{d}y
\end{equation}

Where $L(y,f(\mathrm{x}))$ is the user-defined loss function. The choice of the
loss function has a direct relation to the kind of function obtained
\cite{kolassa_2019}. The most common loss function, the squared loss where
$L(y,f(\mathrm{x})) = (y-f(\mathrm{x}))^2$ yields the conditional mean
$\mathbb{E}_y[y|\mathrm{x}]$ (Eq \ref{seq:conditionalmean}) as the
$f(\mathrm{x})$ \cite{bishop_book}. Discussion on other loss functions and the
consequent effect on the properties of $f^*(\mathrm{x})$ is presented in
section \ref{ssec:metrics}.

\begin{equation} \label{seq:conditionalmean}
    \mathbb{E}_y[y|\mathrm{x}] = \int y \, p(y|\mathrm{x}) \,\mathrm{d}y
\end{equation}

However, to compute the expected loss/risk, the underlying joint probability
density $p(\mathrm{x},t)$ has to be known which is hard to do in practice.
Hence, the expected loss is approximated by empirical loss
$\mathbb{E}_{emp}[L]$ 

\begin{equation} \label{seq:emploss}
    f^*(\mathrm{x}) = \arg \min_{f(\mathrm{x})} \mathbb{E}_{emp}[L] =  \arg \min_{f(\mathrm{x})} \left( \frac{1}{N} \sum_{i=1}^{N} L(y_i^{true},f(\mathrm{x_i})) \right )
\end{equation}

where, $N$ is the number of data points, $y_i^{true}$ are the target values
corresponding to the feature vector $\mathrm{x}_i$. Under certain conditions,
the empirical loss asymptotically (like $N \to \infty$) converges to the
expected loss and so does the corresponding $f(\mathrm{x})$
\cite{muller_intro_2001}. However, for most practical cases (especially for
small data sets), the empirical loss can show large deviations from the expected
loss and so does the corresponding $f(\mathrm{x})$. Generally, the empirical
loss tends to be much lesser on the data set used to infer $f^*(\mathrm{x})$
(called the training set) than on new/unseen data set(s). This is because the
minimization of empirical loss (\textit{per se}) incentivizes the learning
machine to learn the idiosyncrasies (like noise) of the particular training
data sample rather than the trends in the underlying model that generated that
data set \cite{muller_intro_2001,bishop_book,Goodfellow_book}. Hence, the goal
of the learning protocol should be to minimize the error on new/unseen
data set(s) called the generalization error. This phenomena of ML methods having
significantly lesser training error than the generalization error is called
overfitting and is especially relevant for models on small data sets
\cite{muller_intro_2001,bishop_book,Goodfellow_book}.

The most common way to alleviate the problem of overfitting is to reduce the
complexity/capacity of the learning machine thereby reducing its ability to
learn the noise associated with the training data sample. However, the
complexity should not be reduced to such an extent that the general trends in
the data are lost, resulting in underfitting. Hence, the ML model should choose
an \textit{optimal} complexity corresponding to the general trends in the data.
A popular method to control the complexity of the models is called
regularization in which a penalty term (called the regularizer, Eq
\ref{seq:regularizer}) which penalises complex models is added to the empirical
loss \cite{muller_intro_2001,Goodfellow_book}. The common forms of the
regularizer are based on the norm of the weights ($w$) of the model like -
$L^2$ norm (called as ridge regression or Tikhonov regularization), $L^1$ norm,
or a combination of both (for example in LASSO model) \cite{Goodfellow_book}.
We also note that there are many other regularization techniques that are
specific to Deep Learning (DL) methods like - early stopping, dropout, soft
weight sharing, etc \cite{dropout_2014}.

\begin{equation} \label{seq:regularizer}
  f^*(\mathrm{x}) = \arg \min_{f(\mathrm{x})} J = \arg \min_{f(\mathrm{x})} \left( \frac{1}{N} \sum_{i=1}^{N} L(y_i^{true},f(\mathrm{x_i}))  + \sum_{j} \lambda_j \, \Omega_j(f) \right)
\end{equation}
Where $\Omega_j(f)$ are the regularizers and $\lambda_j$ are the parameters
that control the amount of regularization. Now that the ML models have a
mechanism to control the complexity through regularization, the natural next
step would be to choose the values of regularization parameters. This task
falls under the purview of model selection \cite{Goodfellow_book}. In the
following section, we discuss various model selection criteria and a closely
related topic of performance evaluation.

\subsection{Model Selection and Performance Evaluation} \label{ssec:model-selection}
It is a common practice to distinguish the parameters of ML and DL models into
model parameters and hyperparameters \cite{bishop_book,Goodfellow_book}. The
model parameters are learnt during the training phase on the training data.
Examples of model parameters include - slope and intercept in linear
regression, coefficients of kernel expansion in kernel methods (like KRR),
weights of neurons in Neural Networks (NNs), etc. The hyperparameters are
generally the high level settings of ML algorithms which are either set by the
user or inferred during model selection procedure. Examples of hyperparameters
include - regularization parameters, the degree of polynomial in polynomial
regression, choice of the kernel in kernel methods, choice of activation
function in Neural Networks, number of neurons in NNs, etc. This division is
done for two main purposes - (1) computational efficiency, and (2) need for
disjoint data sets \cite{Goodfellow_book,cawley_2010}. For most ML methods, many
efficient optimization algorithms based on gradient descent can be used to
obtain the model parameters. Moreover, some ML methods like Least Squares
Linear Regression, kernel ridge regression (KRR) have closed form solutions for
obtaining model parameters \cite{cawley_2010}. Hyperparameters on the other
hand, are generally obtained using heuristic/search based methods due to the
lack of analytical gradients. Another crucial reason for the distinction is
that hyperparameters that control the regularization should generally be
inferred from a different data set (called the validation set) disjoint from the
training data so as to avoid overfitting \cite{Goodfellow_book}. This process
of obtaining model parameters and hyperparameters using different criterion is
called multi-level inference \cite{cawley_2010}. 

Now, the task of selecting the model with the \textit{optimal} complexity is
reduced to the estimation of values of hyperparameters and the criteria used
for such selection are called model selection criteria. As stated earlier, the
goal of ML models is to minimize the generalization error which is the average
error over \textit{all} unseen data. However, generalization error cannot be
obtained in most practical situations and hence estimators on finite data sets
are constructed to approximate it. The process of estimating the generalization
error by using estimators on finite data sets is called performance evaluation
and is a prerequisite for model selection. It is crucial to note that the error
estimates are obtained over finite data sets and hence depend on the size of the
data set especially for small data sets.  A simple example of such an estimator
is the split sample estimator where the whole data set is split into two parts
(generally unequal) and the error is computed on the split that was not used
for training \cite{cawley_2010}. Split sample estimator is known to be unbiased
i.e., the average split sample error over multiple independent realisations of
unseen data asymptotically converges to the generalization error. Hence,
minimizing the split sample error can in principle reduce the generalization
error. However, it was recently shown that the unbiasedness \textit{per se} is
not as important as the variance of the estimator when it is used for model
selection \cite{cawley_2010}. When an estimator has high variance (occurs with
small data sets\cite{cawley_2010}) the value of the estimated error on any one
particular unseen data sample can be very different from the generalization
error; hence the hyperparameters that minimize the estimated error can be far
off from the \textit{optimal} ones. Cawley et al. showed (on a synthetic
data set) that hyperparameters selected based on split sample estimators can
severely overfit or underfit the data \cite{cawley_2010}. In practice,
users rarely have the capability of generating multiple independent
realizations of the data and hence the variance of the estimator plays a major
role. Therefore, for small data sets, it is not considered a good practice to
estimate the error on a single realisation of the data set \cite{cawley_2010}.
In order to mitigate this problem, various cross-validation schemes are
generally used.

The core idea of k-fold cross-validation (CV) is to use split the entire
data set into k equal disjoint sets, train the ML models on k-1 sets and
estimate the error on the remaining one set. This process is repeated k times,
each time with a different hold-out set \cite{bishop_book,Goodfellow_book}. The
average error over k folds is used as the estimate for the generalization
error. It is a common practice to use 5 or 10 folds during CV
\cite{Goodfellow_book}. The extreme case when the number of folds is equal to
the number of data points ($k = N$) is called Leave One Out (LOO) method. LOO
is also popular because for some ML models, it is possible to compute the LOO
error without training the ML model k times thereby reducing the computational
cost \cite{cawley_2010,Guyon_2006}. 

The error estimates from k-fold CV are often used for model selection by
searching over the space of hyperparameters and choosing the one that yields
minimum CV error. But once the k-fold CV error is used to optimize the
hyperparameters, it is no longer unbiased
\cite{nestedcv_ex_2006,nestedcv_ex_2014,Goodfellow_book}. Typically another
unseen data set (called the test set) is used to estimate the generalization
error of the models with optimized hyperparameters \cite{Goodfellow_book}.
Using a single realization of the test set, however, suffers from the high
variance issue discussed above. Nested cross-validation or double
cross-validation improves upon k-fold CV by doing performance evaluation and
model selection in two nested loops
\cite{cawley_2010,muller_2013,nestedcv_plosone,nestedcv_ex_2006,nestedcv_ex_2014}.
The outer loop is used to estimate the generalization error and the inner loop
is used to select the hyperparameters. Also, we note that there are many
methods of splitting the data set into train/validation/test sets like -
Monte-Carlo CV, bootstrapping, Kennard-Stone splitting, and combinations
thereof \cite{goodacre_2018,Guyon_2006}. Xu and Goodacre compared the
performance (in terms of their ability to predict the generalization error) of
various data splitting methods including k-fold CV, Monte-Carlo CV,
bootstrapping, etc and found that a single best method could not be found
\textit{a priori} and suggest that the choice of the method should be tuned to
the kind of data (No Free Lunch again) \cite{goodacre_2018}.

Most of the methods mentioned above require partitioning the data and then
training the ML models multiple times which can become prohibitively expensive
for large models. In such cases, model selection criterion based on
\textit{information theoretic} approaches can be used like - Akaike Information
Criterion (AIC), Bayesian Information Criterion (BIC), etc
\cite{bishop_book,modelselection_book1}. A thorough comparison of all these
criteria is beyond the scope of this work and it is still an area of active
research.

Finally, we note that model selection and performance evaluation are big and
unsolved challenges on small data sets
\cite{cawley_2010,Guyon_2006,Goodfellow_book}. Guyon \textit{et al} organized a
performance prediction challenge in which the participants (more than 100) are
asked to predict the generalization error on finite data sets of real world
importance like medical diagnosis, speech recognition, text categorization, etc
\cite{Guyon_2006}. They observed that most submissions were overconfident about
their ML models i.e., their prediction of generalization error is less than the
true generalization error. They also noted that the performance of the ML
models truly improved in the first 45 days of the 180 day challenge after which
overfitting set in. It is now a common belief that when a data set is worked
upon repeatedly, even careful performance prediction protocols can result in
optimistic performance predictions over time \cite{Goodfellow_book}.  

\subsection{Performance Metrics for Regression} \label{ssec:metrics}
Performance metrics are generally used in two critical areas of ML model
development workflow - model training and model comparison. Though the choice
of the metric can significantly alter the \textit{kind} of ML model developed
and consequently its real-world performance, there is no clear consensus on
this topic \cite{viswakarma_metrics,kolassa_2019,collopy_1992}. As is the case
with model selection criterion, there is no single best metric that can be used
across all ML tasks \cite{collopy_1992}. In this section, we summarize some of
the principles that can be used to choose a relevant metric to the particular
ML task at hand and also consider the particular case of viscosity data set.

As discussed in section \ref{ssec:struct}, models that learn a function
$f(\mathrm{x})$ (inference level 3) lose some information contained in the
joint density $p(\mathrm{x},y)$ from which the data is assumed to be generated.
Such models generally aim to predict point estimates related to a central
tendency of the conditional density $p(y|\mathrm{x})$ like the conditional mean
or the median \cite{kolassa_2019}. It can be seen that the choice of the loss
functions can be used to decide the kind of central tendency of
$p(y|\mathrm{x})$ to be captured by the $f(\mathrm{x})$
\cite{gneiting_2011,kolassa_2019}. For example, the $f(\mathrm{x})$  that
minimizes the expected value of the squared loss results is the mean of the
$p(y|\mathrm{x})$. Hence the $f^*(\mathrm{x})$  can be thought of as a
functional of the conditional density $p(y|\mathrm{x})$ which can be chosen by
choosing the loss function $L$ \cite{gneiting_2011}. Table \ref{stab:metrics}
summarizes some of the popular loss functions and their corresponding
functionals.

\begin{table}[!ht]
    \caption{\label{stab:metrics} Popular loss functions and the  corresponding regression functionals.}
    \centering
    \begin{tabular}{|l|c|c|c|}
    \hline
     loss name      & loss formula & functional & refs \\
      \hline
    SE       & $(y^{pred}-y^{true})^2$ & mean & \cite{bishop_book,gneiting_2011} \\
    AE       & $|y^{pred}-y^{true}|$ & median & \cite{bishop_book,gneiting_2011} \\
    APE      & $|(y^{pred}-y^{true})/y^{true}|$ & $\beta-$median & \cite{gneiting_2011} \\
    RE       & $|(y^{pred}-y^{true})/y^{pred}|$ & $\beta-$median &  \cite{gneiting_2011} \\
    quantile  & $\tau (y^{true}-y^{pred})$ if $y^{true}-y^{pred} > 0$  & quantile & \cite{koenker_2001,koenker_2005,UQ_tavazza_ACSOmega_2021} \\
    quantile  & $(\tau-1) (y^{true}-y^{pred})$ if $y^{true}-y^{pred} < 0$  & quantile & \cite{koenker_2001,koenker_2005,UQ_tavazza_ACSOmega_2021} \\
     \hline
      \hline
    \end{tabular}
\end{table}

From the perspective of loss functions in model training, model developers have
two approaches - (1) choose the loss function first and work out the kind of
functional (if possible) from it, or (2) choose the desired functional and work
out the corresponding loss function \cite{gneiting_2011}. Most ML practitioners
use the first approach because of its simplicity and the availability of
efficient numerical optimizers for the minimization of the common loss
functions. The latter approach is generally used in quantile regression models
which can be used to predict the confidence intervals around the mean
\cite{UQ_tavazza_ACSOmega_2021}. However, it must be noted that the connection between the
loss function and the corresponding functional is clear only when the models
minimize the expected loss function without any additional constraints. As most
ML methods rely on regularization to avoid overfitting, the resulting
functional can be different from that of the expected loss minimization. Also,
not all ML models use expected loss minimization and hence it might not be
straightforward to work out their corresponding functionals.

Another area in which loss functions are used in ML workflow is model
comparison, in which models are ranked based on their generalization
performance. Ideally, the generalization performance of ML models should also
be measured using the same metric used in their training phase
\cite{kolassa_2019}. For example, an ML model trained by minimizing MSE should
be compared to other models using MSE generalization error. The consequences of
not using consistent metrics during training and comparison are demonstrated by
Tilman Gneiting in which generalization MAPE error showed a value of $~10^5$
for a model trained with MSE loss, whereas the same model trained with MAPE
loss showed the MAPE generalization error to be $1.00$ \cite{gneiting_2011}.

However, in many cases the choice of the loss functions cannot be controlled by
the model developers and hence it is difficult to choose just one metric to
compare such models. For example, Makridakis \textit{et al}  use a weighted
average of sMAPE and MASE to compare the models in the M4 forecasting
competition citing a lack of agreement on the advantages and drawbacks of
various metrics \cite{m4_competition}. Hence, it is generally recommended to
report the estimates of generalization error using multiple metrics
\cite{muller_2013,alam_jpcb_2021,collopy_1992}. Also, given the proliferation
of various metrics, it is important to choose the set of metrics that are
relevant to the ML task at hand and preferably containing complementary
information to each other. Armstrong and Callopy compared six commonly used
metrics and ranked them qualitatively (good,fair,poor) according to five
characteristics - reliability, construct validity, sensitivity, outlier
protection, and their relationship to decision making \cite{collopy_1992}. They
conclude that there is no single metric that can be considered the best in all
situations and that they should be selected based on the kind of data set. 

We use some of the arguments presented in their work to select metrics based on
their suitability to the viscosity data set. [see section data for discussion
about the characteristics of the viscosity data set used in this study]. First,
we look at compatibility of metrics to a data set that spans many orders of
magnitude. All metrics that have units i.e., are not scaled, tend to be
dominated by the error from the highest order of magnitude and hence do not
give information about the contributions of the errors from low orders of
magnitude \cite{collopy_1992}. Metrics based on scaled error like MAPE are more
suited to such a situation. Next, we look at the level of outlier protection of
various metrics. All metrics that take an average of individual errors suffer
from outlier problem because the mean itself is sensitive to large outliers.
Median based metrics like MedAE are better suited to such a situation. However
median based metrics are not sensitive to small changes in the errors and also
do not have clearly defined gradients with respect to model parameters.
Finally, we look at metrics that can capture systematic biases (over or
underestimation) in the ML models. Metrics based on error function with
strictly positive range like SE, AE, APE etc, cannot distinguish between
systematic over or under prediction by the ML models. Metrics based on Mean
Error (ME) or Mean Percentage Error (MPE) can be used to gauge the bias in the
models. Therefore, we rank the ML models developed in this work based on the
following metrics - MSE, MAE, MAPE, MedSE, MedAE, MedAPE, ME, MPE, and R$^2$.

\clearpage
\section{Model description}
\subsubsection{LASSO}
Linear models have a closed form solution and are hence fast. Theoretical
properties of Linear models are also well understood making them amenable for
analysis.

Most practical problems do not have a linear dependence of the labels on the
input features. And all the interesting ones are highly non-linear. An elegant
solution is to use a non-linear mapping of input features into an abstract
space and then use linear models on that space.  A variety of non-linear
functions like polynomial functions are generally employed for this purpose.
However without suitable knowledge about the embedding into the abstract space,
it becomes tedious to select the number and type of the non-linear functions.

\subsubsection{Kernel based methods}
Kernel methods make use of the kernel trick to alleviate the problem of
explicitly choosing the set of non-linear mapping functions. Kernel trick
recasts the algorithm so that the explicit conversion between the input
features and the abstract is not needed. Any kernel function that gives
symmetric positive definite Gram matrix can be used by the kernel trick. Some of the commonly used kernels are:

\begin{equation}
    \kappa(\mathrm{x},\mathrm{x}') = \phi(\mathrm{x})^T \phi(\mathrm{x}') = \langle  \phi(\mathrm{x}), \phi(\mathrm{x}')  \rangle
\end{equation}

Linear kernel:
\begin{equation}
    \kappa(\mathrm{x},\mathrm{x}') = \mathrm{x}^T \mathrm{x}'
\end{equation}

Polynomial kernel
\begin{equation}
    \kappa(\mathrm{x},\mathrm{x}') = (\gamma\mathrm{x}^T \mathrm{x}' + r)^M
\end{equation}

Gaussian (radial basis function) kernel
\begin{equation}
    \kappa(\mathrm{x},\mathrm{x}') = \frac{1}{(2\pi\sigma^2)^{D/2}}\mathrm{exp} \left(-\frac{1}{2\sigma^2} \|\mathrm{x}-\mathrm{x}' \|^2  \right)
\end{equation}

Laplacian kernel
\begin{equation}
    \kappa(\mathrm{x},\mathrm{x}') = \mathrm{exp} \left(-\frac{1}{\sigma} \|\mathrm{x}-\mathrm{x}' \|  \right)
\end{equation}

Matern kernel
\begin{equation}
    \kappa(\mathrm{x},\mathrm{x}') = \frac{2^{1-\nu}}{\Gamma(\nu)} \left(\frac{\sqrt{2\nu \|\mathrm{x}-\mathrm{x}' \|}}{l} \right)^{\nu} \mathrm{K_{\nu}}\left(\frac{\sqrt{2\nu \|\mathrm{x}-\mathrm{x}' \|}}{l} \right)
\end{equation}

$\nu > 0, l > 0, \mathrm{K}_{\nu}$ is a modified Bessel function

\clearpage
\section{Vlugt Data Set}

\begin{figure}[h!]
\centering
    \includegraphics[width=3.25in]{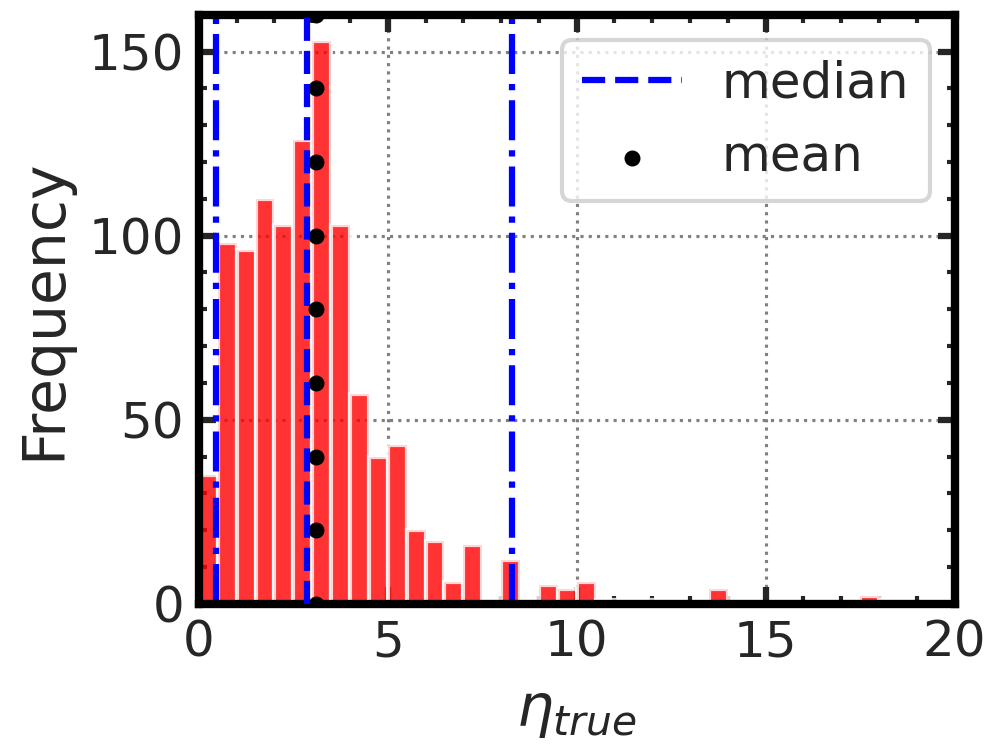}
    \caption{\textbf{Data Distribution:} The distribution of the viscosity 
    from the Vlugt data set~\cite{vlugt_jctc_2018} across decades of viscosity. The blue 
    vertical dashed line represents the median and the two blue 
    vertical dash-dot lines represent the 95 percentile range around the 
    median. The black vertical dotted line represents the mean of the data. }
\label{sfig:viscdist}
\end{figure}

\begin{figure}[h!]
\centering
    \includegraphics[width=7in]{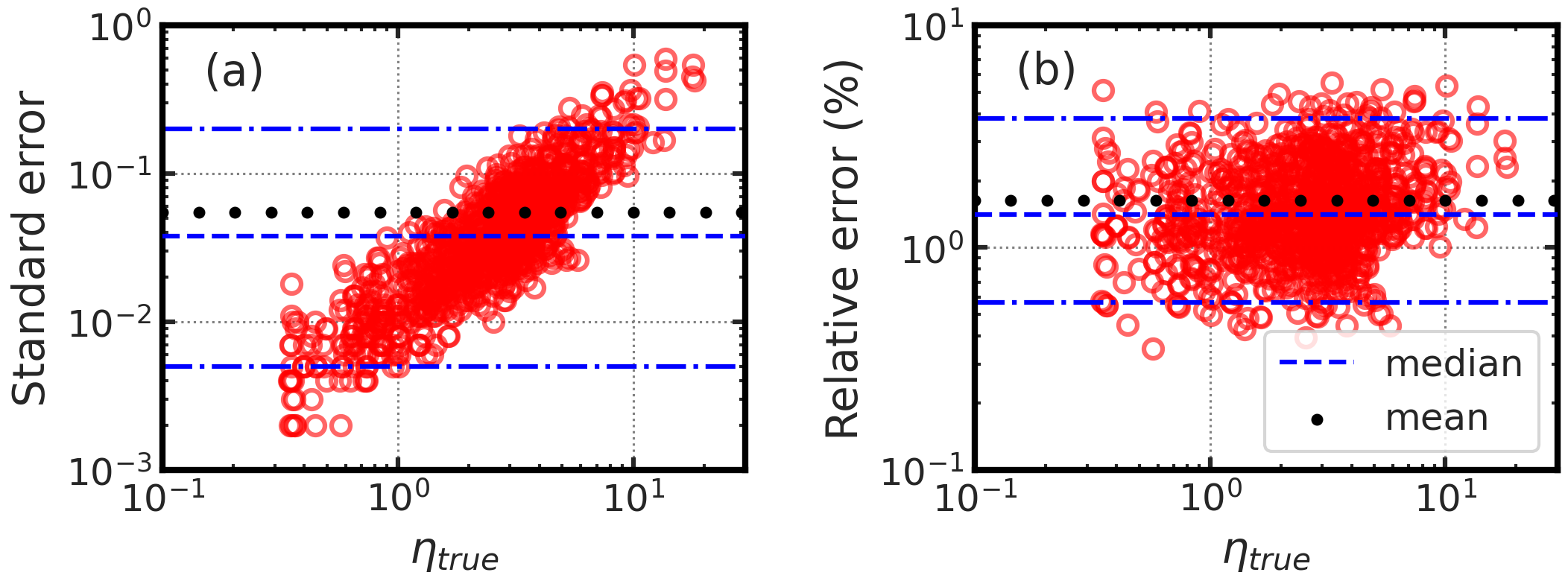}
    \caption{\textbf{Standard Error Distribution:} The distribution of the
	standard error of the viscosity from the Vlugt data
	set~\cite{vlugt_jctc_2018} across decades of viscosity. }
	\label{sfig:viscerrdist}
\end{figure}

\clearpage
\subsection{Feature Dependence} \label{ssec:data-features}

\begin{figure}[h!]
\centering
    \includegraphics[width=7in]{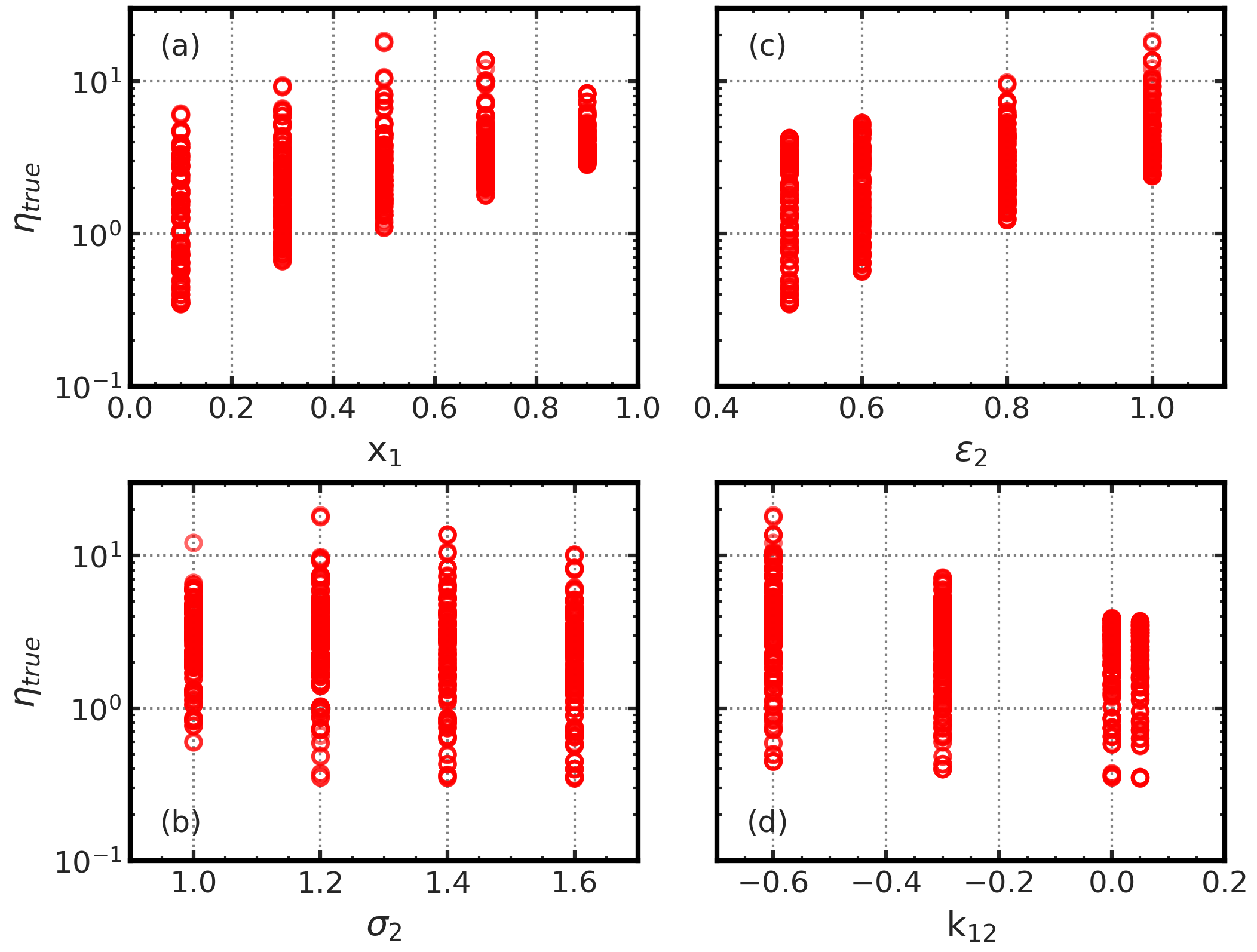}
    \caption{\textbf{Feature Dependence:} Viscosity plotted against individual preMD features. }
\label{sfig:features}
\end{figure}

\begin{figure}[h!]
\centering
    \includegraphics[width=7in]{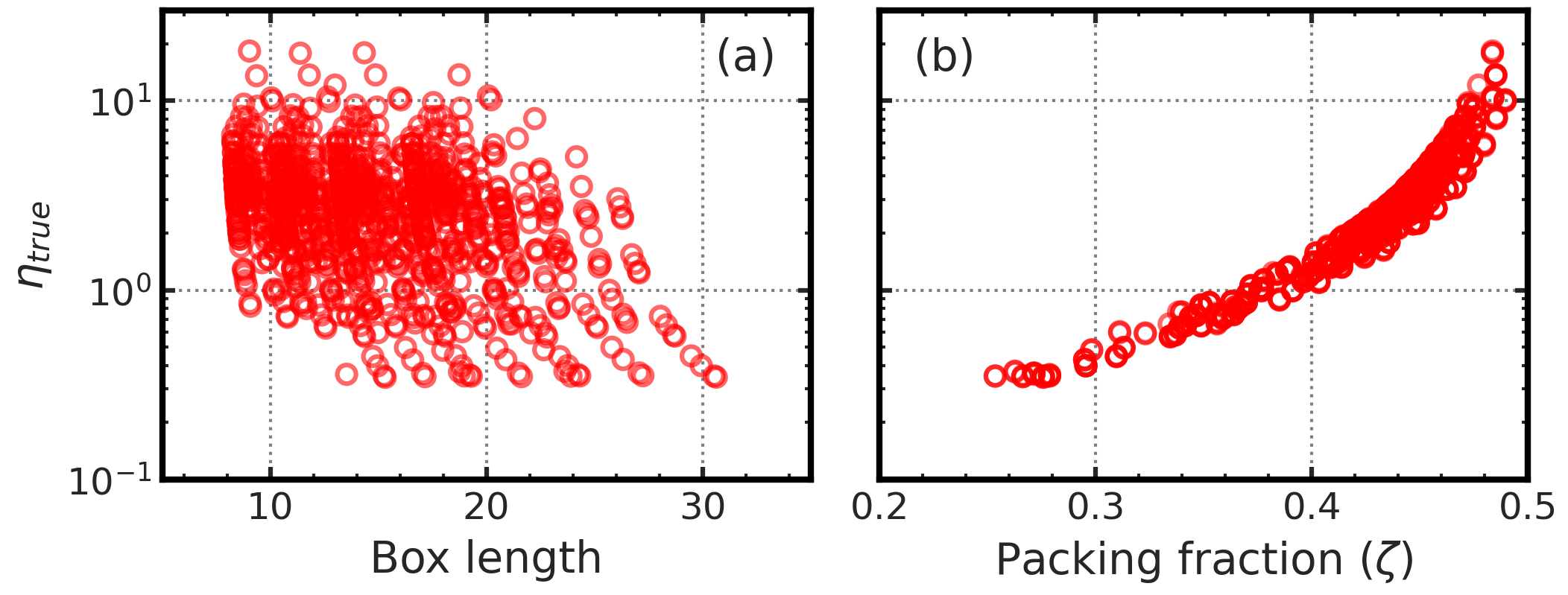}
    \caption{\textbf{Feature Dependence:} Viscosity plotted against box length and packing fraction. }
\label{sfig:box-zeta}
\end{figure}

\begin{figure}[h!]
\centering
    \includegraphics[width=7in]{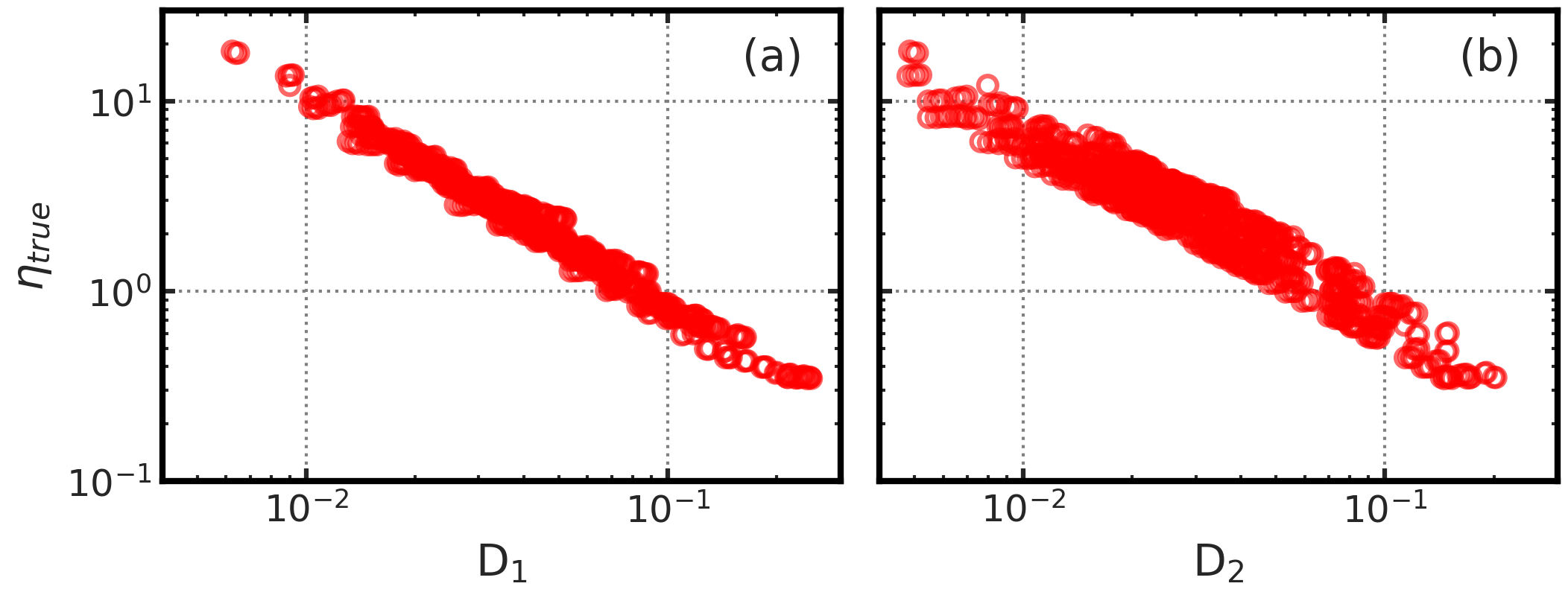}
    \caption{\textbf{Feature Dependence:} Viscosity plotted against
	self-diffusion coefficients of particle types 1 and 2. }
	\label{sfig:d1-d2}
\end{figure}

\clearpage
\section{Computational Details} \label{ssec:comp-details}

\subsection{Model Selection and Performance Estimation} \label{ssec:comp-details-model-sel}

\subsubsection{SS-CV} \label{ssec:comp-details-sscv}
\begin{algorithm}
\caption{The Shuffle Split Cross validation (SS-CV) procedure used in this work }\label{salg:ss}
\begin{algorithmic}
\State  \textbf{Define} SSCV($\mathbb{D}$,A,L,N$_{sp}$,fv$_{sp}$,ft$_{sp}$,$\Omega$):
\Require $\mathbb{D}$, the Vlugt data set, with elements z$^{(i)}$
\Require A, the ML algorithm
\Require $\Omega$, the hyperparameter set of the ML algorithm
\Require L, the loss function
\Require N$_{sp}$, the number of splits
\Require {fv$_{sp}$, the fraction of the $\mathbb{D}$ to be considered as a validation set}
\Require {ft$_{sp}$, the fraction of the $\mathbb{D}$ to be considered as a test set}

\For{$i=1$ to N$_{sp}$}

\State Split $\mathbb{D}$ into three mutually exclusive subsets $\mathbb{F}_i$, $\mathbb{V}_i$ and $\mathbb{T}_i$. The size of $\mathbb{T}_i$ being a fraction ft$_{sp}$ of the size of $\mathbb{D}$. The size of $\mathbb{V}_i$ being a fraction fv$_{sp}$ of the size of $\mathbb{D}$.

\State f$_{i}$ = A($\{\mathbb{F}_i\},\Omega$)

\For{z$^{(l)}$ in $\mathbb{V}_{i}$}
\State $ve_{i,l} =  L(f_{i},z^{(l)}) $
\EndFor

\For{z$^{(l)}$ in $\mathbb{T}_i$}
\State $te_{i,l} =  L(f_{i},z^{(l)}) $
\EndFor

\State Shuffle the points in $\mathbb{D}$

\EndFor

\Return $\mathbf{ve},\mathbf{te}$

\end{algorithmic}
\end{algorithm}

\clearpage
\subsubsection{KFS-CV} \label{ssec:comp-details-kfscv}

\begin{algorithm}
\caption{The K-fold Split Cross validation (KFS-CV) procedure used in this work }\label{salg:kfs}
\begin{algorithmic}
\State  \textbf{Define} KFSCV($\mathbb{D}$,A,L,k,N$_{sp}$,f$_{sp}$,$\Omega$):
\Require $\mathbb{D}$, the Vlugt data set, with elements z$^{(i)}$
\Require A, the ML algorithm
\Require $\Omega$, the hyperparameter set of the ML algorithm
\Require L, the loss function
\Require N$_{sp}$, the number of splits
\Require {f$_{sp}$, the fraction of the $\mathbb{D}$ to be considered as a test set}
\Require k, the number of folds

\For{$i=1$ to N$_{sp}$}

\State Split $\mathbb{D}$ into two mutually exclusive subsets $\mathbb{F}_i$ and $\mathbb{T}_i$. The size of $\mathbb{T}_i$ being a fraction f$_{sp}$ of the size of $\mathbb{D}$.

\State Split $\mathbb{F}_i$ into k mutually exclusive subsets $\mathbb{V}_{i,j}$, whose union is  $\mathbb{F}_i$

\For{$j=1$ to k}
\State f$_{i,j}$ = A($\{\mathbb{F}_i\setminus\mathbb{V}_{i,j}\},\Omega$)

\For{z$^{(l)}$ in $\mathbb{V}_{i,j}$}
\State $ve_{i,j,l} =  L(f_{i,j},z^{(l)}) $
\EndFor

\For{z$^{(l)}$ in $\mathbb{T}_i$}
\State $te_{i,j,l} =  L(f_{i,j},z^{(l)}) $
\EndFor

\EndFor

\State Shuffle the points in $\mathbb{D}$

\EndFor

\Return $\mathbf{ve},\mathbf{te}$

\end{algorithmic}
\end{algorithm}

\clearpage
\subsection{Empirical Correlation} \label{ssec:emp-model}
We could not find an empirical model (also called as empirical correlation)
which is applicable directly on the binary LJ fluids. Rather, we first map the
binary LJ fluids into an effective one-component fluid using the van der Waals
one-fluid model. The effective thermodynamic conditions thus obtained are then
used to predict the shear viscosity using the empirical correlations developed
on pure LJ systems. Meyer et al demonstrated this approach on a small subset of
thermodynamic conditions and found good agreement with the simulated data
\cite{meyer_jcp_2018}. Here we use the empirical correlation of shear viscosity
of pure LJ systems developed by Hasse et al using NEMD simulations
\cite{hasse_FPE_2019}. Their "semi-empirical" correlation used 300 data points
spanning a wide temperature and density range. Moreover, they compared the shear
viscosity estimated from their model with the literature data and found an AAD
(Absolute Average Deviation or MAPE) less than 3 \%. 

\begin{figure}[h!]
\centering
    \includegraphics[width=4in]{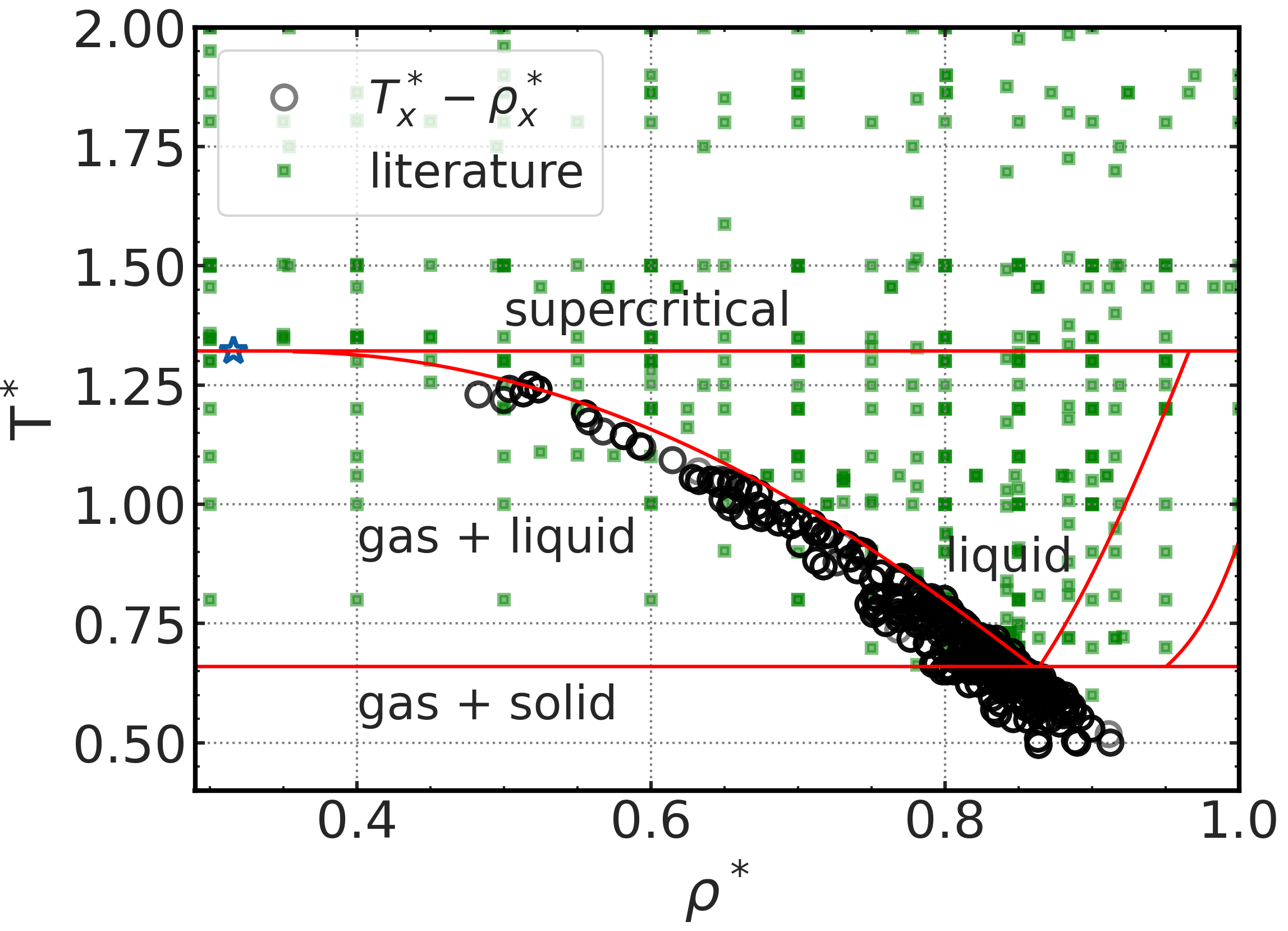}
    \caption{\textbf{LJ phase diagram:} The effective thermodynamic conditions
	of the binary LJ fluids studied by the Vlugt's group, plotted against
	the phase diagram of pure LJ system. The green square indicate the
	thermodynamic conditions at which shear viscosity of pure LJ systems
	was reported in literature. These points were collected by Hasse et al
	and were used to compare their empirical correlation \cite{hasse_FPE_2019}}
	\label{sfig:emp-ljphase}
\end{figure}

\begin{figure}[h!]
\centering
    \includegraphics[width=4in]{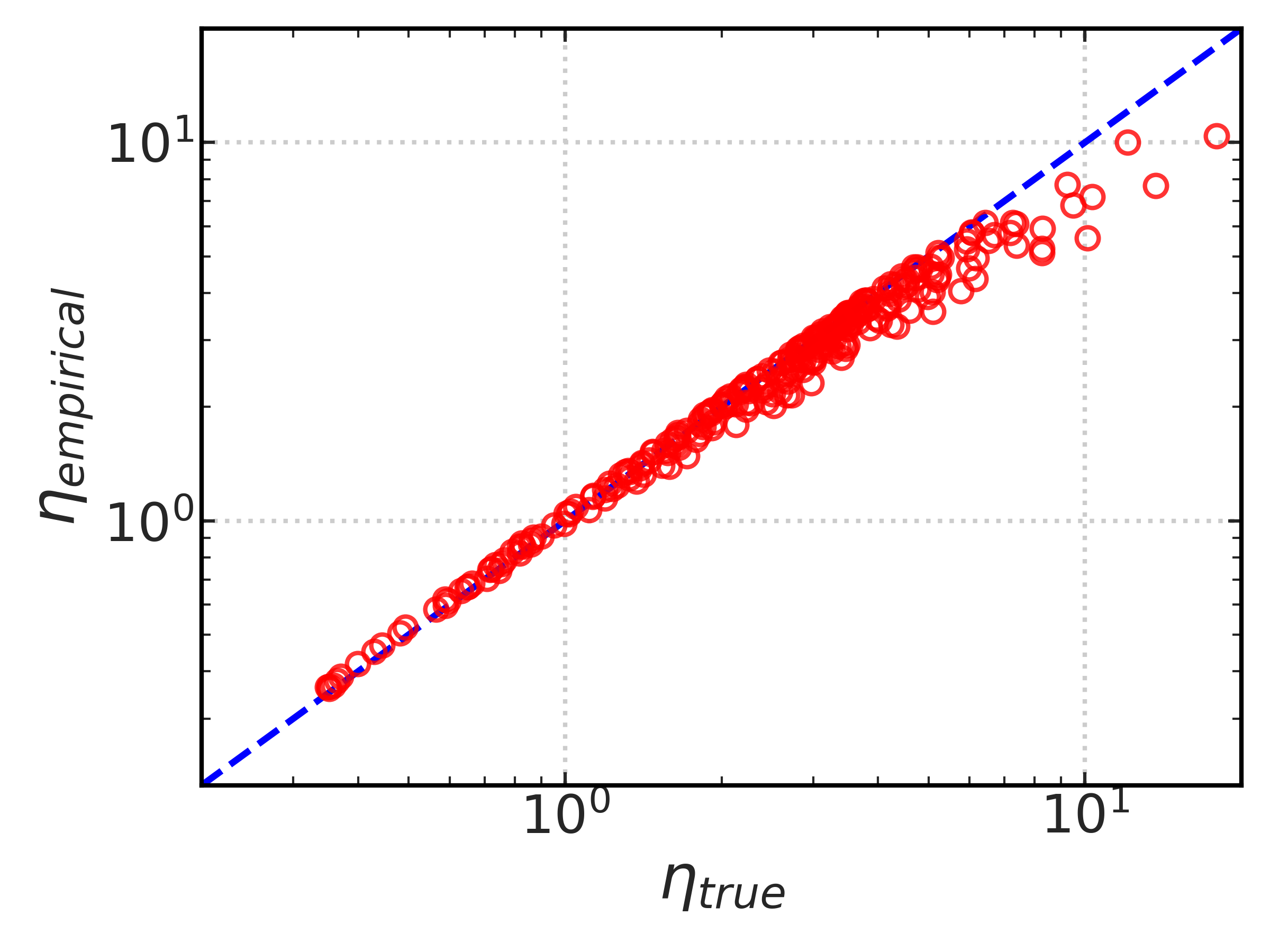}
    \caption{\textbf{Empirical correlation comparison:} Comparison of the shear
	viscosity predicted by the Hasse's empirical correlation \cite{hasse_FPE_2019} and Vlugt's
	simulation data \cite{vlugt_jctc_2018}. The blue dashed line indicates y=x and is drawn as a
	guide to the eye.} \label{sfig:emp-vs-vlugt}
\end{figure}

\clearpage
\subsection{Interpolation Data Set} \label{ssec:interp-data-set}
A small data set to test the predictive performance of the ML models away from the Vlugt data grid called the interpolation data set was created in this work. This data set consists of 17 randomly chosen points in the interpolation region of the Vlugt data grid. We note that the entire interpolation space is not in a single liquid phase and some data points tend to phase separate. For example, we started out by simulating 20 interpolation data points out of which 3 points were not in single liquid phase. Similar observations were reported by Vlugt et al \cite{vlugt_jctc_2018}. The interpolation data set is attached as a csv file in the Supporting Information of this work. 

The procedure to estimate the viscosity was the same as the one reported in the work of Vlugt et al \cite{vlugt_jctc_2018}. The LAMMPS input scripts (suitably modified) provided in the Supporting Information of their work  \cite{vlugt_jctc_2018} were used to run the MD simulations in this work. A brief description of the procedure is given below and the readers should refer to the work of Vlugt et al for a complete description \cite{vlugt_jctc_2018}. All the simulations consist of binary LJ systems with a total of 2000 particles. All the parameters are reported in dimensionless units with $\sigma_1=1$, $\epsilon_1=1$, $m_1=1$, and $m_2=\sigma_2^3$ . The three interaction parameters $\sigma_2$, $\epsilon_2$, and $k_{12}$ and the composition parameter (the mole fraction $X_1$) are the adjustable parameters. First, an NPT simulation at temperature of 0.65 and a pressure of 0.05 was run for 5 million steps out of which the last 4 million steps were used to estimate the average box length. At the average box length, an NVT simulation at temperature of 0.65 was carried out for 5 million steps out of which the last 4 million steps were used to estimate the average total energy. At this average box length and total energy, twelve independent NVE simulations of 200 million steps were carried out. The pressure tensor was stored every five steps and the Einstein-Helfand equation was used to compute the viscosity \cite{allen2017,grest_jcp_1997,livejournal}. The 95\% confidence intervals were computed as two times the standard deviation of the shear viscosity values estimated across the independent runs. A time step of 0.001 and a cutoff radius of 4$\sigma_1$ was used for all simulations. LAMMPS package (7 Aug 2019 version) was used to carry out all the MD simulations \cite{LAMMPS2022}.

\clearpage
\section{Results}

\subsection{Model Selection and Performance Estimation} \label{ssec:results-model-sel}

\subsubsection{SS-CV vs KFS-CV} \label{ssec:results-sscv-kfscv}

\begin{figure}[h!]
\centering
    \includegraphics[width=\linewidth]{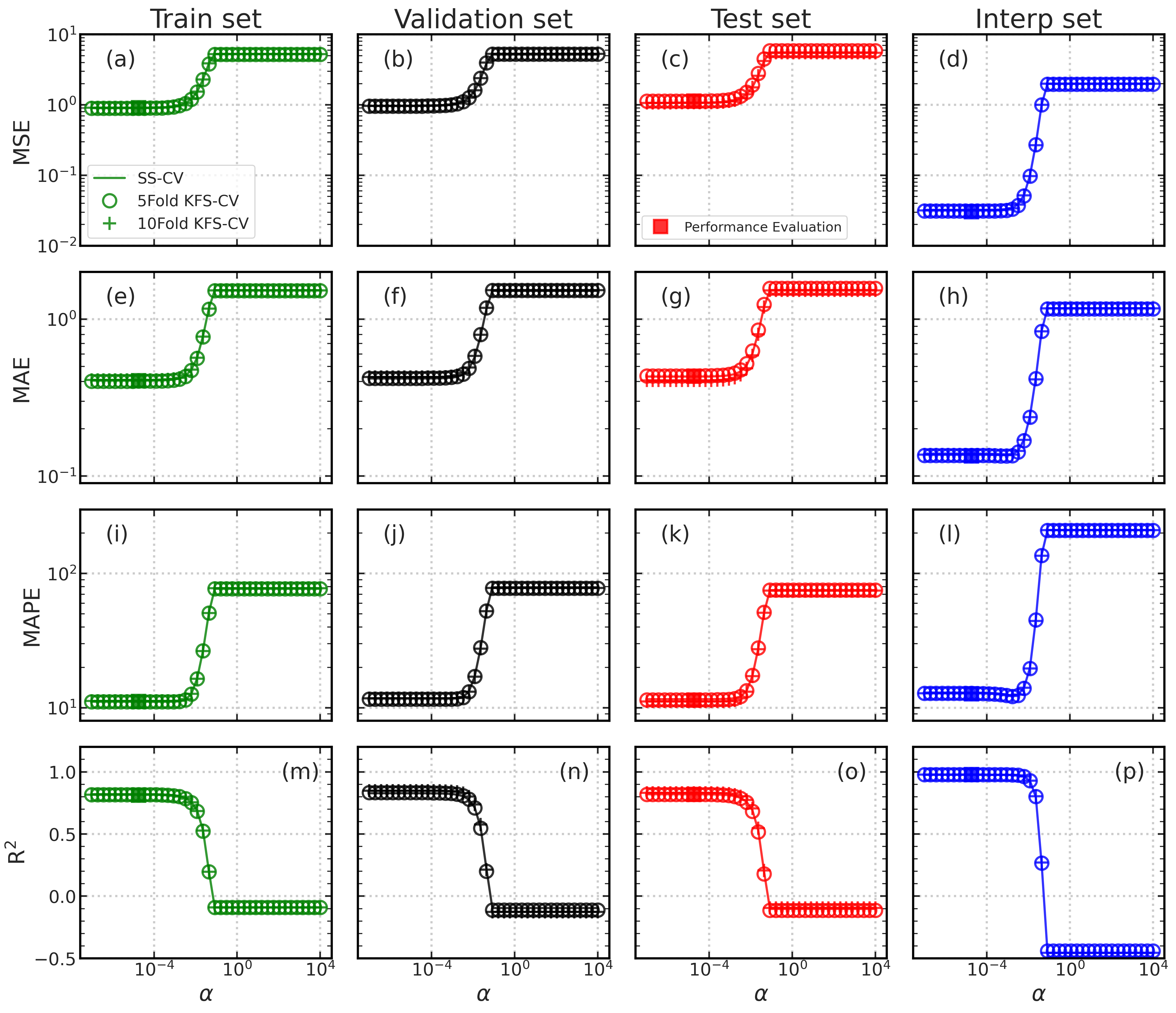}
    \caption{\textbf{Comparison of mean errors from SS-CV vs KFS-CV on LASSO model:} (a)-(d) show the \textit{mean} MSE values of Train, Validation, Test and Interpolation sets respectively, plotted against the LASSO hyperparameter ($\alpha$). (e)-(h) show the \textit{mean} MAE values of Train, Validation, Test and Interpolation sets respectively, plotted against the LASSO hyperparameter ($\alpha$). (i)-(l) show the \textit{mean} MAPE values of Train, Validation, Test and Interpolation sets respectively, plotted against the LASSO hyperparameter ($\alpha$). (m)-(p) show the \textit{mean} R$^2$ values of Train, Validation, Test and Interpolation sets respectively, plotted against the LASSO hyperparameter ($\alpha$). In all plots, the bold line corresponds to the SS-CV estimates, the circles correspond to 5fold KFS-CV estimates, the plus symbols correspond to 10fold KFS-CV estimates. Also, the estimates of the various metrics from an explicit performance evaluation step are shown as filled squares. See section \ref{ssec:comp-details-model-sel} for details on the computation of these estimates. The estimates from SS-CV, 5fold and 10fold KFS-CV of the mean errors seem to behave almost identically across different metrics.}
\label{sfig:ss-vs-kfs-mean}
\end{figure} 

\begin{figure}[h!]
\centering
    \includegraphics[width=\linewidth]{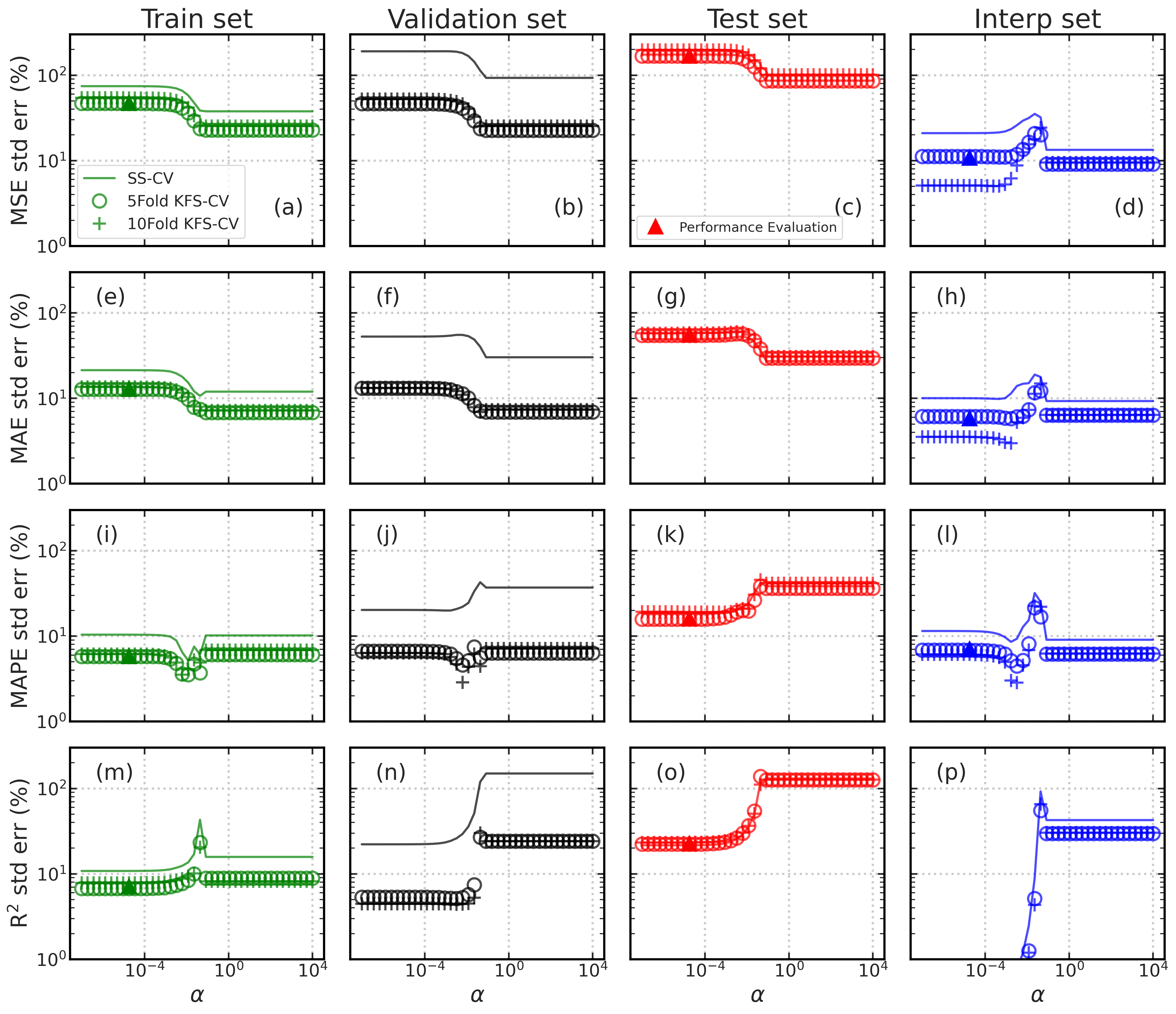}
    \caption{\textbf{Comparison of variance of errors from SS-CV vs KFS-CV on LASSO model:} (a)-(d) show the \textit{relative standard errors} of MSE of Train, Validation, Test and Interpolation sets respectively, plotted against the LASSO hyperparameter ($\alpha$). (e)-(h) show the \textit{relative standard errors} of MAE of Train, Validation, Test and Interpolation sets respectively, plotted against the LASSO hyperparameter ($\alpha$). (i)-(l) show the \textit{relative standard errors} of MAPE of Train, Validation, Test and Interpolation sets respectively, plotted against the LASSO hyperparameter ($\alpha$). (m)-(p) show the \textit{relative standard errors} of R$^2$ of Train, Validation, Test and Interpolation sets respectively, plotted against the LASSO hyperparameter ($\alpha$). In all plots, the bold line corresponds to the SS-CV estimates, the circles correspond to 5fold KFS-CV estimates, the plus symbols correspond to 10fold KFS-CV estimates. Also, the estimates of the various metrics from an explicit performance evaluation step are shown as filled squares. See section \ref{ssec:comp-details-model-sel} for details on the computation of these estimates. In contrast to the mean estimates (Figure \ref{sfig:ss-vs-kfs-mean}), the estimates of relative standard errors (variance) from SS-CV are larger than those from 5fold and 10fold KFS-CV computed on Train and Validation sets. However, the relative standard errors on the Test set behave almost identically across the three CV procedures and different metrics. Also, the variance of the Validation error across different metrics seems to vary widely (5-200\%) with MSE having the highest variance followed by MAE, MAPE, and R$^2$.}
\label{sfig:ss-vs-kfs-relerr}
\end{figure}

\begin{figure}[h!]
\centering
    \includegraphics[width=0.75\linewidth]{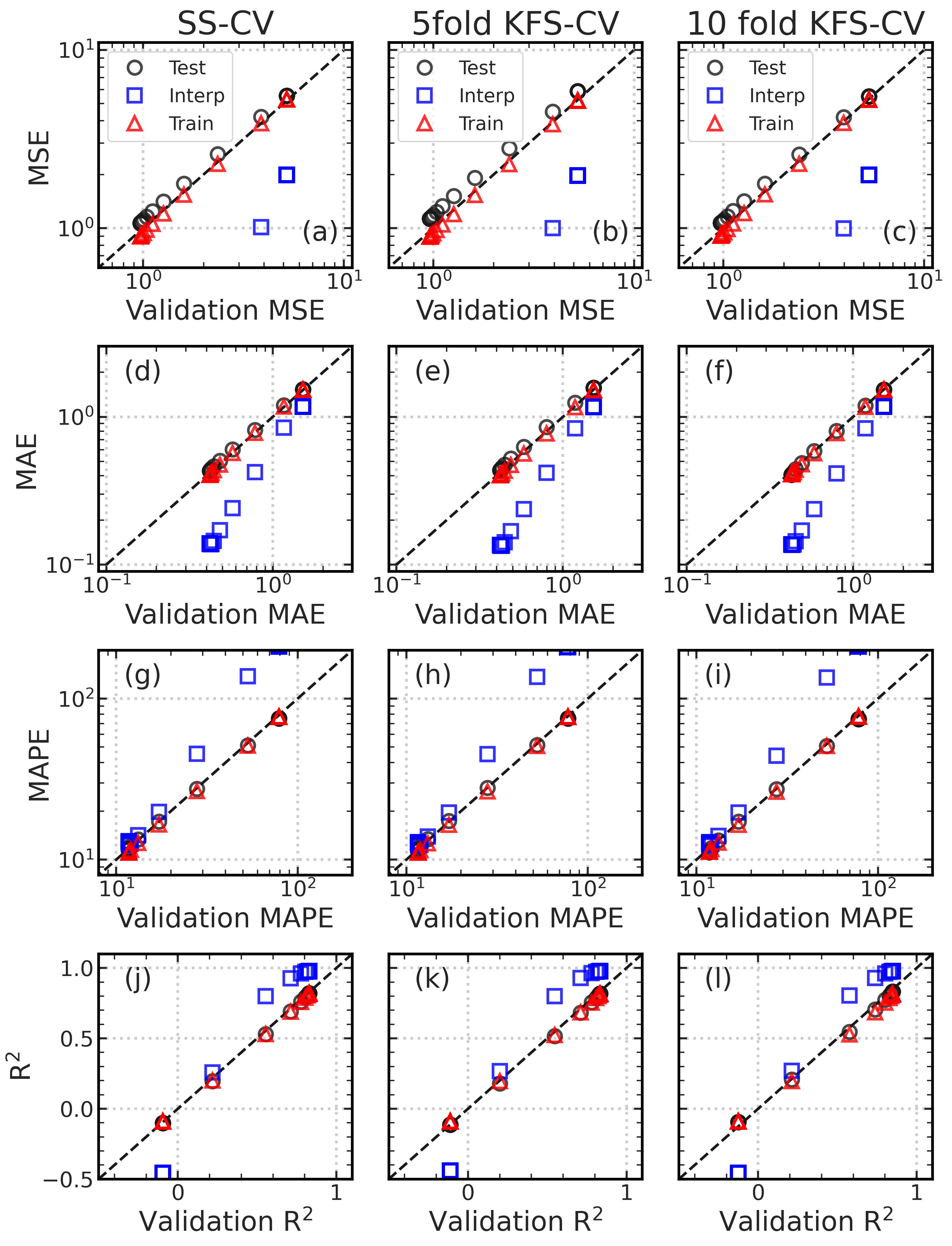}
    \caption{\textbf{Comparison of mean Test and Validation errors from SS-CV vs KFS-CV on LASSO model:} (a)-(c) show the comparison of Train, Test, and Interpolation MSE against the Validation MSE estimated using SS-CV, 5fold and 10fold KFS-CV respectively. (d)-(f) show the comparison of Train, Test, and Interpolation MAE against the Validation MAE estimated using SS-CV, 5fold and 10fold KFS-CV respectively. (g)-(i) show the comparison of Train, Test, and Interpolation MAPE against the Validation MAPE estimated using SS-CV, 5fold and 10fold KFS-CV respectively. (j)-(l) show the comparison of Train, Test, and Interpolation R$^2$ against the Validation R$^2$ estimated using SS-CV, 5fold and 10fold KFS-CV respectively. The average Test and Validation errors are well correlated across different CV procedures and error metrics.}
\label{sfig:ss-vs-kfs-test-val}
\end{figure}

\clearpage
\subsubsection{Selected Hyperparameters} \label{ssec:results-hyperpar-sel}

\begin{table}
\centering
\renewcommand{\arraystretch}{1.5}
\begin{tabular}{|c|c|c|c|c|c|} 
\hline
\multirow{2}{*}{Feature set} & \multirow{2}{*}{Hyperparameter} & \multicolumn{4}{c|}{Metrics for Validation landscape}  \\ 
\cline{3-6}
                             &                                 & MSE      & \textbf{MAE}      & MAPE     & R$^{2}$           \\ 
\hline\hline
AllMD                        & $\alpha$       & 1.00E-07 & \textbf{1.00E-07} & 5.99E-04 & 1.00E-07             \\
\hline
PreMD                        & $\alpha$       & 7.74E-03 & 5.99E-03 & 1.00E-07 & 1.00E-02             \\
\hline
\end{tabular}
\caption{\textbf{LASSO hyperparameters:} Values of the selected hyperparameters based on optimizing the average Validation error of the corresponding metric estimated using 5fold KFS-CV. MAE metric is used to select the final hyperparameters at which the ensemble models are trained (shown in bold).}
\label{stab:best-hyp-lasso}
\end{table}

\begin{table}
\centering
\renewcommand{\arraystretch}{1.5}
\begin{tabular}{|c|c|c|c|c|c|} 
\hline
\multirow{2}{*}{Feature set} & \multirow{2}{*}{Hyperparameter} & \multicolumn{4}{c|}{Metrics for Validation landscape}  \\ 
\cline{3-6}
                             &                                 & MSE      & \textbf{MAE}      & MAPE     & R$^{2}$            \\ 
\hline\hline
\multirow{2}{*}{AllMD }      & $\alpha$       & 5.99E-04 & \textbf{1.67E-04} & 4.64E-05 & 5.99E-04              \\
                             & $\gamma$       & 7.85E-01 & \textbf{3.79E-01} & 2.64E-01 & 7.85E-01              \\ 
\cline{1-6}
\multirow{2}{*}{PreMD}       & $\alpha$       & 2.15E-03 & 5.99E-04 & 1.67E-04 & 2.15E-03              \\
                             & $\gamma$       & 1.13E+00 & 7.85E-01 & 5.46E-01 & 1.13E+00              \\
\hline
\end{tabular}
\caption{\textbf{KRR hyperparameters:} Values of the selected hyperparameters based on optimizing the average Validation error of the corresponding metric estimated using 5fold KFS-CV. MAE metric is used to select the final hyperparameters at which the ensemble models are trained (shown in bold).}
\label{stab:best-hyp-krr}
\end{table}

\begin{table}
\centering
\renewcommand{\arraystretch}{1.5}
\begin{tabular}{|c|c|c|c|c|c|} 
\hline
\multirow{2}{*}{Feature set} & \multirow{2}{*}{Hyperparameter} & \multicolumn{4}{c|}{Metrics for Validation landscape}  \\ 
\cline{3-6}
                             &                                 & MSE      & \textbf{MAE}      & MAPE     & R$^2$            \\ 
\hline\hline
\multirow{3}{*}{AllMD }      & $\epsilon$                         & 1.00E-03 & \textbf{1.00E-03} & 1.00E-03 & 1.00E-03              \\
                             & $\gamma$                           & 5.46E-01 & \textbf{3.79E-01} & 2.64E-01 & 3.79E-01              \\
                             & C                               & 1.00E+02 & \textbf{1.00E+02} & 1.00E+02 & 1.00E+02              \\ 
\hline
\multirow{3}{*}{PreMD}       & $\epsilon$                         & 1.00E-04 & 1.00E-03 & 1.00E-03 & 1.00E-03              \\
                             & $\gamma$                           & 1.62E+00 & 7.85E-01 & 5.46E-01 & 1.13E+00              \\
                             & C                               & 3.16E+00 & 1.00E+01 & 1.00E+02 & 1.00E+01              \\
\hline
\end{tabular}
\caption{\textbf{SVR hyperparameters:} Values of the selected hyperparameters based on optimizing the average Validation error of the corresponding metric estimated using 5fold KFS-CV. MAE metric is used to select the final hyperparameters at which the ensemble models are trained (shown in bold).}
\label{stab:best-hyp-svr}
\end{table}

\begin{table}
\centering
\renewcommand{\arraystretch}{1.5}
\begin{tabular}{|c|c|c|c|c|c|} 
\hline
\multirow{2}{*}{Feature set} & \multirow{2}{*}{Hyperparameter} & \multicolumn{4}{c|}{Metrics for Validation landscape}  \\ 
\cline{3-6}
                             &                                 & MSE  & \textbf{MAE}  & MAPE & R$^2$                        \\ 
\hline\hline
\multirow{5}{*}{AllMD    }   & N estimators                    & 200  & \textbf{200}  & 200  & 200                               \\
                             & Min samples split               & 2    & \textbf{2}    & 2    & 2                                 \\
                             & Min samples leaf                & 1    & \textbf{1}    & 1    & 1                                 \\
                             & Max depth                       & None & \textbf{None} & 40   & 40                                \\
                             & Random seed                     & 8253 & \textbf{8253} & 1    & 8253                              \\ 
\hline
\multirow{5}{*}{PreMD}       & N estimators                    & 100  & 200  & 200  & 100                               \\
                             & Min samples split               & 2    & 2    & 2    & 2                                 \\
                             & Min samples leaf                & 1    & 1    & 1    & 1                                 \\
                             & Max depth                       & 10   & 40   & None & 10                                \\
                             & Random seed                     & 8253 & 8253 & 8253 & 8253                              \\
\hline

\end{tabular}
\caption{\textbf{RF hyperparameters:} Values of the selected hyperparameters based on optimizing the average Validation error of the corresponding metric estimated using 5fold KFS-CV. MAE metric is used to select the final hyperparameters at which the ensemble models are trained (shown in bold).}
\label{stab:best-hyp-rf}
\end{table}

\begin{table}
\centering
\renewcommand{\arraystretch}{1.5}
\begin{tabular}{|c|c|c|c|c|c|} 
\hline
\multirow{2}{*}{Feature set} & \multirow{2}{*}{Hyperparameter} & \multicolumn{4}{c|}{Metrics for Validation landscape}  \\ 
\cline{3-6}
                             &                                 & MSE      & \textbf{MAE}      & MAPE     & R$^2$            \\ 
\hline\hline
\multirow{3}{*}{PreMD}       & Neighbors                       & 4        & \textbf{5}        & 5        & 4                     \\
                             & P                               & 2        & \textbf{1}        & 1        & 2                     \\
                             & Weights                         & distance & \textbf{distance} & distance & distance              \\ 
\hline
\multirow{3}{*}{PreMD}       & Neighbors                       & 4        & 4        & 4        & 4                     \\
                             & P                               & 2        & 1        & 1        & 2                     \\
                             & Weights                         & distance & distance & distance & distance              \\
\hline
\end{tabular}
\caption{\textbf{KNN hyperparameters:} Values of the selected hyperparameters based on optimizing the average Validation error of the corresponding metric estimated using 5fold KFS-CV. MAE metric is used to select the final hyperparameters at which the ensemble models are trained (shown in bold).}
\label{stab:best-hyp-knn}
\end{table}

\clearpage
\subsubsection{Performance Estimation} \label{ssec:results-perf-estimation}

\begin{sidewaystable}
\centering
\caption{\label{stab:perf-est-final} \textbf{Performance estimation:} The mean and standard errors on the test sets by various models evaluated using KFS-CV procedure. The standard errors are calculated as two times the standard deviation.}
\setlength{\tabcolsep}{8pt}
\renewcommand{\arraystretch}{2}
\begin{tabular}{|c|c|c|c|c|c|c|c|c|} 
\hline
      & \multicolumn{4}{c|}{Perfromance with postMD features   } & \multicolumn{4}{c|}{Perfromance with preMD features}  \\ 
\cline{2-9}
      & MSE       & MAE         & MAPE      & R$^2$                 & MSE       & MAE         & MAPE      & R$^2$              \\ 
\hline\hline
GPR   & 0.1 (0.2) & 0.08 (0.06) & 1.4 (0.5) & 0.99 (0.02)        & 0.2 (0.3) & 0.11 (0.08) & 2.3 (0.8) & 0.98 (0.04)     \\
KRR   & 0.1 (0.2) & 0.08 (0.06) & 1.5 (0.5) & 0.99 (0.02)        & 0.2 (0.3) & 0.12 (0.08) & 2.5 (0.9) & 0.98 (0.03)     \\
SVR   & 0.1 (0.2) & 0.07 (0.06) & 1.4 (0.4) & 0.99 (0.02)        & 0.2 (0.3) & 0.11 (0.08) & 2.4 (0.8) & 0.98 (0.03)     \\
ANN   & 0.2 (0.4) & 0.12 (0.08) & 2.7 (0.7) & 0.98 (0.04)        & 0.1 (0.3) & 0.11 (0.07) & 2.6 (0.8) & 0.98 (0.03)     \\
RF    & 0.6 (1)   & 0.3 (0.2)   & 7 (3)     & 0.9 (0.1)          & 0.8 (1)   & 0.4 (0.2)   & 11 (2)    & 0.9 (0.1)       \\
KNN   & 1 (2)     & 0.4 (0.3)   & 11 (3)    & 0.8 (0.2)          & 1 (2)     & 0.4 (0.2)   & 14 (3)    & 0.8 (0.2)       \\
LASSO & 1 (2)     & 0.4 (0.2)   & 12 (2)    & 0.8 (0.2)          & 2 (2)     & 0.8 (0.3)   & 24 (6)    & 0.7 (0.2)       \\
\hline
\end{tabular}
\end{sidewaystable}

\clearpage
\subsection{Model bias} \label{ssec:model-bias}



\clearpage
\subsubsection{Gaussian Process Regression}

\begin{figure}[h!]
\centering
    \includegraphics[width=0.9\linewidth]{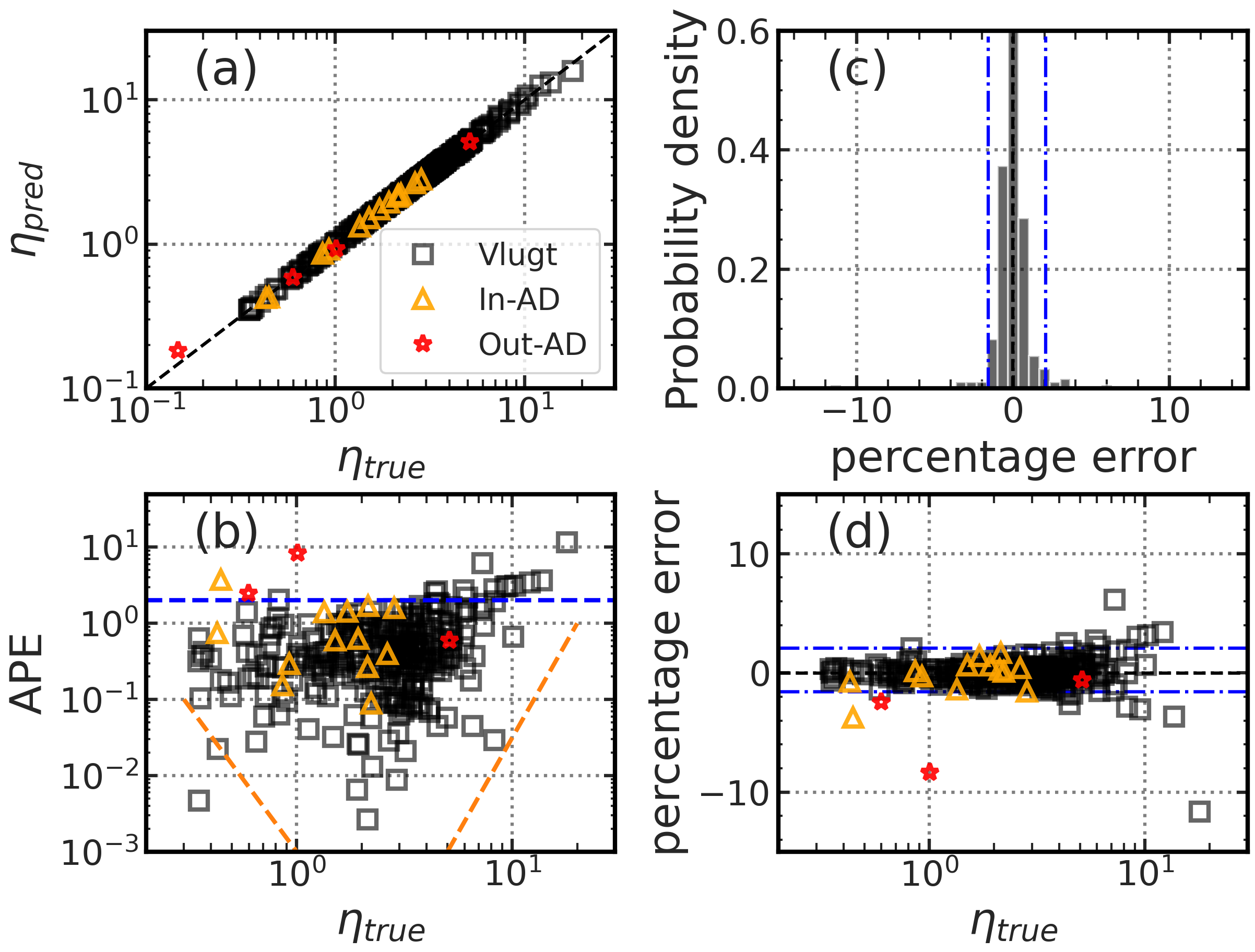}
    \caption{\textbf{AllMD-GPR model performance and bias:} (a) predicted viscosity plotted against true viscosity values. (b) Absolute Percentage Errors (APE) plotted against the corresponding true viscosity values. The relatively poor performance of the model at the extremal decades is highlighted by the orange dashed lines. The black horizontal dashed  line indicates the average APE of the Vlugt data. (c) Probability density of Percentage Errors (PE) on the entire Vlugt data shown in black . The black vertical dashed line indicates the value of the median PE and the blue vertical dashed lines indicate the 95    percentile range around the median. (d) Percentage Errors (PE) plotted against their corresponding true viscosity values. The black horizontal  dashed line indicates the value of the median PE and the blue horizontal dashed lines indicate the 95 percentile range around the median. All the estimates are from the ensemble of GPR models obtained after 5fold KFS-CV procedure. These GPR models are trained using six features called allMD features (See Computation Details section). The black squares represent the predictions on the entire Vlugt data, the orange triangles represent the predictions on the interpolation data points that are within the Applicability Domain, the red stars represent the predictions on the interpolation data points that are outside the Applicability Domain.} \label{sfig:modelperf-gpr}
\end{figure} 

\begin{figure}[h!]
\centering
    \includegraphics[width=0.9\linewidth]{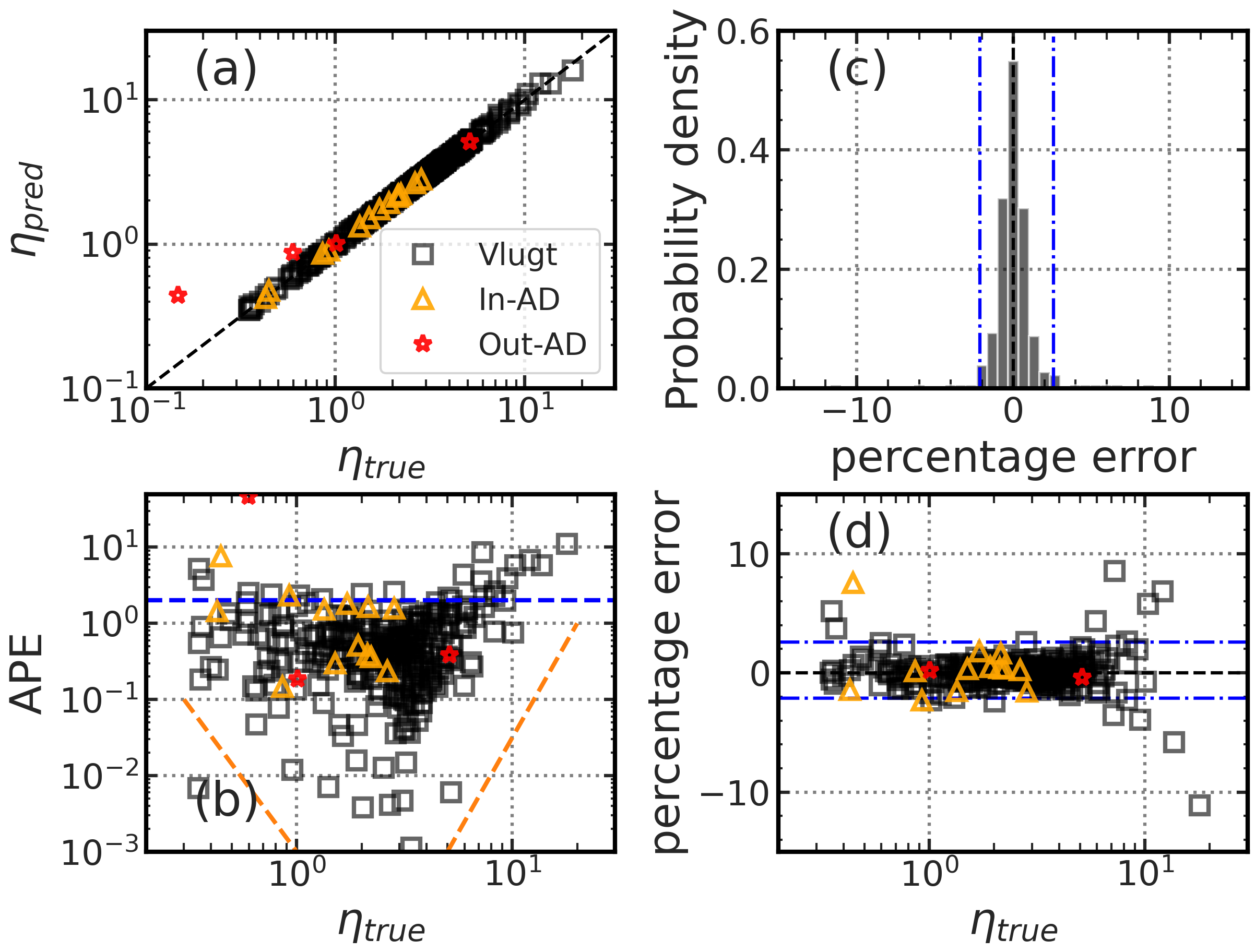}
    \caption{\textbf{PreMD-GPR model performance and bias:} (a) predicted viscosity plotted against true viscosity values. (b) Absolute Percentage Errors (APE) plotted against the corresponding true viscosity values. The relatively poor performance of the model at the extremal decades is highlighted by the orange dashed lines. The black horizontal dashed  line indicates the average APE of the Vlugt data. (c) Probability density of Percentage Errors (PE) on the entire Vlugt data shown in black . The black vertical dashed line indicates the value of the median PE and the blue vertical dashed lines indicate the 95    percentile range around the median. (d) Percentage Errors (PE) plotted against their corresponding true viscosity values. The black horizontal  dashed line indicates the value of the median PE and the blue horizontal dashed lines indicate the 95 percentile range around the median. All the estimates are from the ensemble of GPR models obtained after 5fold KFS-CV procedure. These GPR models are trained using four features called preMD features (See Computation Details section). The black squares represent the predictions on the entire Vlugt data, the orange triangles represent the predictions on the interpolation data points that are within the Applicability Domain, the red stars represent the predictions on the interpolation data points that are outside the Applicability Domain.} \label{sfig:modelperf-gpr-preMD}
\end{figure}

\clearpage
\subsubsection{Kernel Ridge Regression}

\begin{figure}[h!]
\centering
    \includegraphics[width=0.9\linewidth]{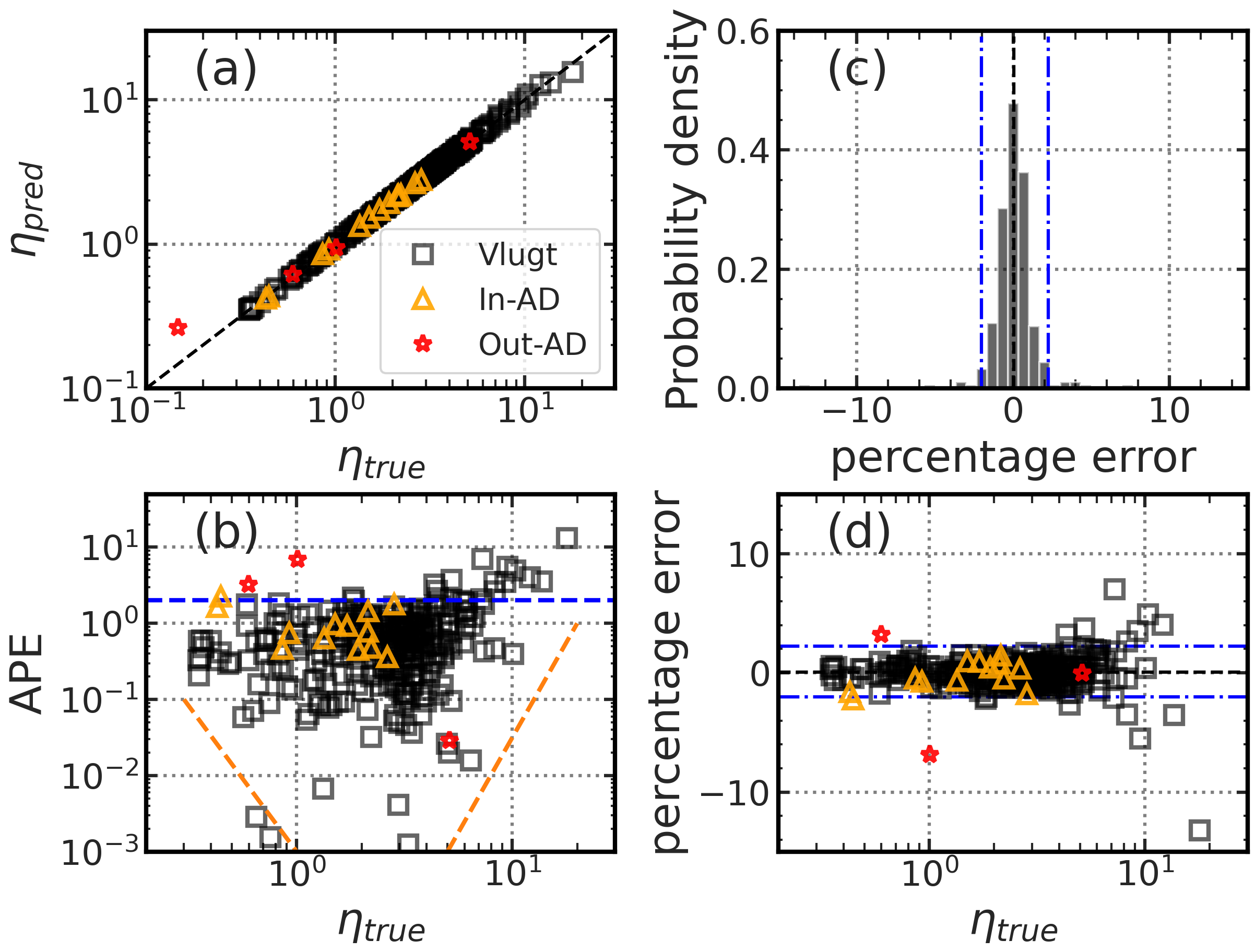}
    \caption{\textbf{AllMD-KRR model performance and bias:} (a) predicted viscosity plotted against true viscosity values. (b) Absolute Percentage Errors (APE) plotted against the corresponding true viscosity values. The relatively poor performance of the model at the extremal decades is highlighted by the orange dashed lines. The black horizontal dashed  line indicates the average APE of the Vlugt data. (c) Probability density of Percentage Errors (PE) on the entire Vlugt data shown in black . The black vertical dashed line indicates the value of the median PE and the blue vertical dashed lines indicate the 95    percentile range around the median. (d) Percentage Errors (PE) plotted against their corresponding true viscosity values. The black horizontal  dashed line indicates the value of the median PE and the blue horizontal dashed lines indicate the 95 percentile range around the median. All the estimates are from the ensemble of KRR models obtained after 5fold KFS-CV procedure. These KRR models are trained using six features called allMD features (See Computation Details section). The black squares represent the predictions on the entire Vlugt data, the orange triangles represent the predictions on the interpolation data points that are within the Applicability Domain, the red stars represent the predictions on the interpolation data points that are outside the Applicability Domain.} \label{sfig:modelperf-krr}
\end{figure} 

\begin{figure}[h!]
\centering
    \includegraphics[width=0.9\linewidth]{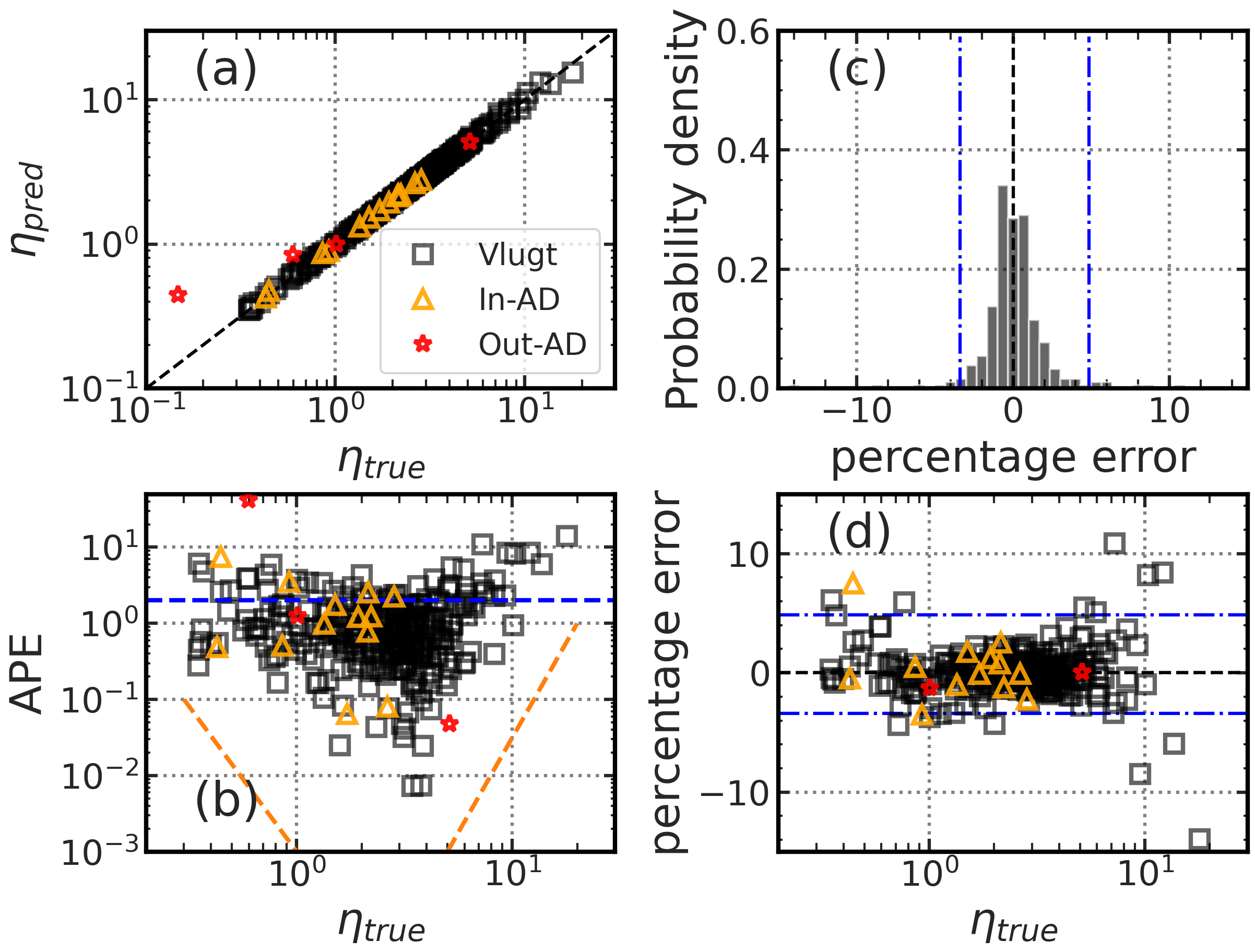}
    \caption{\textbf{PreMD-KRR model performance and bias:} (a) predicted viscosity plotted against true viscosity values. (b) Absolute Percentage Errors (APE) plotted against the corresponding true viscosity values. The relatively poor performance of the model at the extremal decades is highlighted by the orange dashed lines. The black horizontal dashed  line indicates the average APE of the Vlugt data. (c) Probability density of Percentage Errors (PE) on the entire Vlugt data shown in black . The black vertical dashed line indicates the value of the median PE and the blue vertical dashed lines indicate the 95    percentile range around the median. (d) Percentage Errors (PE) plotted against their corresponding true viscosity values. The black horizontal  dashed line indicates the value of the median PE and the blue horizontal dashed lines indicate the 95 percentile range around the median. All the estimates are from the ensemble of KRR models obtained after 5fold KFS-CV procedure. These KRR models are trained using four features called preMD features (See Computation Details section). The black squares represent the predictions on the entire Vlugt data, the orange triangles represent the predictions on the interpolation data points that are within the Applicability Domain, the red stars represent the predictions on the interpolation data points that are outside the Applicability Domain.} \label{sfig:modelperf-krr-preMD}
\end{figure}

\clearpage
\subsubsection{Support Vector Regression}

\begin{figure}[h!]
\centering
    \includegraphics[width=0.9\linewidth]{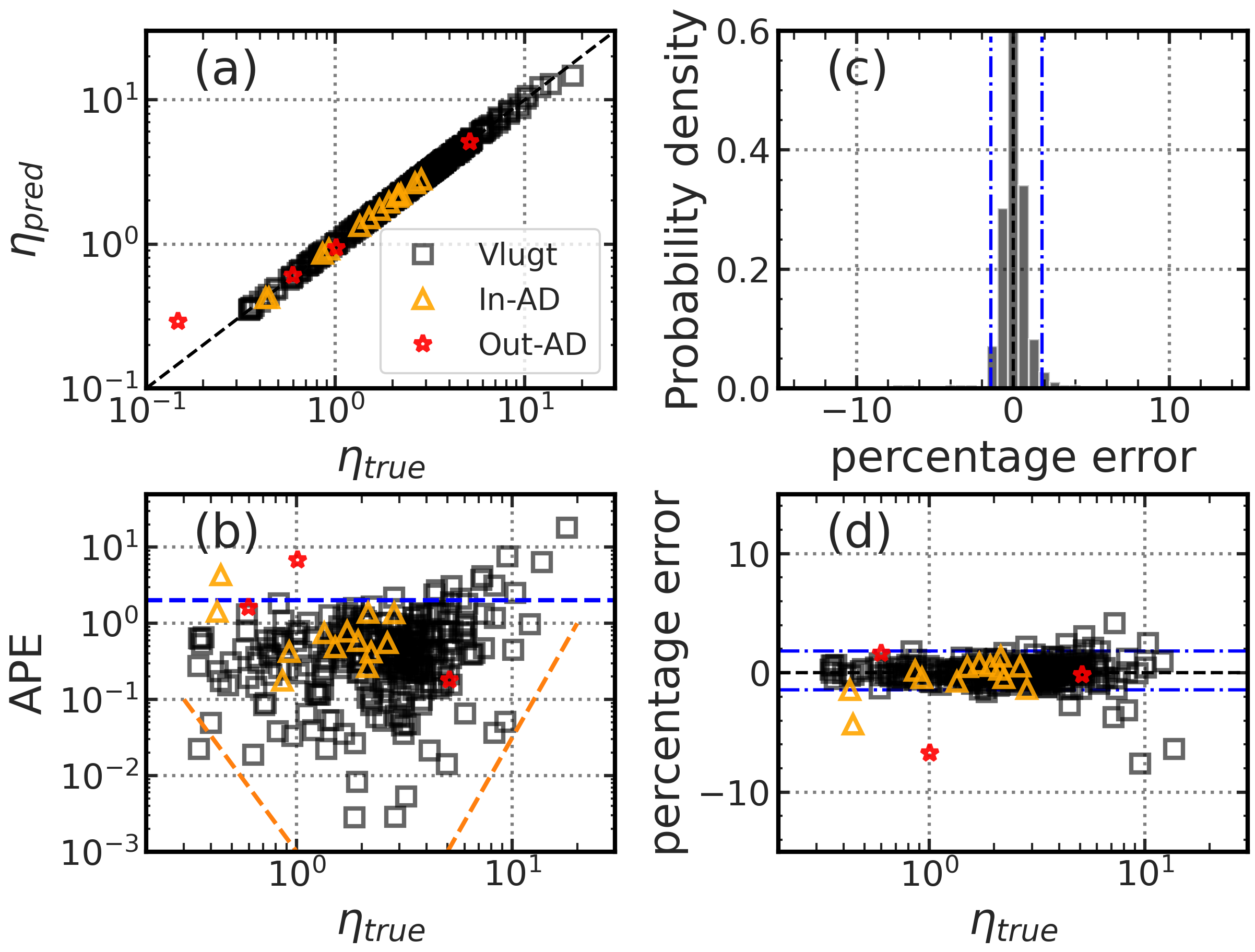}
    \caption{\textbf{AllMD-SVR model performance and bias:} (a) predicted viscosity plotted against true viscosity values. (b) Absolute Percentage Errors (APE) plotted against the corresponding true viscosity values. The relatively poor performance of the model at the extremal decades is highlighted by the orange dashed lines. The black horizontal dashed	line indicates the average APE of the Vlugt data. (c) Probability density of Percentage Errors (PE) on the entire Vlugt data shown in black . The black vertical dashed line indicates the value of the median PE and the blue vertical dashed lines indicate the 95	percentile range around the median. (d) Percentage Errors (PE) plotted against their corresponding true viscosity values. The black horizontal	dashed line indicates the value of the median PE and the blue horizontal dashed lines indicate the 95 percentile range around the median. All the estimates are from the ensemble of SVR models obtained after 5fold KFS-CV procedure. These SVR models are trained using six features called allMD features (See Computation Details section). The black squares represent the predictions on the entire Vlugt data, the orange triangles represent the predictions on the interpolation data points that are within the Applicability Domain, the red stars represent the predictions on the interpolation data points that are outside the Applicability Domain.} \label{sfig:modelperf-svr}
\end{figure} 

\begin{figure}[h!]
\centering
    \includegraphics[width=0.9\linewidth]{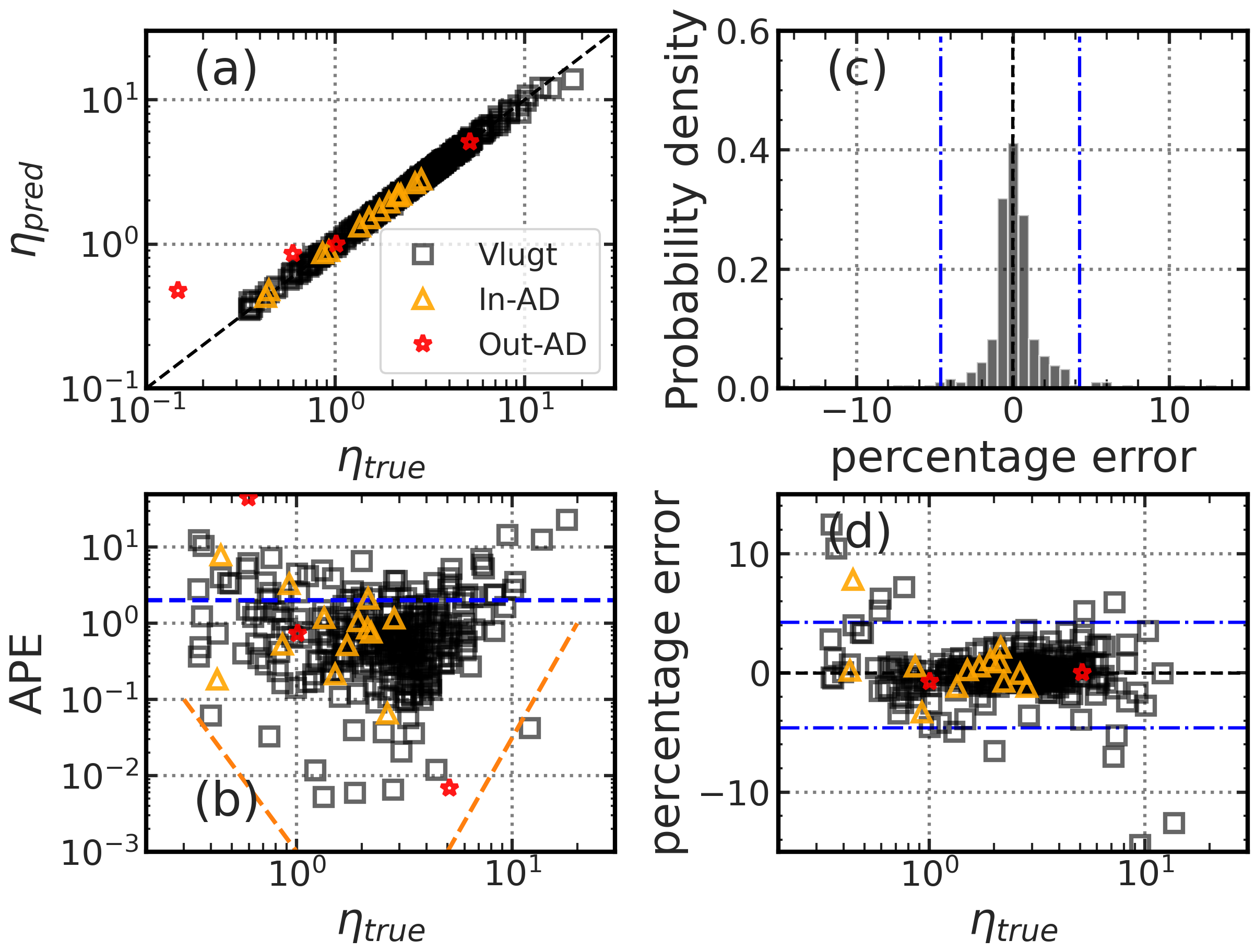}
    \caption{\textbf{PreMD-SVR model performance and bias:} (a) predicted viscosity plotted against true viscosity values. (b) Absolute Percentage Errors (APE) plotted against the corresponding true viscosity values. The relatively poor performance of the model at the extremal decades is highlighted by the orange dashed lines. The black horizontal dashed	line indicates the average APE of the Vlugt data. (c) Probability density of Percentage Errors (PE) on the entire Vlugt data shown in black . The black vertical dashed line indicates the value of the median PE and the blue vertical dashed lines indicate the 95	percentile range around the median. (d) Percentage Errors (PE) plotted against their corresponding true viscosity values. The black horizontal	dashed line indicates the value of the median PE and the blue horizontal dashed lines indicate the 95 percentile range around the median. All the estimates are from the ensemble of SVR models obtained after 5fold KFS-CV procedure. These SVR models are trained using four features called preMD features (See Computation Details section). The black squares represent the predictions on the entire Vlugt data, the orange triangles represent the predictions on the interpolation data points that are within the Applicability Domain, the red stars represent the predictions on the interpolation data points that are outside the Applicability Domain.} \label{sfig:modelperf-svr-preMD}
\end{figure}

\clearpage
\subsubsection{Random Forest}

\begin{figure}[h!]
\centering
    \includegraphics[width=0.9\linewidth]{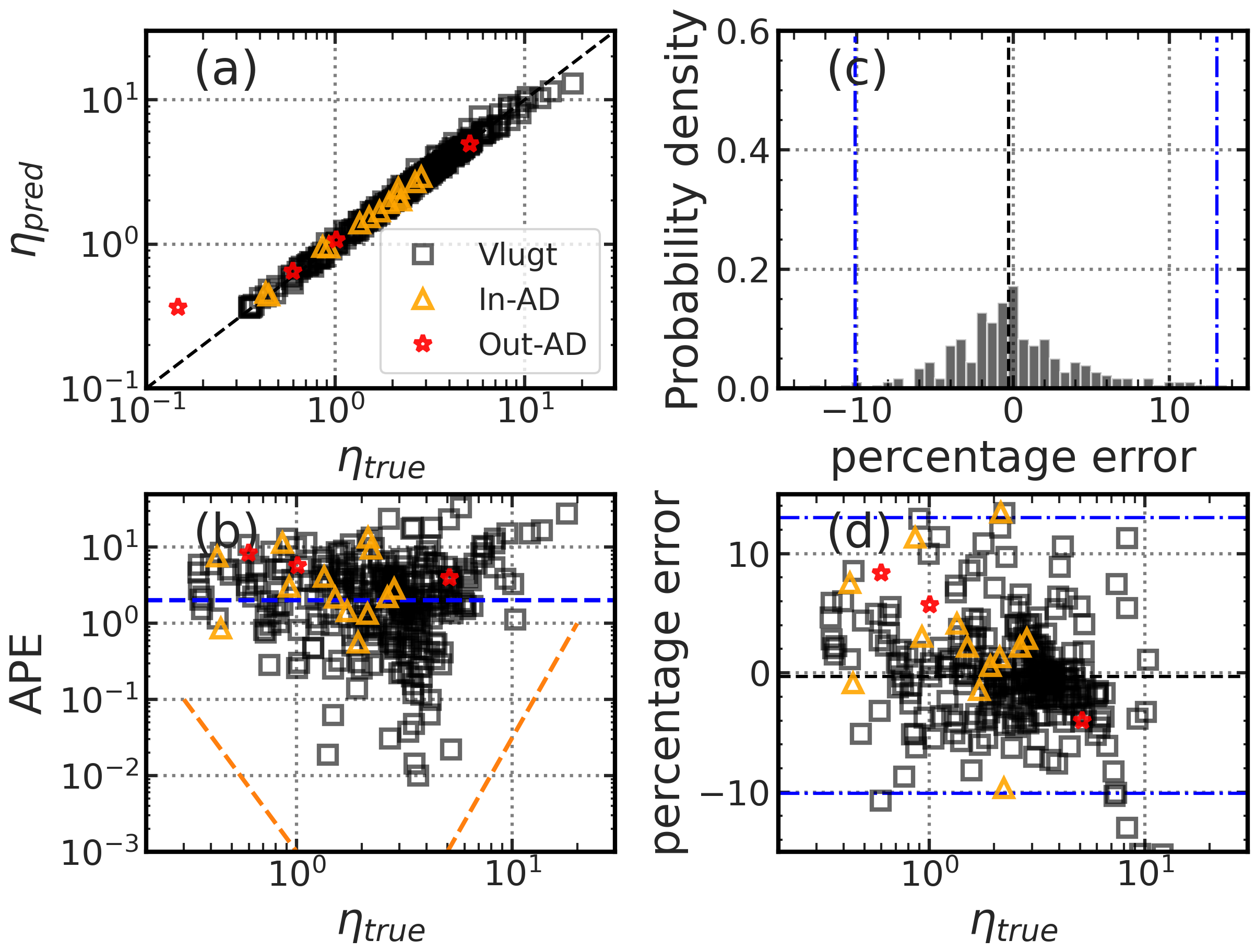}
    \caption{\textbf{AllMD-RF model performance and bias:} (a) predicted viscosity plotted against true viscosity values. (b) Absolute Percentage Errors (APE) plotted against the corresponding true viscosity values. The relatively poor performance of the model at the extremal decades is highlighted by the orange dashed lines. The black horizontal dashed	line indicates the average APE of the Vlugt data. (c) Probability density of Percentage Errors (PE) on the entire Vlugt data shown in black . The black vertical dashed line indicates the value of the median PE and the blue vertical dashed lines indicate the 95	percentile range around the median. (d) Percentage Errors (PE) plotted against their corresponding true viscosity values. The black horizontal	dashed line indicates the value of the median PE and the blue horizontal dashed lines indicate the 95 percentile range around the median. All the estimates are from the ensemble of RF models obtained after 5fold KFS-CV procedure. These RF models are trained using six features called allMD features (See Computation Details section). The black squares represent the predictions on the entire Vlugt data, the orange triangles represent the predictions on the interpolation data points that are within the Applicability Domain, the red stars represent the predictions on the interpolation data points that are outside the Applicability Domain.} \label{sfig:modelperf-rf}
\end{figure} 

\begin{figure}[h!]
\centering
    \includegraphics[width=0.9\linewidth]{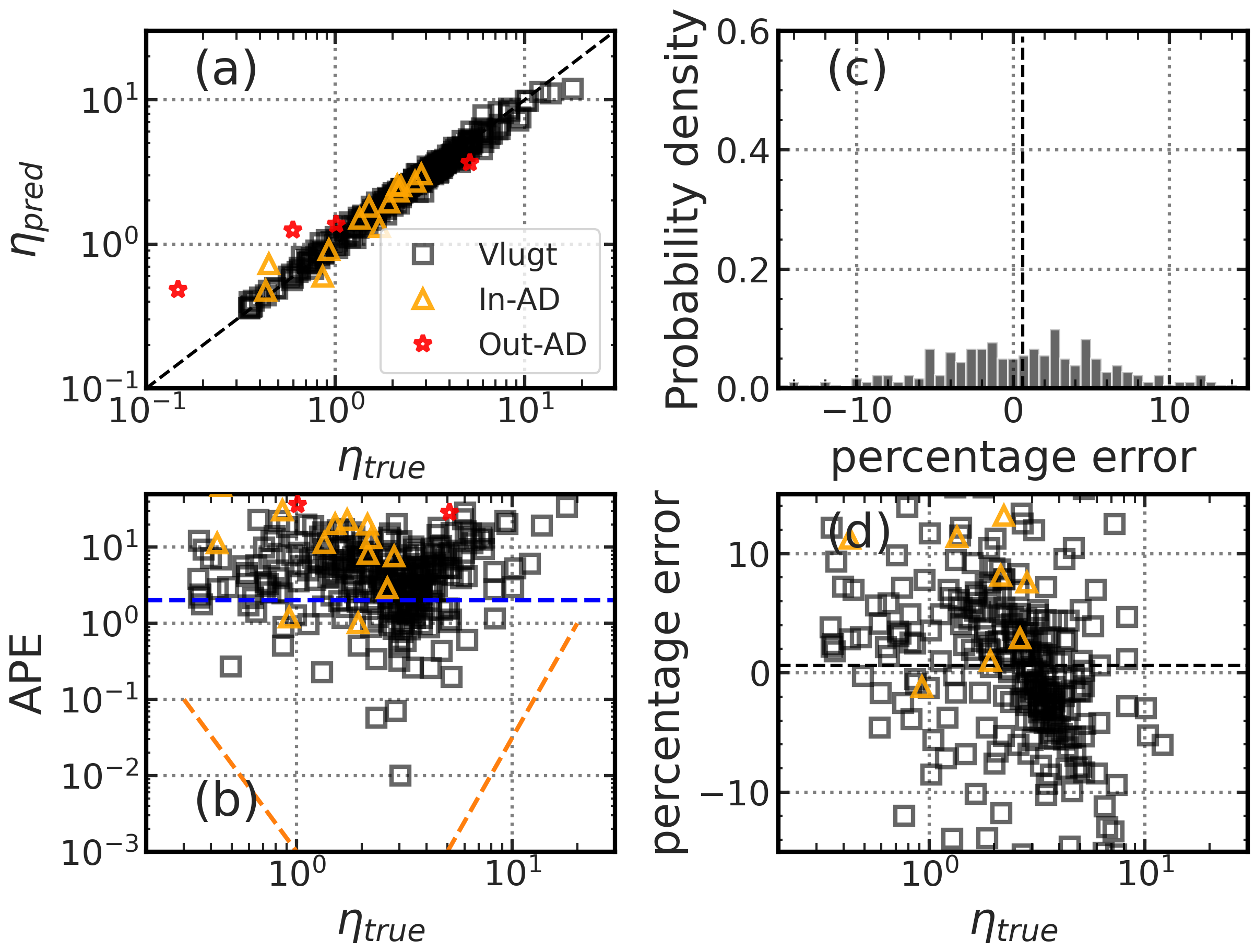}
    \caption{\textbf{PreMD-RF model performance and bias:} (a) predicted viscosity plotted against true viscosity values. (b) Absolute Percentage Errors (APE) plotted against the corresponding true viscosity values. The relatively poor performance of the model at the extremal decades is highlighted by the orange dashed lines. The black horizontal dashed	line indicates the average APE of the Vlugt data. (c) Probability density of Percentage Errors (PE) on the entire Vlugt data shown in black . The black vertical dashed line indicates the value of the median PE and the blue vertical dashed lines indicate the 95	percentile range around the median. (d) Percentage Errors (PE) plotted against their corresponding true viscosity values. The black horizontal	dashed line indicates the value of the median PE and the blue horizontal dashed lines indicate the 95 percentile range around the median. All the estimates are from the ensemble of RF models obtained after 5fold KFS-CV procedure. These RF models are trained using four features called preMD features (See Computation Details section). The black squares represent the predictions on the entire Vlugt data, the orange triangles represent the predictions on the interpolation data points that are within the Applicability Domain, the red stars represent the predictions on the interpolation data points that are outside the Applicability Domain.} \label{sfig:modelperf-rf-preMD}
\end{figure}

\clearpage
\subsubsection{k-Nearest Neighbors}

\begin{figure}[h!]
\centering
    \includegraphics[width=0.9\linewidth]{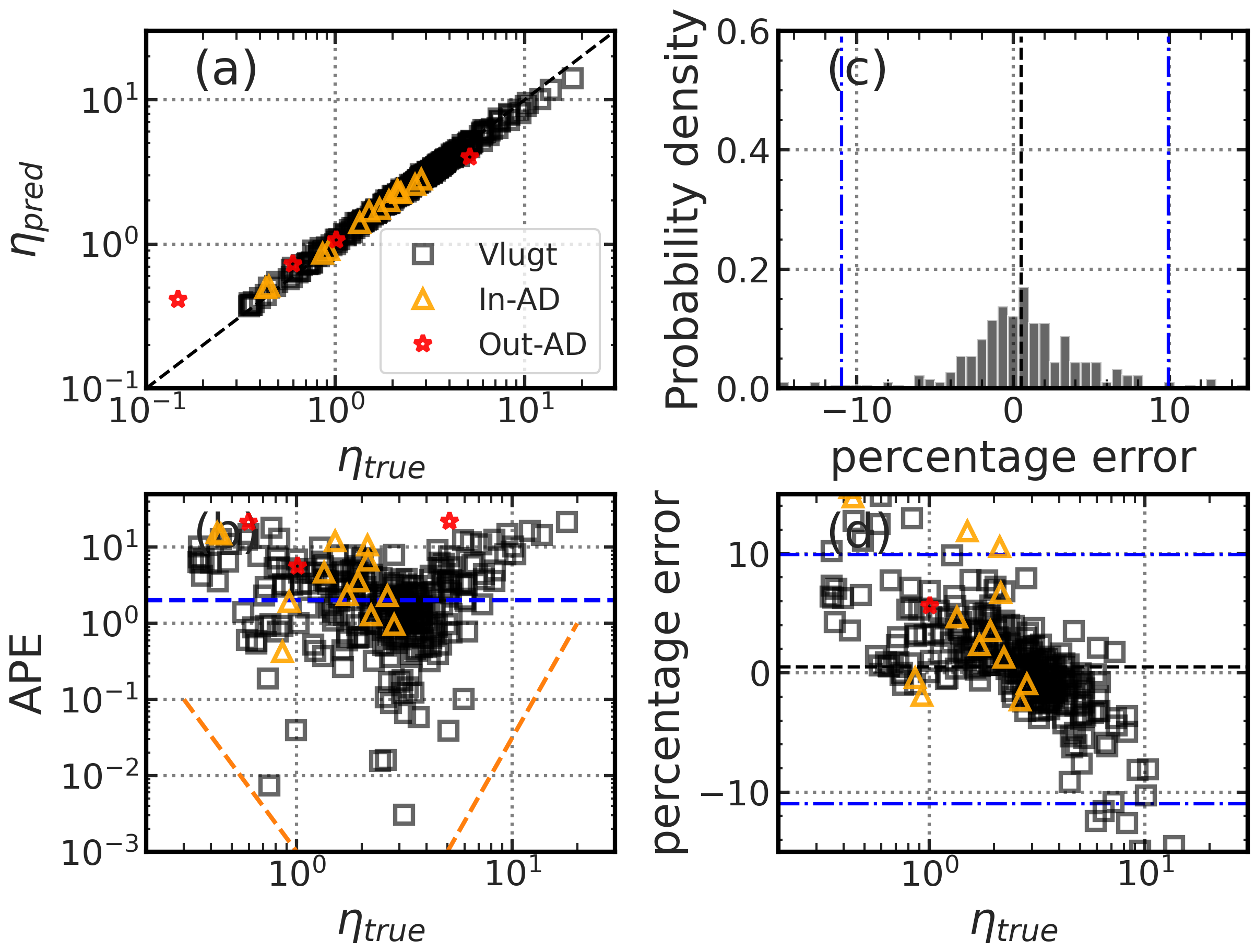}
    \caption{\textbf{AllMD-KNN model performance and bias:} (a) predicted viscosity plotted against true viscosity values. (b) Absolute Percentage Errors (APE) plotted against the corresponding true viscosity values. The relatively poor performance of the model at the extremal decades is highlighted by the orange dashed lines. The black horizontal dashed	line indicates the average APE of the Vlugt data. (c) Probability density of Percentage Errors (PE) on the entire Vlugt data shown in black . The black vertical dashed line indicates the value of the median PE and the blue vertical dashed lines indicate the 95	percentile range around the median. (d) Percentage Errors (PE) plotted against their corresponding true viscosity values. The black horizontal	dashed line indicates the value of the median PE and the blue horizontal dashed lines indicate the 95 percentile range around the median. All the estimates are from the ensemble of KNN models obtained after 5fold KFS-CV procedure. These KNN models are trained using six features called allMD features (See Computation Details section). The black squares represent the predictions on the entire Vlugt data, the orange triangles represent the predictions on the interpolation data points that are within the Applicability Domain, the red stars represent the predictions on the interpolation data points that are outside the Applicability Domain.} \label{sfig:modelperf-knn}
\end{figure} 

\begin{figure}[h!]
\centering
    \includegraphics[width=0.9\linewidth]{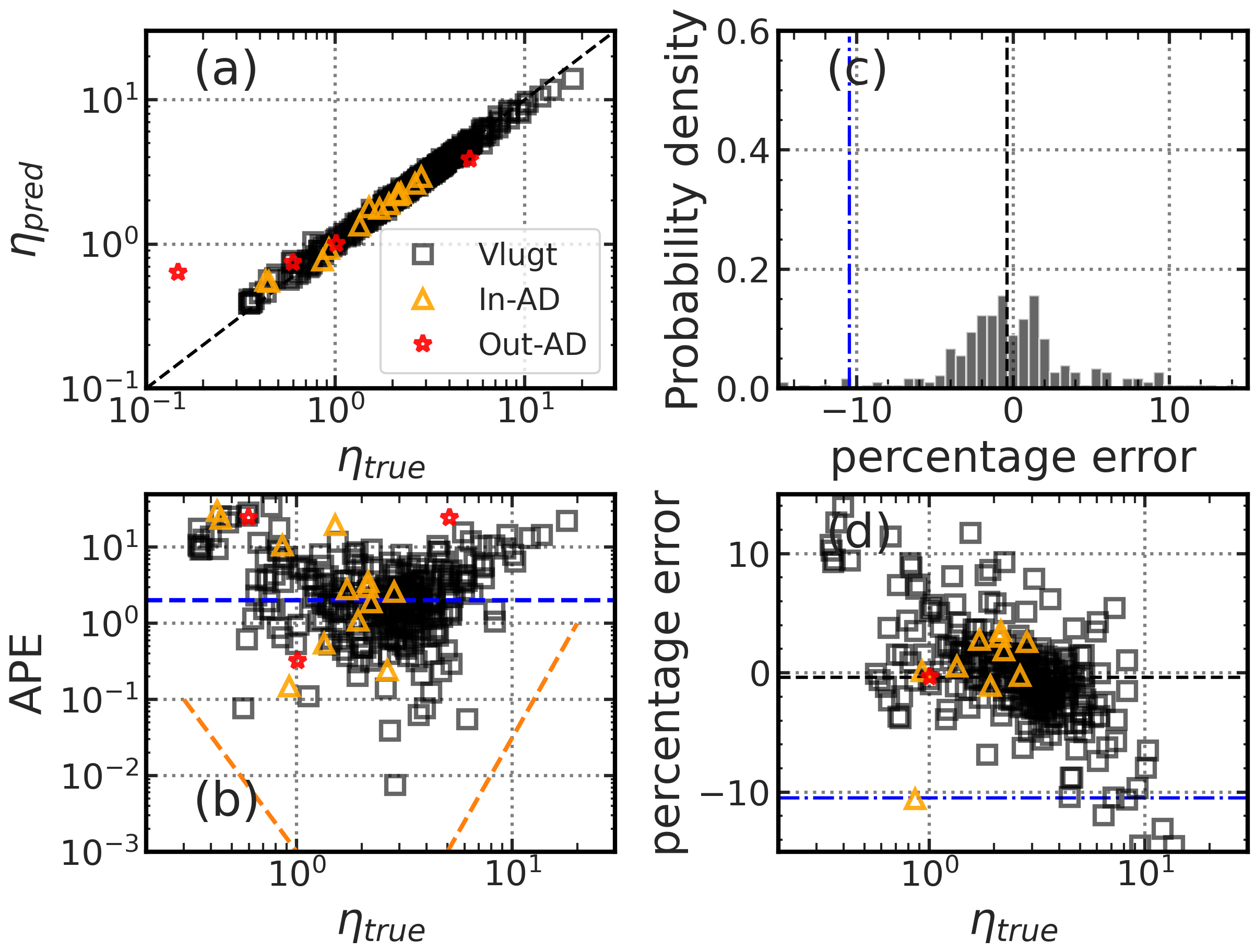}
    \caption{\textbf{PreMD-KNN model performance and bias:} (a) predicted viscosity plotted against true viscosity values. (b) Absolute Percentage Errors (APE) plotted against the corresponding true viscosity values. The relatively poor performance of the model at the extremal decades is highlighted by the orange dashed lines. The black horizontal dashed	line indicates the average APE of the Vlugt data. (c) Probability density of Percentage Errors (PE) on the entire Vlugt data shown in black . The black vertical dashed line indicates the value of the median PE and the blue vertical dashed lines indicate the 95	percentile range around the median. (d) Percentage Errors (PE) plotted against their corresponding true viscosity values. The black horizontal	dashed line indicates the value of the median PE and the blue horizontal dashed lines indicate the 95 percentile range around the median. All the estimates are from the ensemble of KNN models obtained after 5fold KFS-CV procedure. These KNN models are trained using four features called preMD features (See Computation Details section). The black squares represent the predictions on the entire Vlugt data, the orange triangles represent the predictions on the interpolation data points that are within the Applicability Domain, the red stars represent the predictions on the interpolation data points that are outside the Applicability Domain.} \label{sfig:modelperf-knn-preMD}
\end{figure}

\clearpage
\subsubsection{LASSO}

\begin{figure}[h!]
\centering
    \includegraphics[width=0.9\linewidth]{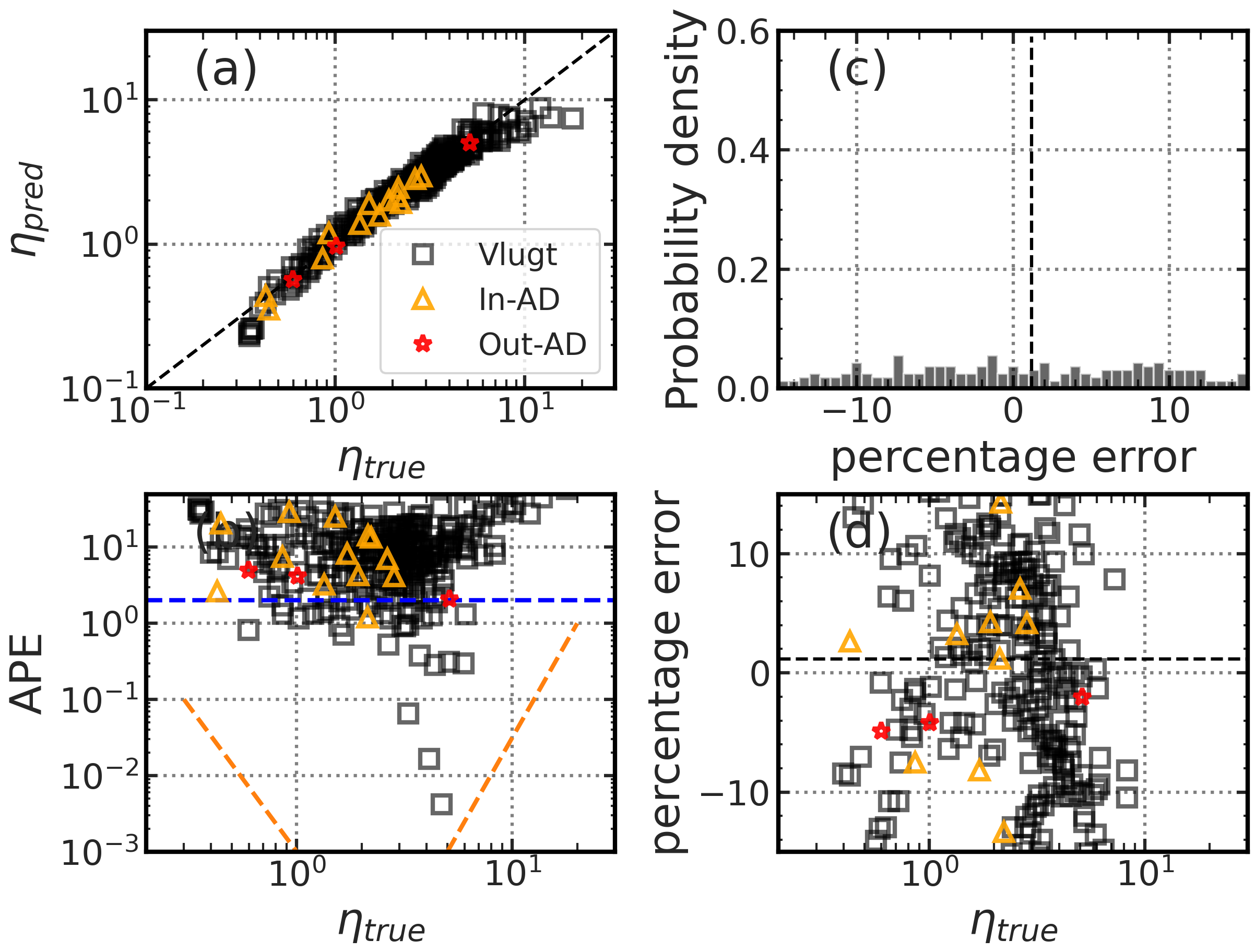}
    \caption{\textbf{AllMD-LASSO model performance and bias:} (a) predicted viscosity plotted against true viscosity values. (b) Absolute Percentage Errors (APE) plotted against the corresponding true viscosity values. The relatively poor performance of the model at the extremal decades is highlighted by the orange dashed lines. The black horizontal dashed	line indicates the average APE of the Vlugt data. (c) Probability density of Percentage Errors (PE) on the entire Vlugt data shown in black . The black vertical dashed line indicates the value of the median PE and the blue vertical dashed lines indicate the 95 percentile range around the median. (d) Percentage Errors (PE) plotted against their corresponding true viscosity values. The black horizontal	dashed line indicates the value of the median PE and the blue horizontal dashed lines indicate the 95 percentile range around the	median. All the estimates are from the ensemble of LASSO models obtained after 5fold KFS-CV procedure. These LASSO models are trained using six features called allMD features (See Computation Details section). The black squares represent the predictions on the entire Vlugt data, the orange triangles represent the predictions on the interpolation data points that are within the Applicability Domain, the red stars represent the predictions on the interpolation data points that are outside the Applicability Domain.} \label{sfig:modelperf-lasso}
\end{figure}

\begin{figure}[h!]
\centering
    \includegraphics[width=0.9\linewidth]{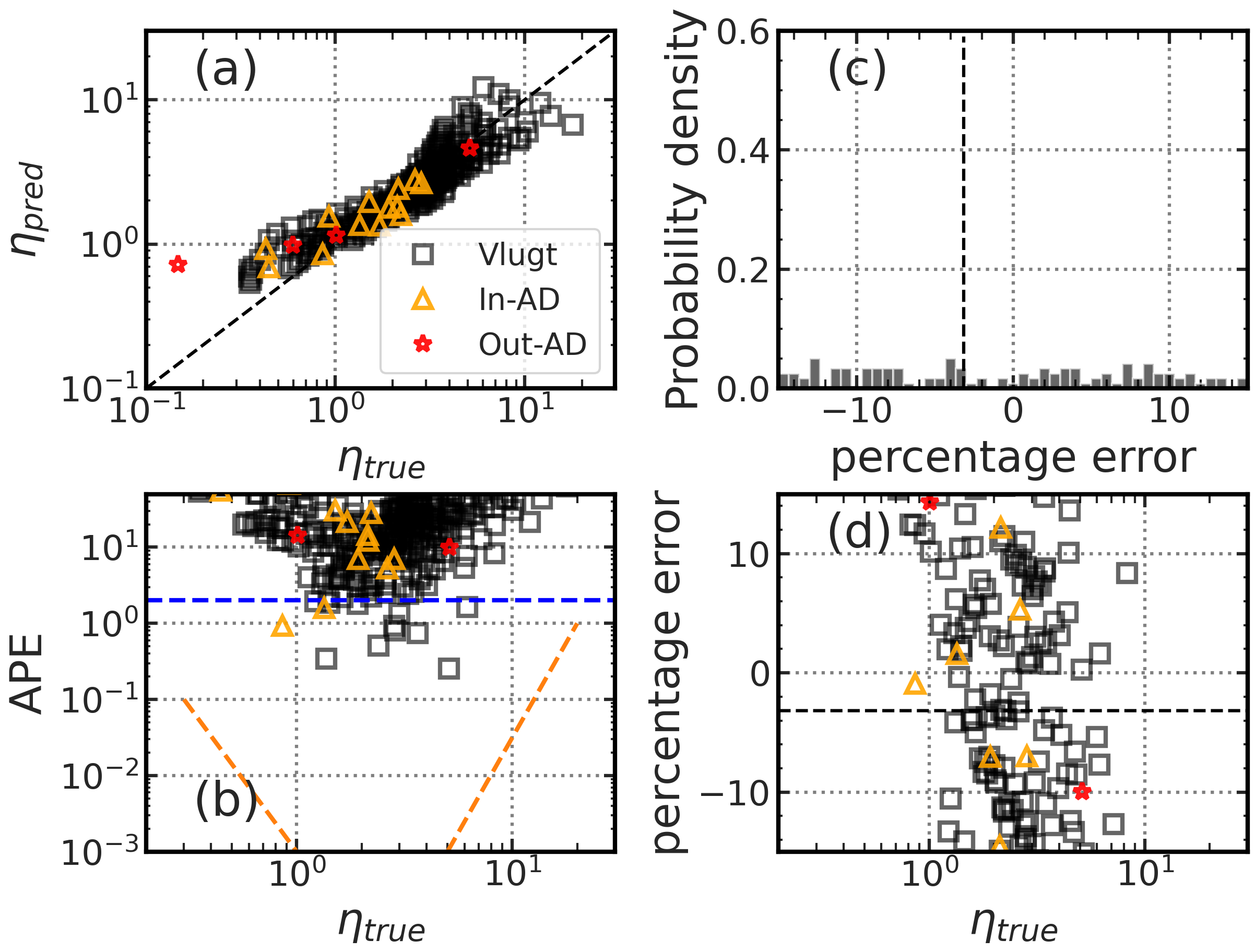}
    \caption{\textbf{PreMD-LASSO model performance and bias:} (a) predicted viscosity plotted against true viscosity values. (b) Absolute Percentage Errors (APE) plotted against the corresponding true viscosity values. The relatively poor performance of the model at the extremal decades is highlighted by the orange dashed lines. The black horizontal dashed	line indicates the average APE of the Vlugt data. (c) Probability density of Percentage Errors (PE) on the entire Vlugt data shown in black . The black vertical dashed line indicates the value of the median PE and the blue vertical dashed lines indicate the 95 percentile range around the median. (d) Percentage Errors (PE) plotted against their corresponding true viscosity values. The black horizontal	dashed line indicates the value of the median PE and the blue horizontal dashed lines indicate the 95 percentile range around the	median. All the estimates are from the ensemble of LASSO models obtained after 5fold KFS-CV procedure. These LASSO models are trained using four features called preMD features (See Computation Details section). The black squares represent the predictions on the entire Vlugt data, the orange triangles represent the predictions on the interpolation data points that are within the Applicability Domain, the red stars represent the predictions on the interpolation data points that are outside the Applicability Domain.} \label{sfig:modelperf-lasso-preMD}
\end{figure}

\clearpage
\subsection{Model Interpolation}
\subsubsection{ANN}
\begin{figure}[h!]
\centering
    \includegraphics[width=\linewidth]{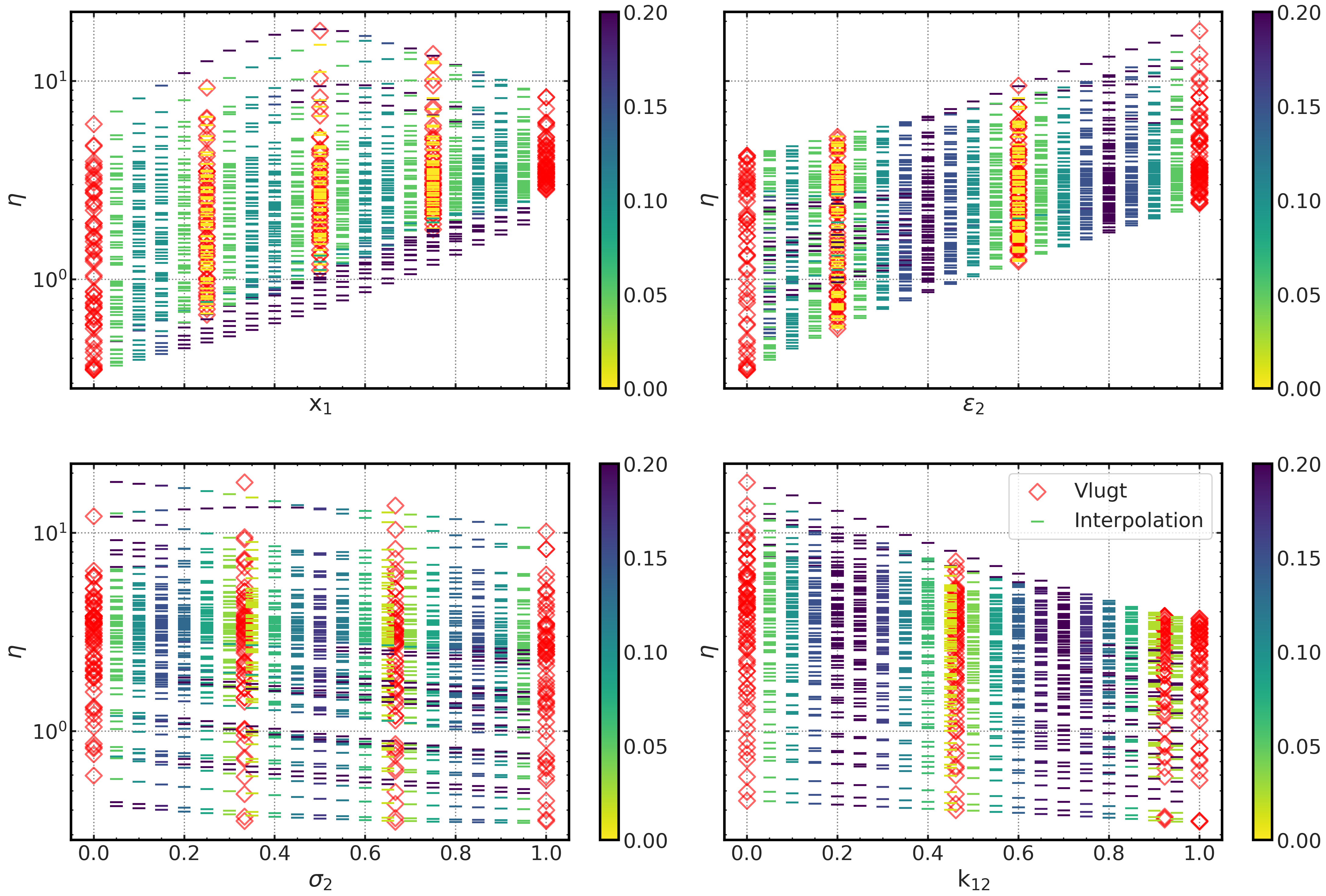}
    \caption{\textbf{ANN model interpolation:} Predicted viscosity values from the
ensemble of ANN in the X$_1$, $\sigma_2$, $\epsilon_2$, and k$_{12}$ interpolation range. The red diamonds are viscosity values from the Vlugt data set. The color of the dashes indicate the distance from the nearest Vlugt data point in the scaled feature space. See Computational Methods section for details on feature scaling and the construction of the interpolation grid.} \label{sfig:modelinterp-ann}
\end{figure}

\clearpage
\subsubsection{GPR}
\begin{figure}[h!]
\centering
    \includegraphics[width=\linewidth]{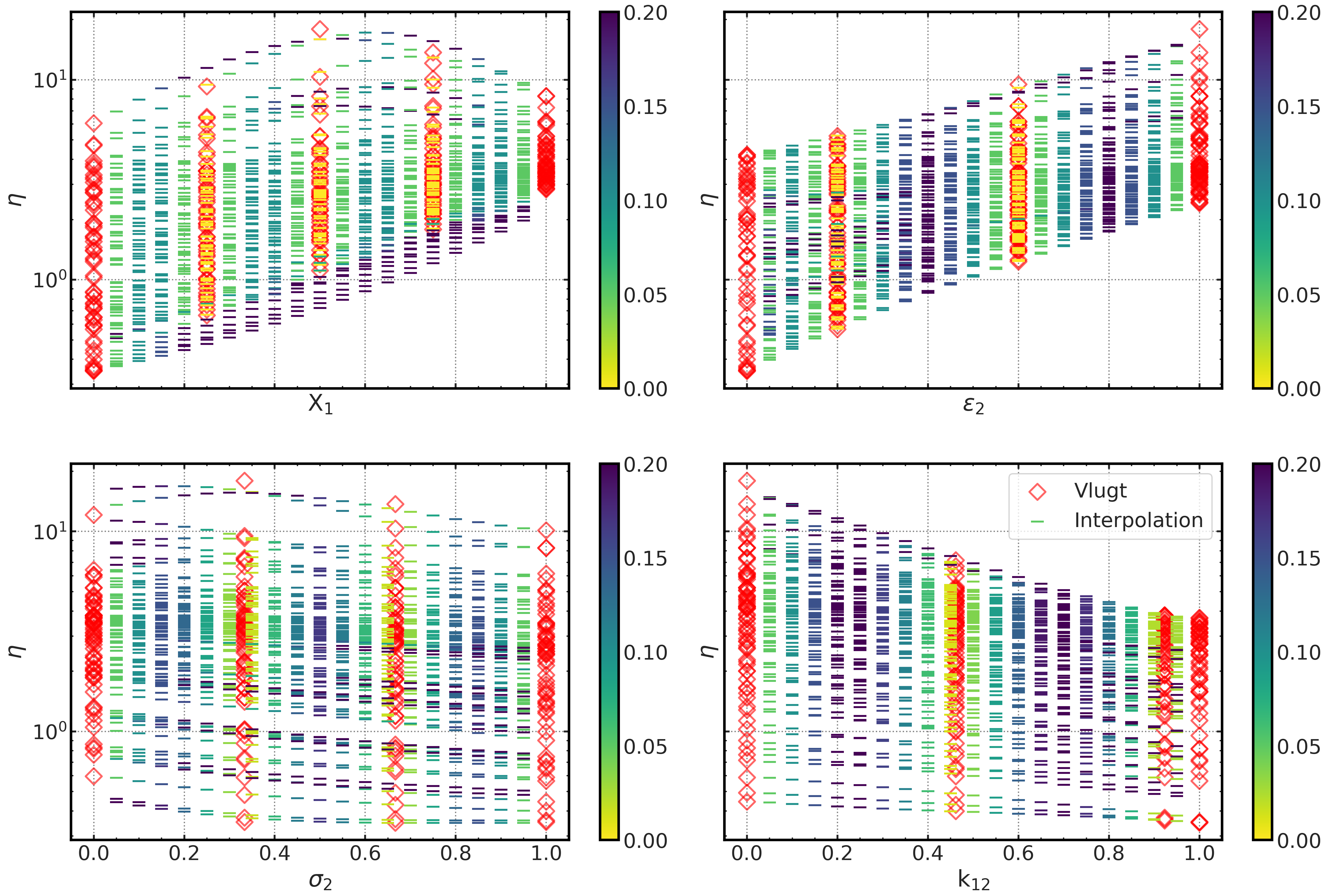}
    \caption{\textbf{GPR model interpolation:} Predicted viscosity values from the
ensemble of GPR in the X$_1$, $\sigma_2$, $\epsilon_2$, and k$_{12}$ interpolation range. The red diamonds are viscosity values from the Vlugt data set. The color of the dashes indicate the distance from the nearest Vlugt data point in the scaled feature space. See Computational Methods section for details on feature scaling and the construction of the interpolation grid.} \label{sfig:modelinterp-gpr}
\end{figure}

\clearpage
\subsubsection{KRR}
\begin{figure}[h!]
\centering
    \includegraphics[width=\linewidth]{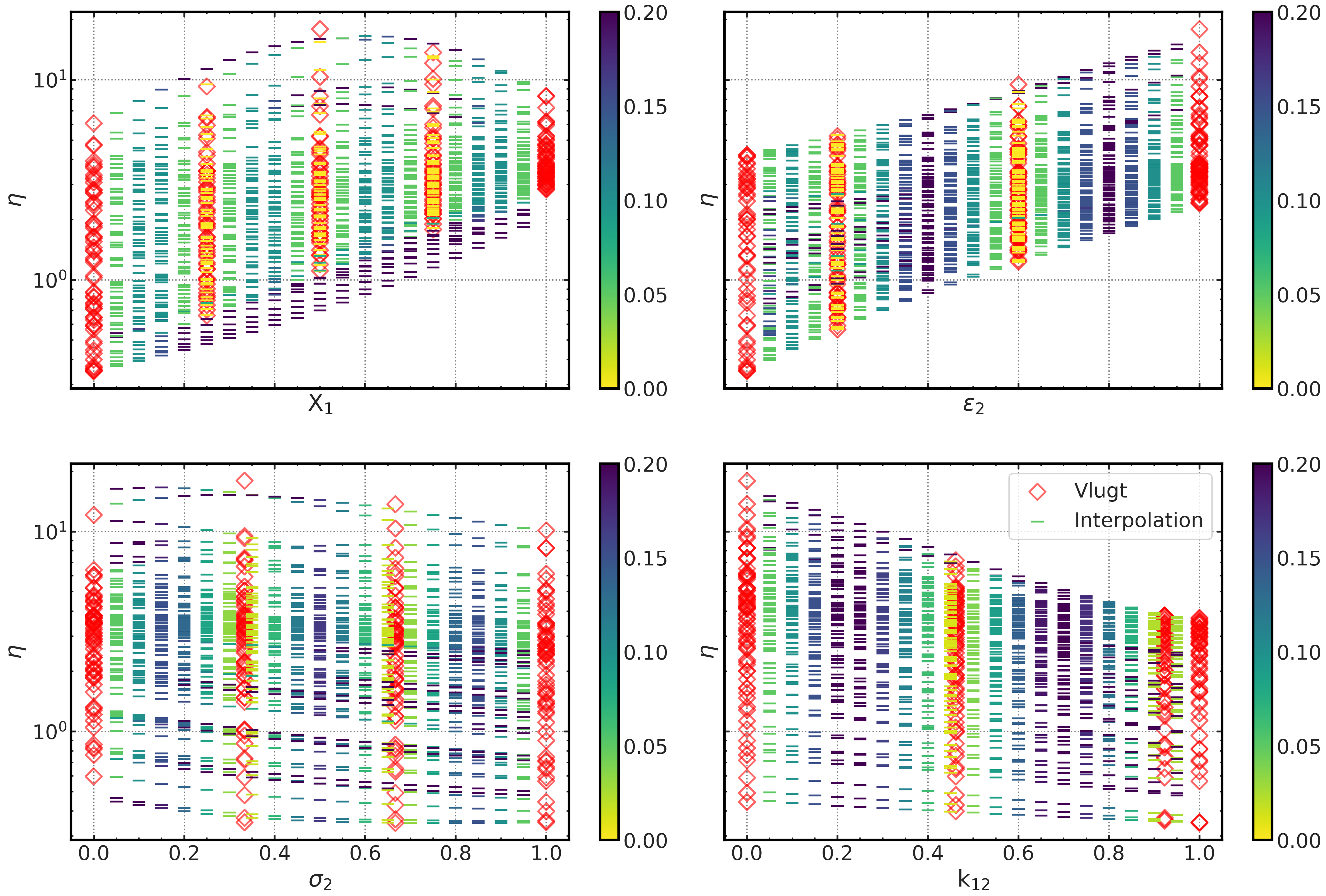}
    \caption{\textbf{KRR model interpolation:} Predicted viscosity values from the
ensemble of KRR in the X$_1$, $\sigma_2$, $\epsilon_2$, and k$_{12}$ interpolation range. The red diamonds are viscosity values from the Vlugt data set. The color of the dashes indicate the distance from the nearest Vlugt data point in the scaled feature space. See Computational Methods section for details on feature scaling and the construction of the interpolation grid.} \label{sfig:modelinterp-krr}
\end{figure}

\clearpage
\subsubsection{SVR}
\begin{figure}[h!]
\centering
    \includegraphics[width=\linewidth]{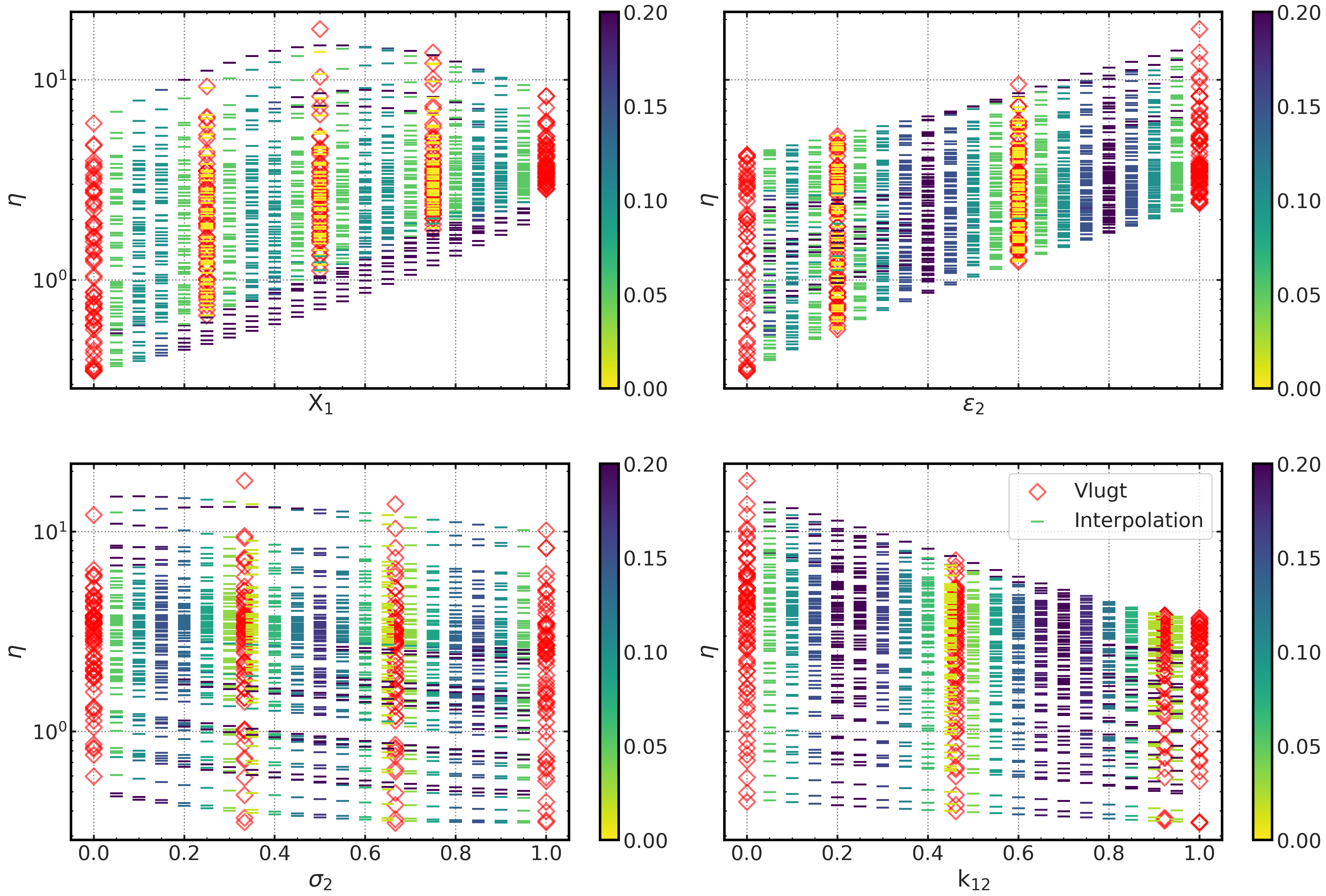}
    \caption{\textbf{SVR model interpolation:} Predicted viscosity values from the
ensemble of SVR in the X$_1$, $\sigma_2$, $\epsilon_2$, and k$_{12}$ interpolation range. The red diamonds are viscosity values from the Vlugt data set. The color of the dashes indicate the distance from the nearest Vlugt data point in the scaled feature space. See Computational Methods section for details on feature scaling and the construction of the interpolation grid.} \label{sfig:modelinterp-svr}
\end{figure}

\clearpage
\subsubsection{RF}
\begin{figure}[h!]
\centering
    \includegraphics[width=\linewidth]{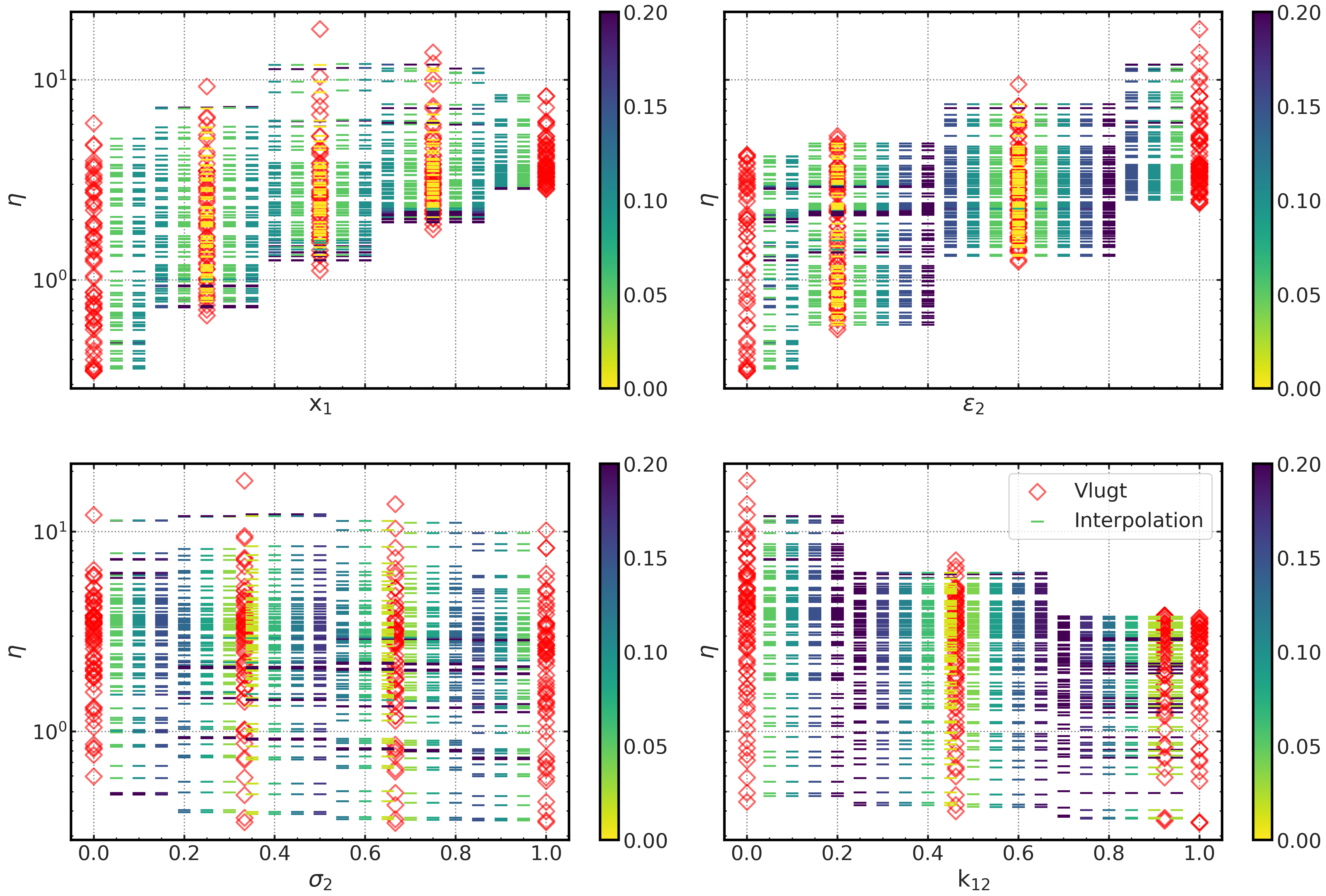}
    \caption{\textbf{RF model interpolation:} Predicted viscosity values from the
ensemble of RF in the X$_1$, $\sigma_2$, $\epsilon_2$, and k$_{12}$ interpolation range. The red diamonds are viscosity values from the Vlugt data set. The color of the dashes indicate the distance from the nearest Vlugt data point in the scaled feature space. See Computational Methods section for details on feature scaling and the construction of the interpolation grid.} \label{sfig:modelinterp-rf}
\end{figure} 

\clearpage
\subsubsection{KNN}
\begin{figure}[h!]
\centering
    \includegraphics[width=\linewidth]{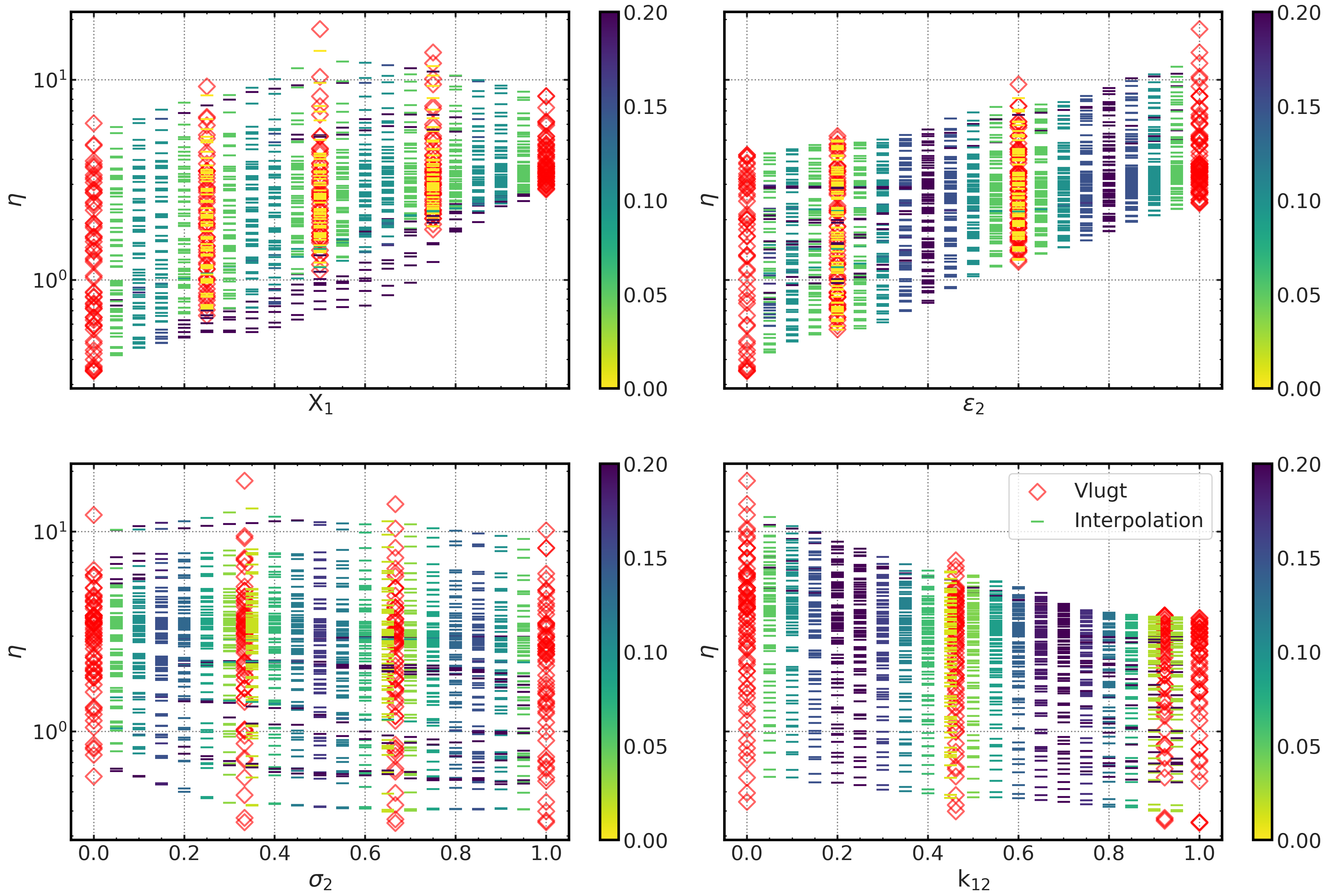}
    \caption{\textbf{KNN model interpolation:} Predicted viscosity values from the
ensemble of KNN in the X$_1$, $\sigma_2$, $\epsilon_2$, and k$_{12}$ interpolation range. The red diamonds are viscosity values from the Vlugt data set. The color of the dashes indicate the distance from the nearest Vlugt data point in the scaled feature space. See Computational Methods section for details on feature scaling and the construction of the interpolation grid.} \label{sfig:modelinterp-knn}
\end{figure} 

\clearpage
\subsubsection{LASSO}
\begin{figure}[h!]
\centering
    \includegraphics[width=\linewidth]{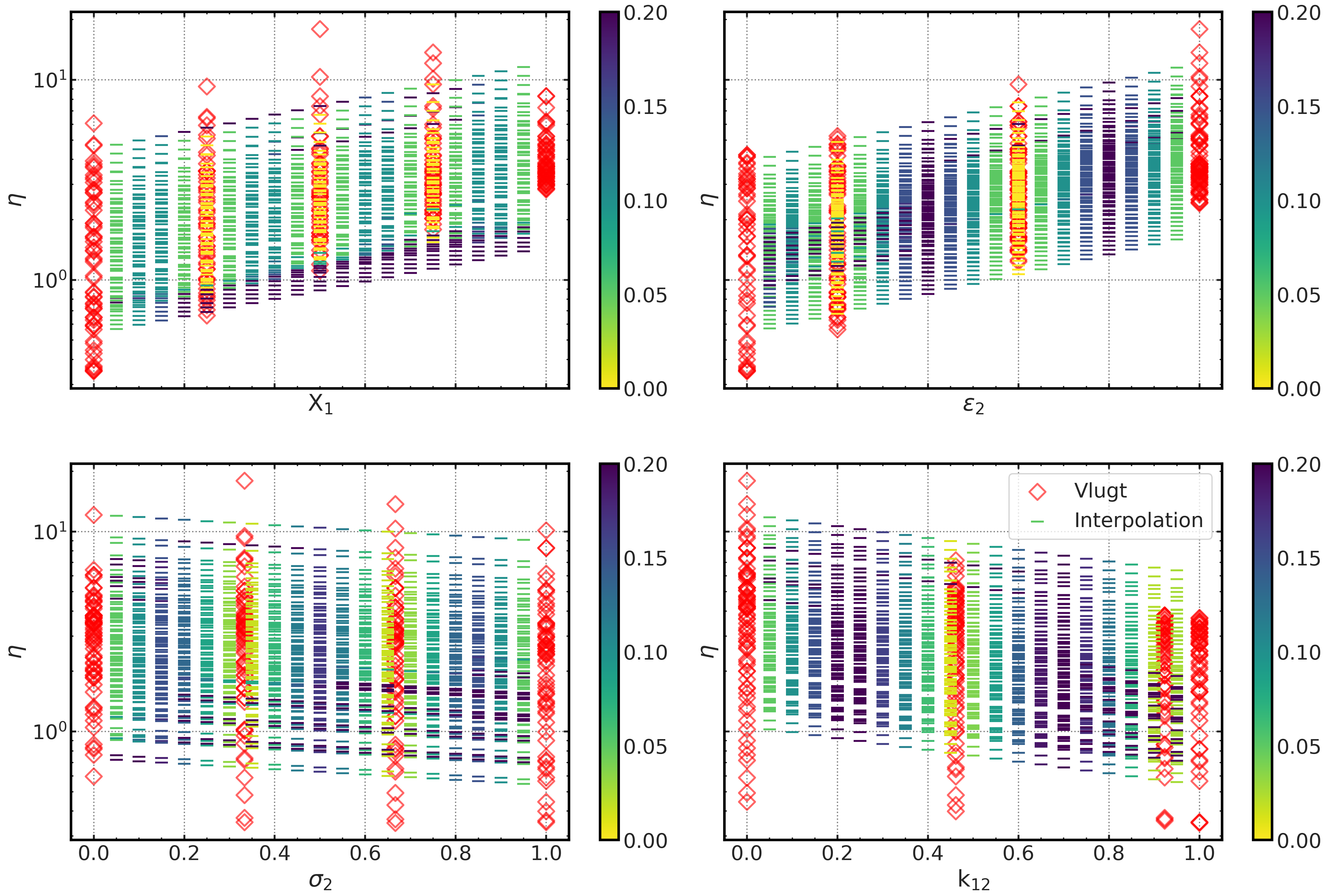}
    \caption{\textbf{LASSO model interpolation:} Predicted viscosity values from the
ensemble of LASSO in the X$_1$, $\sigma_2$, $\epsilon_2$, and k$_{12}$ interpolation range. The red diamonds are viscosity values from the Vlugt data set. The color of the dashes indicate the distance from the nearest Vlugt data point in the scaled feature space. See Computational Methods section for details on feature scaling and the construction of the interpolation grid.} \label{sfig:modelinterp-lasso}
\end{figure}

\clearpage

\bibliography{supplement}